\documentclass[onecolumn]{aastex6} 
\usepackage{multirow} 
\usepackage{booktabs}
\usepackage{lineno}  
\newcounter{magicrownumbers}

\slugcomment{The Astrophysical Journal, in press as of March 3rd, 2022}  
\shorttitle{White Dwarfs With and Without a Mask  
}
\shortauthors{Godon \& Sion}
\watermark{Arxiv Version} 
\setwatermarkfontsize{0.9in}

\begin{document}

\title{{\bf 
White Dwarf Photospheric Abundances in Cataclysmic Variables - II.
White Dwarfs With and Without a Mask 
}} 

\author{Patrick Godon\altaffilmark{1,2} }

\and

\author{Edward M. Sion\altaffilmark{1} } 

\email{patrick.godon@villanova.edu}

\altaffiltext{1}{Department of Astrophysics \& Planetary Science, 
Villanova University, Villanova, PA 19085, USA}
\altaffiltext{2}{Henry A. Rowland Department of Physics \& Astronomy,
The Johns Hopkins University, Baltimore, MD 21218, USA}

\begin{abstract}

Taking advantage of the now available Gaia EDR3 parallaxes, 
we carry out an archival {\it Hubble Space Telescope} (HST) far ultraviolet spectroscopic analysis of 10 
cataclysmic variable systems, including 5 carefully selected eclipsing systems. 
We obtain accurate white dwarf (WD) masses and temperatures, 
in excellent agreement with the masses for 4 of the eclipsing systems.    
For three systems in our sample, BD Pav, HS 2214, and TT Crt,  we report
the first robust masses for their WDs. 
We modeled the absorption lines to derive the WD chemical abundances and 
rotational velocities for each of the ten systems. 
As expected, for five higher inclination ($i \gtrsim 75^{\circ}$) 
systems, the model fits are improved with the  
inclusion of a cold absorbing slab (an iron curtain masking the WD)  
with $N_{\rm H} \approx 10^{20}-10^{22}$cm$^{-2}$.
Modeling of the metal lines in the HST spectra reveals that 7
of the 10 systems have significant subsolar carbon abundance, and six have 
subsolar silicon abundance, thereby providing  further evidence that CV WDs 
exhibit subsolar abundances of carbon and silicon. 
We suggest that strong aluminum absorption lines (and iron absorption features)   
in the spectra of some CV WDs (such as IR Com) may be due to the presence of a 
{\it thin} iron curtain ($N_{\rm H}\approx 10^{19}$cm$^{-2}$) rather than 
to suprasolar aluminum and iron abundances in the WD photosphere. 
The derived WD (projected) rotational velocities all fall in the range 
$\approx 100-400$~km/s, all sub-Keplerian similar to the values obtained in 
earlier studies.

\end{abstract}


\section{{\bf Introduction}}

Cataclysmic variables (CVs) are compact binaries  
in which a white dwarf (WD) accretes matter and angular momentum from
a main-sequence-like star. 
In nonmagnetic CVs, the matter is transferred, at continuous 
or sporadic rates, by means of an accretion disk around the WD. 
Dwarf novae (DNe; a class of nonmagnetic CVs) release gravitational energy when 
a thermal-viscous instability in the accretion disk around the WD           
leads to rapid accretion at a high rate (the dwarf nova {\it outburst}, 
lasting days to weeks) until the disk has largely emptied and
the system returns to {\it quiescence}, at which time the buildup of 
disk gas begins again (lasting weeks to months). 
When a DN system is in quiescence,
the WD is often revealed in the ultraviolet (UV), as its emission greatly 
outshines the disk \citep{war95,fra02}.
\\

Accretion is expected to spin up the WD envelope relative to the rotation of its core \citep{nar89}, 
and over the lifetime of a CV (e.g. $\sim 10^9$yrs), the accretion of 0.1-0.2 $M_{\odot}$ of gas with angular momentum
should spin up the CV WD to its critical {\it breakup} Keplerian rotational velocity \citep{liv98}.  
Interestingly enough, however, most exposed CV WDs reveal a (projected) stellar rotation rate
of only a fraction of the breakup velocity \citep[usually 200-400~km/s, see e.g.][]{sio98,sio99,god12}.  
The efficiency of the coupling between the core and envelope, however, remains poorly understood \citep{yoo04}.  
While nova explosions may spin down the rotation rates to well
below the critical rotation, if the accretion rate is low and the WD
of average or moderate mass, then the recurrence time between novae is longer and 
an accreting WD should possibly have a faster rotation rate than
higher mass accreting WDs. 
Clearly, accurate WD stellar rotational velocities are required to achieve an understanding 
of CV evolution and spin-up of the WDs due to accretion with angular momentum. 
\\ 

On the other hand, the UV spectra of accreted material on the WD can be regarded as a
{\it mass spectrometer} for revealing the composition of the donor and its evolutionary history. 
Suprasolar N/C ratio anomaly has been detected in some systems and it is estimated 
that at least 10 to 15\% of CVs might have a suprasolar N/C ratio \citep[][based on anomalous UV line
flux ratios]{gan03}. 
Infrared analyses of CV secondary stars \citep[e.g.][]{har04,har05,how10} 
have further shown, based on depleted levels of $^{12}$C and enhanced levels of $^{13}$C, 
that material that has been processed in the CNO
cycle is finding its way into the photospheres of secondary stars.  
One possible scenario is that the N/C overabundance anomaly arises from the donor star,
a formerly more massive secondary donor star (capable of CNO burning) having been peeled away by mass
transfer down to its CNO-processed core. 
%
%
An alternative explanation is that the anomaly originates in the WD itself due to explosive 
hot CNO burning associated with nova explosions. 
In that case, the N/C anomaly and suprasolar abundances of heavy elements could be due to the contamination 
of the donor star by nova explosion followed by re-accretion of this material by the WD 
\citep{sio14,spa21}.  
\\

Accurate CV WD masses have been needed to help answer fundamental questions
in CV evolution theories, such as whether a CV WD mass can grow,
reach the Chandrasekhar limit, and explode as a supernova (SN) Ia; 
or which of the different CV evolution theories based on different (binary) 
angular momentum braking laws is correct \citep{sch16,nel16,lau18,mca19,zor20}. 
However, knowing the mass and temperature of a WD is also needed to derive  
the WD surface chemical abundances and (projected) stellar rotational velocities.  
This is because the shape and depth of absorption lines in UV spectra  
depend, not only on the chemical abundances and rotational velocity,  
but also on the temperature and gravity of the WD stellar photosphere.
Consequently, in order to derive chemical abundances and rotational velocities one has 
to first derive the WD surface temperature and mass.
\\

With the era of space UV telescopes, and in particular with the {\it International 
Ultraviolet Explorer} (IUE), the {\it Far Ultraviolet Spectroscopic Explorer} (FUSE), 
and the {\it Hubble Space Telescope} (HST),  significant advances have been made in the
the field of CVs \citep[see e.g.][]{lad91,hac93,war95}.  
Thanks to recent Gaia eDR3 parallaxes \citep{bro21}, further progress can {\it now} be made 
to fully exploit the large number of available far-UV  (FUV, $\lambda < 2000$~\AA ) 
spectra of CVs \citep[see, e.g. the current FUV spectral analysis of ][]{pal21}. 
Scaled to the Gaia distances, model spectra can provide accurate temperature, 
WD radius/masses, as well as accurate WD photospheric abundances and 
rotational velocities for a large number of CV WDs.  
\\

Up to now, there are only 5 systems (all DNe) in which detections of suprasolar
heavy element abundances are manifested by FUV photospheric absorption line 
in the spectra of CV WDs exposed during DN quiescence 
\citep{lon99,gan05}.  All five CV WDs have suprasolar abundances
of nuclides like Al with atomic masses A$>$20. 
The best studied systems are U Gem \citep{sioetal98,lon99,god17} and VW Hyi
\citep{sio95a,sio95b,sio97,lon06,lon09}. 
U Gem WD after an outburst and into mid-quiescence 
has abundances C$\approx$0.30 - 0.35, N$\approx$35 - 41,
and Si$\approx$1.4 - 4, S$\approx$6.6 - 10 \citep{lon99}.  
In quiescence subsolar to solar abundances of C and S were found,
together with suprasolar abundance of N and Al (both up to 20 solar) and
subsolar to suprasolar abundances of Si \citep{god17}. 
For VW Hyi, subsolar (0.4-0.8) abundance of carbon was found with 
suprasolar abundance of N (4-6) and Si (2.5-4.4) and about solar
abundance of S (with rotational velocity on the higher side: 400-570~km/s).  
In three SU UMa-type CVs with exposed WDs, BW Scl, SW UMa, and BC UMa, the
detected photospheric features reveal subsolar abundances of C, O and Si and suggest
substantial suprasolar aluminum abundances
of $3.0\pm 0.8$, $1.7\pm 0.5$, and $2.0\pm 0.5$, respectively
\citep{gan05}. 
All three of these CV WDs with UV absorption lines revealing suprasolar abundances of aluminum are 
too cool ($< $20,000K) for radiative acceleration to be a factor
in the overabundance. The central temperatures of any formerly more massive
secondary stars in CVs undergoing hydrostatic
CNO burning are far too low to produce these suprasolar abundances \citep{ili18}.  
           
In the present analysis, we derive chemical abundances and projected stellar rotational 
velocities for 10 CV WDs, by carrying out an FUV spectral analysis of HST spectra.  
Our current archival FUV spectral analysis of CV WDs           
was initiated with the analysis of SS Aur and TU Men 
in \citet{god21}, for which we found subsolar carbon and silicon 
abundances, with suprasolar nitrogen abundance for TU Men.  
We further improved our method and refined our analysis in the study of V386 Ser
\citep{szk21}, and here we now implement our methodology with the inclusion of an iron curtain 
needed for highly inclined and/or eclipsing systems.  
The results of the present analysis indicate that a majority (6-7 out of 10) 
of systems have a WD with subsolar carbon and silicon abundances.  
The iron curtain modeling introduces an additional degree of complexity that  
renders that analysis more difficult, but may point to subsolar phosphorus as well.    
We also discuss the possibility of a very thin iron curtain which may affect
the WD abundances analysis, and must therefore be taken into account.  
 
In the next section we introduce the systems we selected for this analysis, 
the archival HST spectra are reviewed in Section 3, the analysis and tools are 
presented in section 4, the results are discussed in Section 5, and we 
conclude with a summary and further considerations in the last section.

\section{{\bf The Systems}}  

For the present work, we selected 10 DN systems observed in quiescence,  
all with an HST (STIS or COS) spectrum dominated by emission from the WD, 
and all with a Gaia eDR3 parallax.  
This selection of systems covers a large range of orbital periods
(from 1h22 min to 11h; about half of the systems are above the period gap) 
and WD temperatures  (from $\sim$11,000~K to $\sim$35,000~K). 
Five of the systems show WD eclipses: SDSS 1035, IY UMa, DV UMa, IR Com, 
and GY Cnc. BD Pav does not show WD eclipse, but instead it 
displays disk eclipses during outburst \citep{kim18}.   
The systems are listed in Table \ref{syspar} using their {\it Simbad} name, 
together with their parameters and references. 
HS 2214+2845 is also known as V513 Peg.    
CRTS J153817.3+512338 is also known as SDSS J153817.35+512338.0. 
For convenience, in the text we refer to the two SDSS objects
with their 8 or 4 digits only, i.e. SDSS 1538+5123 or just SDSS 1538 for short. 

A few  systems were observed about a week after the end of an outburst 
(HS 2214+2845, TT Crt, V442 Cen) and likely exhibit a still elevated WD temperature. 
The eclipsing systems SDSS 1035 \citep{sav11}, IY UMa, DV UMa 
and GY Cnc  \citep{mca19} were chosen because they have accurate WD masses 
(as derived by the above authors) to check and confirm the WD masses
we derive from the UV fit.  We also analyze IR Com, which is also eclipsing. 

This is the first large sample (10 objects or more) from our current ongoing study, 
which, by including cold and hot WDs above and below the gap,   
is representative of DNe WDs in general. It is also representative of 
most of the HST DATA obtained for CVs, as it consists of 4 STIS spectra and 6 COS spectra.

\begin{deluxetable}{lllccllll}[h!] 
\tablewidth{0pt}
\tablecaption{System Parameters
\label{syspar} 
} 
\tablehead{
~~~(1)       & ~(2)  &  ~~(3)          &  (4)    &  (5)             & ~~(6)                        & ~~~(7)            & ~~~(8)         & (9)     \\  
System       & Type  & ~~$P_{\rm orb}$ &  $i$    & $\Pi$            & ~~~~$d$                      & $E(B-V)$          & ~~$M_{\rm wd}$ &  $K_1$      \\       
~Name        &       & ~(days)         & (deg)   & (mas)            & ~~(pc)                       &                   & ~~($M_{\odot}$) & km~s$^{-1}$        
}
\startdata
SDSS J103533.02+055158.4 & DN & 0.0570067 &$83.98\pm0.08$& $5.1220 \pm 0.2865$ & $195^{+12}_{-10}$ & $0.017 \pm 0.017$ & $0.835\pm0.009$           & 28.5 \\[3pt]   
CRTS J153817.3+512338 & DN  & 0.06466  &             & $1.6322 \pm 0.1161$ & $613^{+47}_{-41}$     & $0.010 \sim0.020$ &                           &      \\[3pt] 
IY UMa & DN SU & 0.0739089282 & $84.9^{+0.1}_{-0.5}$    & $5.4931 \pm 0.0674$ & $182 \pm 2$        & $0.015 \pm 0.015$ & $0.955^{+0.013}_{-0.027}$ & 66   \\[3pt]
DV UMa & DN SU & 0.0858526308 & $83.29^{+0.29}_{-0.10}$ & $2.5859 \pm 0.1544$ & $387^{+24}_{-22}$  & $0.008 \sim0.020$ & $1.09 \pm 0.03$           & 76   \\[3pt] 
IR Com       & DN    & 0.087039 & $ > 75 $    & $4.6001 \pm 0.0646$ & $217 \pm 3$                  & $0.019 \pm 0.022$ & $\approx 0.8-1.0$         & 77   \\[3pt] 
GY Cnc & DN UG & 0.175442399  & $77.06^{+0.29}_{-0.18}$ & $3.6353 \pm 0.0590$ & $275^{+5}_{-4}$    & $0.023 \pm 0.013$ & $0.881\pm0.016$           & 125  \\[3pt]  
BD Pav       & DN UG & 0.179301 &   71-75     & $3.0034 \pm 0.0226$ & $333 \pm 3$                  & $0.057 \pm 0.018$ & $\approx 0.8 - 1.04$      &  95  \\[3pt] 
HS 2214+2845  & DN    & 0.179306 &             & $2.5006 \pm 0.0496$ & $400 \pm 8$                  & $0.052 \pm 0.023$ &                           &      \\[3pt] 
TT Crt       & DN UG & 0.2683522& $52<i<70$   & $1.8271 \pm 0.0438$ & $547^{+14}_{-12}$            & $\sim0.01 \sim0.029$ & $> 0.7$                & 212  \\[3pt]
V442 Cen     & DN UG & 0.46     &             & $2.8758 \pm 0.0437$ & $348 \pm 5$                  & $0.048 \pm 0.015$ &                           &      \\[3pt]
\enddata
\tablecomments{
The system names (column \#1) are the ones by which the systems appear in {\it Simbad}.  
The CV type (column \#2) are as defined in \citet{rit03}.  
The periods (\#3), inclinations (\#4) and WD masses (\#8) were taken from 
\citet[][BD Pav]{fri90},\citet[][IY UMa, DV UMa, GY Cnc]{mca19}, \citet[][IR Com]{man14},
\citet[][TT Crt]{tho04},    
and \citet[][SDSS 1035]{sav11}.  
The Gaia Early Data Release 3 (EDR3) parallaxes \citep{bro21,ram17,lin18,lur18} are listed in column \#5. 
The distances (\#6) were derived from the Gaia parallax as explained in the text. 
The reddening values (\#7) were obtained from \citet{cap17} using the Gaia distances,
or from the NASA/IPAC online Galactic Dust Reddening and Extinction Map \citep{sch98,sch11} 
for those systems beyond the distance range of \citet{cap17}. 
The WD velocity amplitude $K_1$ (column 9) were taken or derived from \citet{har05,sav11,mca19,man14}.   
} 
\end{deluxetable}

To derive the distance from the Gaia parallax, we  
follow \citet{sch18}. Namely,                           
we use the probability distribution of the
distance $d$ (to the system) given by \citet[][equation 18]{bai15}, where 
$\varpi$ is the Gaia parallax, and  
$\sigma_{ \varpi }$ is its error.  
We then integrate the expression (over the distance $r$) to find the 
1-sigma intervals containing the central 68.3\% probability.  
Doing so, we find that for most systems, as long as the Gaia distance
is short with a small error 
(say $\sigma_{\varpi} / \varpi \sim 0.1$ or so), 
the distance with its errors can also be obtained by simply 
inverting the parallax, namely, $d=1000/ (\varpi \pm \sigma_{\varpi})$.

\section{{\bf The Archival Data}} 

Six systems (SDSS 1035, SDSS 1538, IY UMa, IR Com, BD Pav, and HS 2214+2845)
have archival COS spectra, and the remaining 4 (DV UMa, GY Cnc, TT Crt, 
and V442 Cen) have archival STIS spectra (see Table \ref{obslog}).  

The COS instrument was set in the FUV configuration with the
G140L grating centered at 1105~\AA\ (with the \texttt{PSA} aperture), 
generating a spectrum from $\approx 1110$~\AA\ to 2150~\AA ,  
with a resolution of $R \sim 3000$.  The data were collected in
\texttt{TIME-TAG} mode during one or more HST orbits, each consisting of 
4 subexposures (obtained on 4 different {\it positions}  on the
detector).   
Originally, these spectra were obtained as part of a large research
program \citep[see ][]{pal17}.  

The STIS instrument was set up in the FUV configuration with the G140L grating
centered at 1425 \AA\ (with the 52"x0.2" aperture), 
thereby producing a spectrum from $\sim$1140 \AA\ to 
$\sim$1715 \AA\  (with a spectral resolution of $R\sim 1000$).  
The data were collected in \texttt{ACCUM} mode and consist of 
one echelle spectrum only, taken in the \texttt{SNAPSHOT} mode. 
As a consequence, the total good exposure time of the STIS spectra
is about 700-900~s. These STIS spectra were obtained as part of 
previous research program \citep[see ][]{gan03,sio08}.  

The archival data were retrieved directly from \textsc{mast},
and consequently were processed through the pipeline 
with \textsc{calcos} version 3.3.10 or \textsc{calstis} version 3.4.2. 
For the STIS data, we extracted single spectra, for the COS
data, we extracted all the subexposures as well as the combined
spectra. The STIS (single exposure) spectra and the COS
(co-added) spectra are presented in Figures 
\ref{sdss1035cos} -  \ref{v442censtis}.  The COS spectra 
have a higher resolution and are displayed on two panels each,
while the STIS spectra with their lower resolution are 
displayed on a single panel each.

When possible, we determined the orbital phase at which each 
spectrum/subexposure was collected using the ephemerides 
from
\citet[][TT Crt]{tho04},  
\citet[][GY Cnc, IR Com]{fel05},
\citet[][SDSS 1035]{sav11}, 
\citet[][BD Pav]{kim18}, 
and \citet[DV UMa, GY Cnc, IY UMa]{mca19}. 
The orbital phase at which each exposure was obtained is displayed
in the last column of Table \ref{obslog}, it indicates the 
middle of the exposure. We use the usual notation where $\Phi = 0.0$ 
corresponds to the inferior conjunction of the secondary. 
In the results section we address the orbital phase of the exposures
when needed, and which subexposures were used or discarded.  

We use the AAVSO light curve generator to check the state (quiescence/outburst) 
in which the systems were at the time of the observations. 
All the data were collected as the systems were in quiescence.
However, GY Cnc was observed about 2-3 weeks after an outburst, 
while HS 2214+2845 and TT Crt were observed about 1
week after an outburst. V442 Cen was observed about one week
after showing some optical variability. As a consequence, the 
temperature of the WD in these systems might be elevated, but 
we still expect emission from the disk to be minimal or negligible.  

In preparation for the fitting, we deredden the spectra  
assuming the values of the reddening given in Table \ref{syspar}. 
We use the analytical expression of \citet{fit07} for the extinction curve, 
which we slightly modified to agree with an extrapolation 
of the standard extinction curve of \citet{sav79} in the FUV range.
It was shown by \citet{sas02} that the observed extinction curve is actually 
consistent with an extrapolation of the standard extinction curve of \citet{sav79} 
in the FUV range \citep[see also ][]{sel13}. For all the objects in the present
study, the reddening is actually very small ($< 0.06$).  
To assess how the error of the reddening value propagates into an error on the 
derived temperature and gravity of the WD, we need only to consider two
reddening values: the largest and the smallest reddening values, namely, 
the reddening value plus its error and the reddening value minus its error.
We then average the two values (each) of $log(g)$ and $T_{\rm wd}$ (obtained
for the two reddening values assumed) yielding to the final result with
the error bars due to the uncertainty in the reddening (this is explicitly
carried out in the results section for SDSS 1538).   

Before the spectral fitting we mask all the regions of emission lines
and strong absorption lines. For example, the COS spectra are contaminated
with hydrogen Ly$\alpha$ ($\sim$1216) and  O\,{\sc i} ($\sim$1300) airglow emission.  
The spectra also often display broad emission lines from 
C\,{\sc iv} (1548.20 \& 1550,77),  as well as possibly from 
C\,{\sc iii} (1174.93-1176.37), N\,{\sc v} (1238.82 \& 1242,80), 
Si\,{\sc iv} (1393.76 \& 1402.77), and He\,{\sc ii} (1640.33-1640.53). 
Since we wish to derive the temperature and gravity, we fit,
in a first step, the Ly$\alpha$ wings and the continuum slope of the spectra
and mask the prominent emission and absorption lines.
An accurate fit to the absorption lines is carried out in a second step.

\begin{deluxetable}{lcccccccc}[h!] 
\tablewidth{0pt}
\tablecaption{Archival Data Observation Log    
\label{obslog} 
} 
\tablehead{ 
System      & Instrument   & Filter   & Central     & Date       & Time     & ExpTime   &  DataID &  Orb.Phase  \\     
Name        & /Aperture    & Gratings & $\lambda$(\AA)& yyyy-mm-dd & hh:mm:ss & (s)    &  Exposure &  $\Phi$   
}
\startdata
SDSS1035+0551 &  COS/PSA   & G140L     & 1105        & 2013-03-08 & 08:08:51 & 12282     & LC1VA3010 &         \\[1pt]   
               &           &           &             & 2013-03-08 & 08:08:51 &  1680     & lc1va3mfq & 0.428   \\[1pt] 
               &           &           &             & 2013-03-08 & 09:24:56 &  1154     & lc1va3miq & 0.302   \\[1pt] 
               &           &           &             & 2013-03-08 & 09:45:55 &  1470     & lc1va3moq & 0.589   \\[1pt] 
               &           &           &             & 2013-03-08 & 11:00:40 &  1679     & lc1va3msq & 0.521   \\[1pt] 
               &           &           &             & 2013-03-08 & 11:30:24 &   945     & lc1va3mzq & 0.809   \\[1pt] 
               &           &           &             & 2013-03-08 & 12:36:24 &  2204     & lc1va3n8q & 0.740   \\[1pt] 
               &           &           &             & 2013-03-08 & 13:14:53 &   420     & lc1va3nbq & 0.028   \\[1pt] 
               &           &           &             & 2013-03-08 & 14:12:07 &  2729     & lc1va3njq & 0.960   \\[1pt] 
SDSS1538+5123 & COS/PSA   & G140L     & 1105        & 2013-05-16 & 23:52:15 & 4704      & LC1V30010 &          \\[1pt]                            
               &           &           &             & 2013-05-16 & 23:52:15 & 1032      & lc1v30uqq &       \\[1pt] 
               &           &           &             & 2013-05-17 & 00:11:12 & 1032      & lc1v30v2q &       \\[1pt] 
               &           &           &             & 2013-05-17 & 01:21:25 & 1319      & lc1v30w9q &       \\[1pt] 
               &           &           &             & 2013-05-17 & 01:45:09 & 1320      & lc1v30wgq &       \\[1pt] 
IY UMa         & COS/PSA   & G140L     & 1105        & 2013-03-30 & 00:12:57 & 4195      & LC1VA0010 &        \\[1pt] 
               &           &           &             & 2013-03-30 & 00:12:57 & 991       & lc1va0yeq & 0.069   \\[1pt] 
               &           &           &             & 2013-03-30 & 00:31:13 & 555       & lc1va0yjq & 0.206   \\[1pt] 
               &           &           &             & 2013-03-30 & 01:25:56 & 1311      & lc1va0zcq & 0.779   \\[1pt] 
               &           &           &             & 2013-03-30 & 01:49:32 & 1337      & lc1va0zgq & 0.003   \\[1pt]  
DV UMa      & STIS/0.2X0.2 & G140L     & 1425        & 2004-02-08 & 19:14:19 & 900       & O8MZ36010 & 0.204   \\[1pt]                            
IR Com      & COS/PSA      & G140L     & 1105        & 2013-02-08 & 02:08:21 & 6866      & LC1VA6010 &         \\[1pt]                            
            &              &           &             & 2013-02-08 & 02:08:21 & 1627      & lc1va6pqq & 0.629   \\[1pt] 
            &              &           &             & 2013-02-08 & 03:29:01 & 1749      & lc1va6qjq & 0.281   \\[1pt] 
            &              &           &             & 2013-02-08 & 03:59:55 & 875       & lc1va6qoq & 0.470   \\[1pt] 
            &              &           &             & 2013-02-08 & 05:04:43 & 864       & lc1va6qvq & 0.986   \\[1pt] 
            &              &           &             & 2013-02-08 & 05:21:02 & 1750      & lc1va61yq & 0.175   \\[1pt] 
GY Cnc         & STIS/0.2x0.2 & G140L  & 1425        & 2004-04-29 & 17:34:43 & 830       & O8MZ06010 & 0.246   \\[1pt]
BD Pav      & COS/PSA      & G140L     & 1105        & 2013-06-14 & 05:49:13 & 7375      & LC1V17010 &         \\[1pt] 
            &              &           &             & 2013-06-14 & 05:49:13 & 2088      & lc1v17aeq & 0.934   \\[1pt] 
            &              &           &             & 2013-06-14 & 07:17:06 & 1753      & lc1v17aiq & 0.264   \\[1pt] 
            &              &           &             & 2013-06-14 & 07:48:04 & 895       & lc1v17alq & 0.356   \\[1pt] 
            &              &           &             & 2013-06-14 & 08:52:44 & 876       & lc1v17asq & 0.606   \\[1pt] 
            &              &           &             & 2013-06-14 & 09:09:15 & 1762      & lc1v17avq & 0.698   \\[1pt] 
HS 2214+2845 & COS/PSA      & G140L     & 1105        & 2013-07-18 & 21:28:42 & 4680      & LC1V35010 &         \\[1pt]                            
            &              &           &             & 2013-07-18 & 21:28:42 & 1033      & lc1v35uxq & 0.312   \\[1pt] 
            &              &           &             & 2013-07-18 & 22:30:50 & 1034      & lc1v35v4q & 0.522   \\[1pt] 
            &              &           &             & 2013-07-18 & 22:49:59 & 1306      & lc1v35v6q & 0.635   \\[1pt] 
            &              &           &             & 2013-07-19 & 00:34:33 & 1306      & lc1v35veq & 0.041   \\[1pt] 
TT Crt      & STIS/52X0.2  & G140L     & 1425        & 2003-02-12 & 05:12:41 & 700       & O6LI1K010 & 0.293    \\[1pt]                           
V442 Cen    & STIS/52X0.2  & G140L     & 1425        & 2002-12-29 & 20:55:00 & 700       & O6LI1V010 &          \\[1pt]                           
\enddata
\end{deluxetable}

\clearpage

\begin{figure} 
\vspace{-4.5cm} 
\gridline{ 
        \fig{zsdss1035_cos.ps}{0.44\textwidth}{} 
}  
\vspace{-1.0cm} 
\caption{
The COS spectrum (in red) of the eclipsing system SDSS 1035 is 
displayed with flux errors (in black) and line identifications.
The spectrum exhibits emission lines from C\,{\sc iii} ($\sim$1175),
N\,{\sc v} ($\sim$1240), C\,{\sc iv} ($\sim$1550) and He\,{\sc ii} ($\sim$1640).
The hydrogen Ly$\alpha$ (1216) and O\,{\sc i} (1300) 
broad emission lines are due geocoronal (airglow) emission present in most COS spectra, 
they are marked with a cross inside a circle.  
There are also some emission features around 1200~\AA . 
At wavelength $ \lambda < 1200$~\AA , except for emission features, 
the flux is nearly zero and dominated by noise.  
In spite of the long exposure time ($\sim$12,000 s), the overall S/N
is not very high.  
\label{sdss1035cos}
}
\vspace{-4.5cm} 
\gridline{ 
        \fig{zsdss1538_cos.ps}{0.44\textwidth}{} 
}  
\vspace{-1.0cm} 
\caption{
The COS spectrum of CRTS J153817.3+512338 (SDSS 1538) is displayed 
with its error and line identifications. It has some weak  
and moderately broad emission lines of N\,{\sc v} $\sim1240$~\AA , 
Si\,{\sc iv} ($\sim$1400 \AA), He\,{\sc ii} ($\sim$1640 \AA ), 
and strong C\,{\sc iv} ($\sim$1550 \AA ) emission. Many Si and C absorption
lines appear to have some weak emission wings. 
\label{sdss1538cos}  
}
\end{figure}

\clearpage 

\begin{figure} 
\vspace{-4.5cm} 
\gridline{ 
          \fig{ziyuma_cos.ps}{0.44\textwidth}{}  
} 
\vspace{-1.0cm}  
\caption{
The COS spectrum of the eclipsing system IY UMa is 
displayed with its error and some line identifications.
The spectrum does not exhibit any obvious emission lines and shows
all the signs of strong veiling. The presence of an 
``iron curtain'' is recognizable
by its strong Fe\,{\sc ii} absorption bands near 1565~\AA\ and 1635~\AA ,
and a multitude (``a forest'') of absorption lines over the entire spectral
range.  
\label{iyumacos}
}
\vspace{-9.0cm} 
\gridline{ 
          \fig{zdvuma_stis.ps}{0.30\textwidth}{}  
}
\vspace{-1.0cm}  
\caption{
The STIS spectrum of DV UMa is shown with
prominent line identifications. The spectrum is a 900~s single
exposure ``snapshot'' with a rather large noise (in black), 
especially near the detector's edges. Here too, absorption features
produce bands ($\sim1560-1590$~\AA\ and $\sim1635-1655$~\AA ) indicating
the presence of cold veiling material (iron curtain).  
\label{dvumastis}
}
\end{figure}

\clearpage 

\begin{figure} 
\vspace{-4.5cm} 
\gridline{
          \fig{zircom_cos.ps}{0.44\textwidth}{} 
}
\vspace{-1.cm}  
\caption{
The COS (G140L, 1105~\AA ) spectrum of IR Com is displayed 
with its error and line identifications. This spectrum 
does not exhibit any emission lines nor apparent iron curtain absorption.
As such, and with its silicon and carbon absorption lines, as well as 
quasi-molecular hydrogen satellite line ($\sim 1400$), 
it is the {\it clearest} spectrum in this study
representative of a WD. The bottom of the Ly$\alpha$ does not
go down to zero, indicating the presence of a second (hotter)
component (which is also seen in many of the other spectra).  
\label{ircomcos}
}
\vspace{-9.cm} 
\gridline{
          \fig{zgycnc_stis.ps}{0.30\textwidth}{}  
}
\vspace{-1.cm} 
\caption{
The STIS spectrum of the eclipsing system GY Cnc    is 
displayed with flux errors and line identifications.
The spectrum exhibits emission lines from C\,{\sc iii} (1173),
Si\,{\sc iv} ($\sim$1400), and C\,{\sc iv} ($\sim$1550). 
Strong veiling by an ``iron curtain'' is marked by the multitude of
absorption lines and 
by its strong Fe\,{\sc ii} absorption bands near 1565~\AA\ and 1635~\AA .  
\label{gycncstis} 
}
\end{figure}

\clearpage 

\begin{figure} 
\vspace{-4.5cm} 
\gridline{ 
          \fig{zbdpav_cos.ps}{0.44\textwidth}{}  
} 
\vspace{-1.0cm} 
\caption{
The COS spectrum of BD Pav is displayed 
with its error and line identifications. It is characterized
by some strong and broad emission lines (such as N\,{\sc v} $\sim1240$~\AA , 
Si\,{\sc iv} $\sim 1400$~\AA ), and many absorption lines,
including a rather large number of Al\,{\sc ii+iii} lines.  
In order to further identify and differentiate between absorption
and emission lines, one  
needs to carry out a spectral modeling and analysis.  
\label{bdpavcos}  
}
\vspace{-4.5cm} 
\gridline{ 
          \fig{zhs2214+2845_cos.ps}{0.44\textwidth}{}  
} 
\vspace{-1.0cm} 
\caption{
The COS spectrum of HS 2214+2845 is shown 
together with its error. The most prominent absorption
lines are marked. As for IR Com, this spectrum clearly shows 
a WD spectrum. We display this spectrum including the longer
wavelength range to show how the noise (in black) is as large
as the signal (in red) when approaching 2000~\AA . This is
the case for all the COS spectra. 
\label{hs2214cos} 
} 
\end{figure}

\clearpage 

\begin{figure} 
\vspace{-9.cm} 
\gridline{
          \fig{zttcrt_stis.ps}{0.30\textwidth}{}  
}
\vspace{-1.cm} 
\caption{
The STIS snapshot spectrum of TT Crt
is shown with its error and line identifications.
This spectrum does not show any emission line but it does
exhibit absorption lines of higher ionization species in 
addition to the more usual lower ionization species.  
\label{ttcrtstis} 
}
\vspace{-9.cm} 
\gridline{
          \fig{zv442cen_stis.ps}{0.30\textwidth}{}  
}
\vspace{-1.cm} 
\caption{
The STIS snapshot spectrum of V442 Cen  
is shown with its error and line identifications.
This spectrum does not exhibit many absorption lines 
and is rather featureless, except for its Ly$\alpha$
absorption feature and N\,{\sc v} ($\sim$1240) and C\,{\sc iv} ($\sim$1550)
emission lines.  
\label{v442censtis} 
}
\end{figure}

\clearpage

\section{{\bf Analysis Tools and Technique}}

\subsection{{\bf Stellar Atmosphere Models with TLUSTY}}  

Our main spectral analysis tool is \textsc{tlusty} \citep{hub88,hub95}, which we
use to generate theoretical spectra of white dwarfs. 
The code (version 203) includes hydrogen quasi-molecular satellite lines opacities 
(which is required for low temperature high gravity photospheres),  
NLTE approximation, rotational and instrumental broadening, and limb darkening. 
The latest documentation on \textsc{tlusty} is given in \citet{hub17a,hub17b,hub17c}. 
In the following, we only concentrate on the inclusion of an iron curtain
in the modeling of the WD.  We refer the readers to our previous
works \citep{god21,szk21} and to Ivan Hubeny's above publications for further 
details. 

\subsection{{\bf The $\chi^2$ Map}} 

Using \textsc{tlusty}, we built a grid of solar composition stellar photospheric spectra 
in a region of the $(T_{\rm wd},log(g))$ parameter space. 
The temperature $T_{\rm wd}$ ranges from 10,000~K to 40,000~K in steps of 250~K, 
and surface gravity $log(g)$ from $log(g)=7.0$ to $log(g)=9.0$ 
in steps of 0.1.  
We then fit each observed HST spectrum to the above grid of theoretical 
stellar spectra  using the $\chi^2$ minimization technique.
Doing so, we obtain a reduced $\chi^2_{\nu}$ 
($\chi^2$ per degree of freedom $\nu$) for each model in the grid as a 
function of $log(g)$ and  $T_{\rm wd}$.  
A distance $d$ is also obtained for each grid model fit  
by scaling the model spectrum to the observed spectrum, 
assuming a WD radius given by the non-zero temperature C-O WD 
mass-radius relation from \citet{woo95} for a given $log(g)$ and $T_{\rm wd}$. 
Therefore, fitting an observed spectrum to the grid of model
spectra yields values of $\chi^2_{\nu}$ and $d$ as a function of 
$log(g)$ and $T_{\rm}$:  
\begin{equation}
\left \{ 
\begin{array}{ll}  
\chi^2_{\nu} \equiv \chi^2_{\nu} ( log(g),T_{\rm wd} ) , \\ 
d \equiv d ( log(g), T_{\rm wd} ). 
\end{array}
\right. 
\end{equation}         
The results are then summarized as a map of $\chi^2_{\nu}$ in the parameter
space $log(g)$ vs. $T_{\rm wd}$. Such $\chi^2$-maps are presented in
the section A of the Appendix in Figs.A\ref{sdss1035chi} through
A\ref{v442cenchi}. 
The best-fit for the given Gaia distance ($\pm$ errors) is then found
where $\chi^2_{\nu}$ reaches a  minimum along the line $d=d_{\rm Gaia}$
in the $\chi^2$-map. The coordinates of this point in the $\chi^2$-map
gives the best-fit $log(g)$ and $T_{\rm wd}$.   
Namely, we use the $\chi^2$-map to derive the best-fit WD gravity and temperature
when fitting a spectrum.  Further details are given in Appendix A. 

\subsection{{\bf Absorbing Slabs with SYNSPEC and CIRCUS}}  

Some systems have a relatively high inclination and show signs
that the WD is veiled by material above the disk (e.g. 
due to the $L1$-stream flowing over the rim of the disk).  
Such veiling is often referred to as the {\it iron curtain} \citep{hor94}, as it 
produces very strong absorption bands at wavelengths 
$\sim 1565$~\AA\ and $1635$~\AA , due to a forest of Fe\,{\sc ii} absorption
lines (see Fig. \ref{dvumastis}).  
In the extreme case, the spectrum is affected 
by a multitude of absorption lines at almost all wavelengths  
(see Figs. \ref{iyumacos} \& \ref{gycncstis}).  
We found {\it a posteriori} that some absorption bands (due to veiling) 
are also observed at short wavelengths,
near 1130~\AA\ \& 1145~\AA (Fig.\ref{bdpavcos}).   
In order to model the effect of the veiling material, we use \textsc{synspec} 
\citep[][which {\it comes} together with the \textsc{tlusty} package]{hub94,hub11} 
to generate opacity tables for the (cold) veiling material, and   
\textsc{circus} \citep{hub96} to generate a final attenuated spectrum.  
The veiling material is first characterized by its temperature,
electron density, and turbulent velocity input into \textsc{synspec} to obtain
the opacity table. One can also input the chemical
abundance of the veiling material into \textsc{synspec} (which is otherwise
assumed to be solar). The hydrogen column density
is input into \textsc{circus} which allows for partial veiling 
(geometry), as well a as possible radial motion of the veiling
material. Unless otherwise specified, when computing the veiling curtain, 
we assume a complete veiling of the WD  and a zero radial velocity.  
Since most of the absorption due to veiling comes from cold material, 
we assume $T=10,000$~K and $n_e=10^{13}$cm$^{-3}$ as first suggested
by  \citet{hor94}.   
\\

\subsection{{\bf A Second Flat Component}} 

Since the reddening toward all the objects presented here is very small
(as the objects are relatively nearby), the interstellar medium (ISM) absorption is not expected to 
drive the bottom of the Ly$\alpha$ (in the vicinity of 1216~\AA ) 
down to zero (but see the modeling of V442 Cen at end of the Results Section). 
We therefore expect the spectra of cool WDs to be dominated by the Ly$\alpha$
absorption feature going down to zero. 
Though, the center of the Ly$\alpha$ absorption profile is affected
by airglow emission in COS spectra, one can still discern whether  
or not it goes down to zero: see e.g. the adjacent regions on both sides of 
L$\alpha$ where the spectrum flattens in Fig.\ref{ircomcos}.  
If the bottom of the Ly$\alpha$ does not go down to zero, 
it could be due to either an elevated WD temperature or to the 
presence of a second component (the precise nature of which        
is still a matter of debate). 
In the present work, we model such a second component as a flat continuum. 

If we suspect that a second component is present (i.e. if the bottom of the
Ly$\alpha$ does not go down to zero), we carry out the following steps 
to find the best value of the continuum flux level of this second component. 
We first model the spectrum assuming a WD with no second component
(and given the Gaia distance), which yields  
a WD temperature ($T_{\rm wd}$) and gravity ($log(g)$), and a $\chi^2_{\nu}$.  
We then continue and model the same spectrum, but now 
assuming a WD plus a second small (flat) component, 
while we increase, in successive steps, the value of this second component,
until its value matches the bottom of the Ly$\alpha$. 
This yields a series of $\chi^2_{\nu}$ values for all successive values of the second component
we assumed. 
We then chose the value of the second component that gives the smallest $\chi^2_{\nu}$.  
For that reason, the flux values of the second component are round values, 
which do not especially equal the bottom of the Ly$\alpha$ flux level. 
If this value is zero, then there is no need for a second component. 
 
In the modeling, the second component is subtracted from the observed
spectrum, rather than being added to the scaled model. 
This is because the scaling of the model to the distance assumes that the flux     
is emitted all from the WD surface with a given radius.  
Namely, the scaling only takes into account the WD as the source, while 
the size and geometry of the second emitting component are not known.
Therefore, by removing the second component, we only consider emission from the WD.   
\\

\clearpage 

\section{{\bf Results and Discussion}}

\subsection{{\bf SDSS 1538 as a Basic Example}}

We start our spectral analysis with SDSS 1538, 
since this system does not appear to have a high inclination, no eclipses 
are observed, and the spectrum does not show signs of being veiled. Also, 
it appears that the modeling
of the WD does not require the addition of a second component, as is the case for 
7 of the 10 systems. 

\subsubsection{{\bf Deriving the Temperature and Gravity}}  

\begin{figure} 
\epsscale{0.9}  
\plotone{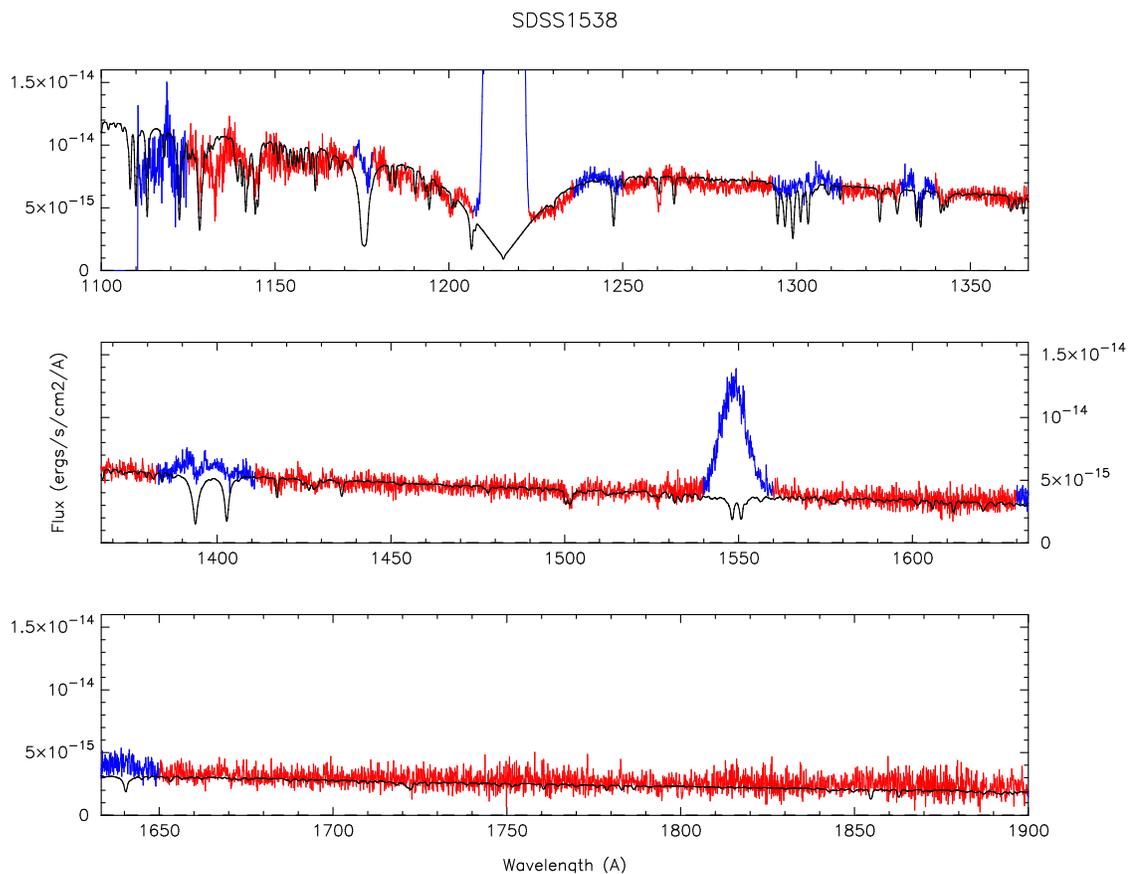} 
\caption{
The spectrum of SDSS 1538 (in red) dereddened assuming E(B-V)=0.010 
has been fitted with a WD model (in black). The model has a temperature
of 35,875~K and gravity $log(g)=8.74$, with solar abundances
and broadening velocity of 200~km/s. This COS spectrum is the  
co-added spectrum of the 4 COS subexposures. 
Regions in blue have been masked before the fitting and correspond to:  
the Ly$\alpha$ (1216) and O\,{\sc i} ($\sim$1300) regions 
with air-glow emission; emission lines (or wings) from  
N\,{\sc iv} ($\sim$1240), C\,{\sc ii} (1335), Si\,{\sc iv} ($\sim$1400), C\,{\sc iv} ($\sim$1550);  
the C\,{\sc iii} ($\sim$1175) line is very weak;  and the lower edge of the
detector. 
The Gaia distance to SDSS 1538 is $613^{+47}_{-41}$~pc.
\label{sdss1538wdfit1} 
}   
\end{figure}

In a first step, we derive the temperature $T_{\rm wd}$ and gravity 
$log(g)$ of the WD, using the combined exposures of the COS spectrum. 
The best fit to the COS spectrum of SDSS 1538 for the known Gaia distance 
of 613~pc is found by finding the least $\chi^2_{\nu}$ along the line $d=613$~pc  
in the $\chi^2$-map, as illustrated in 
Fig.A\ref{sdss1538chifit2} for the lower value of the reddening, $E(B-V)=0.010$. 
This gives $T_{\rm wd}=35,875$~K with $log(g)=8.74$. This model is presented in 
Fig.\ref{sdss1538wdfit1}. This model has solar abundances and none of the
absorption lines are fitted since the COS spectrum exhibits few 
and shallow absorption lines. The absorption lines are fitted in 
the next following step (Sec.5.1.2).

\paragraph{{\bf Distance Uncertainties.}} 
From Fig.A\ref{sdss1538chifit2}, we further have that the error in the 
distance propagates into an error of $\pm 500$~K in $T_{\rm wd}$
and $\pm 0.1$ in $log(g)$, given by the location of the left and right red dots. 

\paragraph{{\bf Reddening Uncertainties.}} 
Since the upper value of the reddening
(Table \ref{syspar}) is $E(B-V)=0.020$, we carry out the same 
fit to the COS spectrum of SDSS 1538 dereddened assuming
$E(B-V)=0.020$. 
Namely, we carry out an analysis and obtain
a $\chi^2$-map for the case $E(B-V)=0.020$, from which  
we find $T_{\rm wd}$=35,610~K with $log(g)=8.676$. 
This temperature is lower than for the smaller reddening of $E(B-V)=0.010$,
which is counterintuitive, since the slope of the continuum flux level,  
when dereddening a spectrum, increases with the value of $E(B-V)$.
However, the spectrum dereddened with $E(B-V)=0.020$ 
has a continuum flux level larger than when dereddened with $E(B-V)=0.010$.
As a consequence the solutions that scale to this larger
flux have a larger radius, and therefore lower gravity (here 
$log(g)=8.676$ vs. 8.74). Since the 
best fit solutions (gray diagonal) have a decreasing temperature
with decreasing gravity, the overall solution becomes colder for 
the larger dereddening value.

We average the results as follows: for a reddening of $E(B-V)=0.015 \pm 0.005$ the
solution (for $d=613$~pc) is $T_{\rm wd}=35,743$ with $log(g)=8.708$, 
and the propagation of the reddening error translates into an error of 
$\mp 133$~K in $T_{\rm wd}$ and $\mp 0.064$ in $log(g)$.  

\paragraph{{\bf Statistical Errors.}} 
Due to the noise in the data, the value of $\chi^2$ is also subject to noise,
and the uncertainty in $\chi^2$ (and therefore $\chi^2_{\rm MIN}$), translates into 
uncertainties on the derived parameters $log(g)$ and $T_{\rm wd}$ - the {\it statistical errors}.  
Details on the treatment of the statistical errors are given in \citet{god21,szk21}.
In the present work we consider the uncertainty $\chi^2_p$ in $\chi^2$ for a one parameter
problem (along the line $d=613$~pc in the $log(g)-T_{\rm wd}$ parameter space) 
for a 90\% confidence level (1.6$\sigma$), which gives 
$\chi^2_p=2.71$ \citep[][or $\chi^2_p/\nu$ for the reduced $\chi^2_{\nu}$]{avn76,lam76}.  
The statistical errors give an uncertainty of $\pm 200$~K in $T_{\rm wd}$ and
$\pm 0.0085$ in $log(g)$ for SDSS 1538. 

\paragraph{{\bf Instrumental Errors.}}  
The amplitude of the systematic errors in the continuum flux level from
instrument calibration are $\approx 2$\% for COS \citep{deb16}  
and $\approx 3$\% for STIS \citet{boh14}. 
While this error is a function of the wavelength
(larger toward the edges), 
we assume here an error of $\pm 3$\% in the entire continuum flux level
for all COS and STIS spectra. This is good enough to assess the order of magnitude
of the instrumental error, since in most cases, the instrumental error is much smaller 
than the error propagating from the error on the reddening value $E(B-V)$. 
We note that this 3\% error intrinsically includes any error possibly  
associated with \textsc{tlusty} (v204) when computing WD models, since it was derived  
using \textsc{tlutsy} 204 WD models \citep[][figure 14]{boh14}.  
The 3\% of error in flux associated with instrument calibration 
yields an error of $\pm 94$~K in $T_{\rm wd}$ and 
$\pm 0.019$ in $log(g)$ for SDSS 1538. 

\paragraph{{\bf Finite Steps Errors.}}  
To these, we add a modeling error of $\pm 125$~K and $0.05$ in $log(g)$,
since the models are in steps of $250$~K and $0.1$ in $log(g)$. 

All the above errors are then added in quadrature, and the final result for 
SDSS 1538  gives $T_{wd}=35,743 \pm 576$~K and $log(g)=8.708\pm 0.130$,
for $d=613^{+47}_{-41}$pc, $E(B-V)=0.015 \pm 0.005$.  
Further details and illustrative graphics on how we compute errors 
are given explicitly in \citet{god21,szk21}. 

\clearpage

\clearpage 

\subsubsection{\bf{ Deriving Abundances and Rotational Velocities}} 

In a second step, after we found the best   
$log(g)$ and $T_{\rm wd}$ fit for the spectrum, 
we vary the abundances of the elements 
Si, S, C, N, Fe, Al,.. one at a time, and vary the broadening velocity 
(assumed at first to be the projected WD stellar rotational  
velocity, $V_{\rm rot} sin(i)$) in the best fit model. 
The abundances are varied from $0.01 \times$ solar (or lower if needed) to 
$50 \times$ solar in steps of about a factor of two or so; 
the broadening velocity is varied  from
50 km/s to 1000 km/ in steps of 50 km/s. 

As in \citet{god21}, \citet{szk21} \citep[see also][]{god17}, 
the results of the abundances/velocity modeling are examined by visual 
inspection of the fitting of the absorption lines for each element. 
The reason we use visual examination rather than the $\chi^2$ minimization
technique is that a visual examination can recognize and distinguish 
real absorption features from the noise,
while the minimum $\chi^2$ is almost always obtained for the largest
velocity model fitting the continuum but missing many absorption lines.  
This is because the spectral
binning size of $\sim 0.58$~\AA\ for STIS  is of the same order of magnitude  
as the width of some of the absorption lines, 
and the depth of some of the absorption features is of the same  
amplitude as the flux errors. 
When the model is unable to reproduce some of the absorption lines, to compensate, 
the $\chi^2$ fitting drags the model continuum down \citep{lon06},  
and a lower $\chi^2$ is obtained for a higher broadening velocity. 
As a consequence, the best-fit in the $\chi^2$ sense doesn't especially 
always provide the best-fit to the absorption lines, nor to the broadening
velocity and can provide a larger distance (as the model continuum is 
dragged down).

Though the COS spectra have a higher resolution
and S/N, we try (when possible) to fit the absorption lines in individual 
subexposures rather than in the co-added spectra (to avoid cumulative
broadening due to the WD motion during the observation), and the subexposures
are very often nearly under exposed.

\begin{figure} 
\epsscale{0.9}  
\plotone{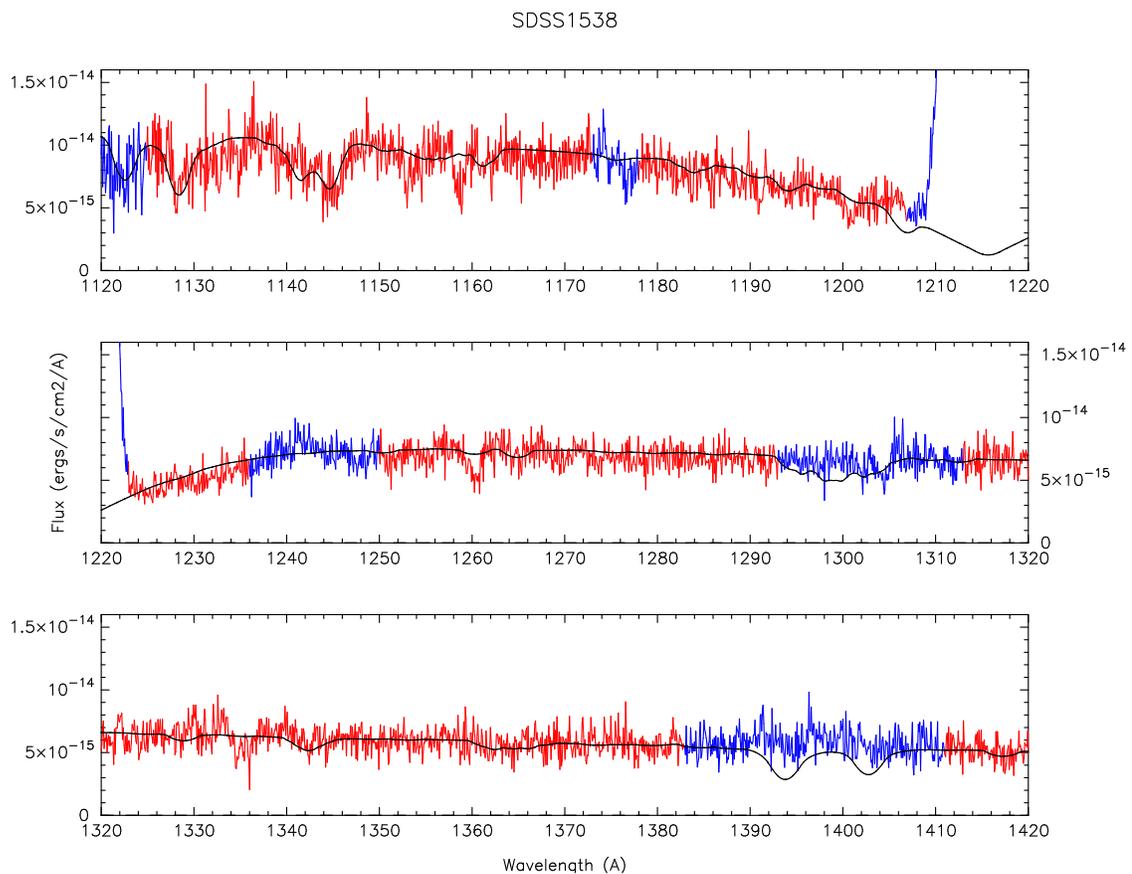}  
\caption{
The 4th subexposure of the COS spectrum of 
SDSS 1538 dereddened assuming E(B-V)=0.010 
has been fitted with a WD model. The model has a temperature
of 35,875~K and gravity $log(g)=8.74$, with a carbon abundance of 
0.0001$\times$ solar and  silicon solar abundance. 
The broadening velocity is 400~km/s. 
What seems to be an absorption feature just below 1160~\AA\ is a 
detector artifact. 
For convenience we only show the short wavelength range where all the 
absorption lines of interest are located. See text for details.  
\label{sdss1538Si} 
}
\end{figure}

The abundance analysis of SDSS 1538 is carried out by
checking only the silicon and carbon lines, since, in any case, not
many lines are present and Si and C lines dominate the spectrum. 
No ephemerides were found for SDSS 1538,
however, the system is not eclipsing and all the 4 subexposures
are similar. Consequently, we carry out the abundance analysis
on the 4th subexposure to avoid line broadening due to the 
orbital motion of the WD during the observation.  
We found that carbon has to be
very low, or the order of $\sim 0.0001$ solar,
with a broadening velocity of 500 km/s, 
to model the absence of the C\,{\sc iii} ($\sim$1175) line. 
Silicon appears to be solar (within a factor of about 2)
with a velocity of $400 \pm 100$~km/s,  
based on the absorption features in the
shorter wavelengths, see Fig.\ref{sdss1538Si}. 
The Si\,{\sc iv} (1393.76,1402.77) doublet seems to be subsolar with 
a higher velocity ($> 500$~km/s), however, this is due to 
some broad and shallow emission (which is better seen
in Fig.\ref{sdss1538cos}). 

At first glance, it seems that the Si lines (at short wavelengths)
in the model are too wide, but a lower broadening velocity produces
sharper lines that do not especially match the shape of the observed
features; a sign that some of these features  might be due to noise. 
Deriving abundances from fitting the absence of absorption lines 
or weak absorption lines is more challenging (than fitting a spectrum with 
strong absorption lines) as the absorption features are of the same order 
as the noise.   

Overall, the rest of the continuum does not disagree with solar abundance 
for the other species and a high velocity of 400 km/s, however,
since the dominant absorption lines 
are from Si and C, this says little on the abundances of the 
other species. 

From Fig.\ref{sdss1538Si}, we see that the model does not reproduce 
the Si\,{\sc ii} 1260~\AA\ line, nor the C\,{\sc ii} 1335~\AA\ line. 
We tried to reproduce these lines with a very thin cold absorbing slab 
with a temperature $T=10,000$~K, an electron density $n_e=10^{13}$cm$^{-3}$, 
and a hydrogen column density $N_{\rm H}\approx 10^{18}-10^{19}$cm$^{-2}$.
We find that the Si\,{\sc ii} (1260) line always appears in the slab model
together with the Si\,{\sc ii} (1265) line, where the Si (1265) 
line is deeper than the Si (1260) line (just as in the WD photosphere
model). Obviously, the Si (1260) line {\it alone} cannot form 
in the cold absorbing slab, nor  in the WD photosphere.  
Similarly, to form in the WD photosphere,  the C\,{\sc ii} (1335) line 
requires such a carbon abundance that a strong C\,{\sc iii} (1175) 
also forms. An extremely thin absorbing slab model with
carbon solar abundance, a hydrogen column density
of $\sim 10^{18}$cm$^{-2}$, and a turbulent dispersion velocity
of $\sim 100$~km/s is, however, able to form such a 
a single C\,{\sc ii} (1335) line alone with no other noticeable
effect on the present spectrum.    
Since the Si (1260) and C (1335) lines both appear 
{\it alone} in the spectrum, it is also possible that they form 
in part in the interstellar medium \citep[ISM, see e.g.][]{red04}. 
In either case, this does not affect the results we obtain here
for the WD abundances as we show that these lines do not form in the WD photosphere.  
\\

For the spectral analysis of the STIS and COS spectra of all the other systems, 
we follow exactly the 
same procedure as described here for SDSS~1538, 
except for the presence of an iron curtain and/or a second component when needed.

\clearpage 

\clearpage

\subsection{{\bf SDSS 1035}: an Eclipsing WD without a Mask}

The COS spectrum of SDSS 1035 is made of 8 subexposures with a rather 
low S/N. A look at the timing of the data 
(see Table \ref{obslog}, where the ephemerides of the system were  
taken from \citet{lit08}) 
indicates that only subexposures \#7 and \#8 cover the eclipse, which lasts
approximately from $\Phi \approx -0.02$ to +0.02 \citep{sav11}. 
Exposure \#7 lasted from $\Phi \approx 0.01 $ to $\Phi \approx 0.05$, 
while exposure \#8 lasted from $\Phi \approx 0.82 $ to $\Phi \approx 1.1$. 
The relative portion of the subexposures obtained during the eclipse is a very 
small fraction of the observing time, and both subexposures  
do not appear to have a flux that is noticeably lower than the
other exposures. Because of the relatively low S/N of all the individual
subexposures, we decided to combine
all the subexposures together for the spectral analysis.  

A preliminary modeling of the spectrum (with solar abundance WD models) 
shows that the WD has a temperature of the order of 11,000~K. 
At this temperature, silicon forms many strong
absorption bands and carbon forms strong wide absorption lines. 
However, no such strong features are present in the COS spectrum
of SDSS 1035. 
Because of that, the modeling of the spectrum with solar abundance
WD models does not give reliable results for the temperature and gravity. 
Consequently, we lower the metallicity to a few percent 
(in solar units)  to provide a best fit to the spectrum. 
We use the low abundance WD models to derive the WD temperature, gravity,
abundances and broadening velocity.  

For SDSS 1035, the final result, including
all the errors, yields $T=11,475 \pm 188$~K, with $log(g)=8.385 \pm 0.166$.  
No second component and no iron curtain were necessary for the analysis. 

\begin{figure}[h!]
\epsscale{0.9}  
\plotone{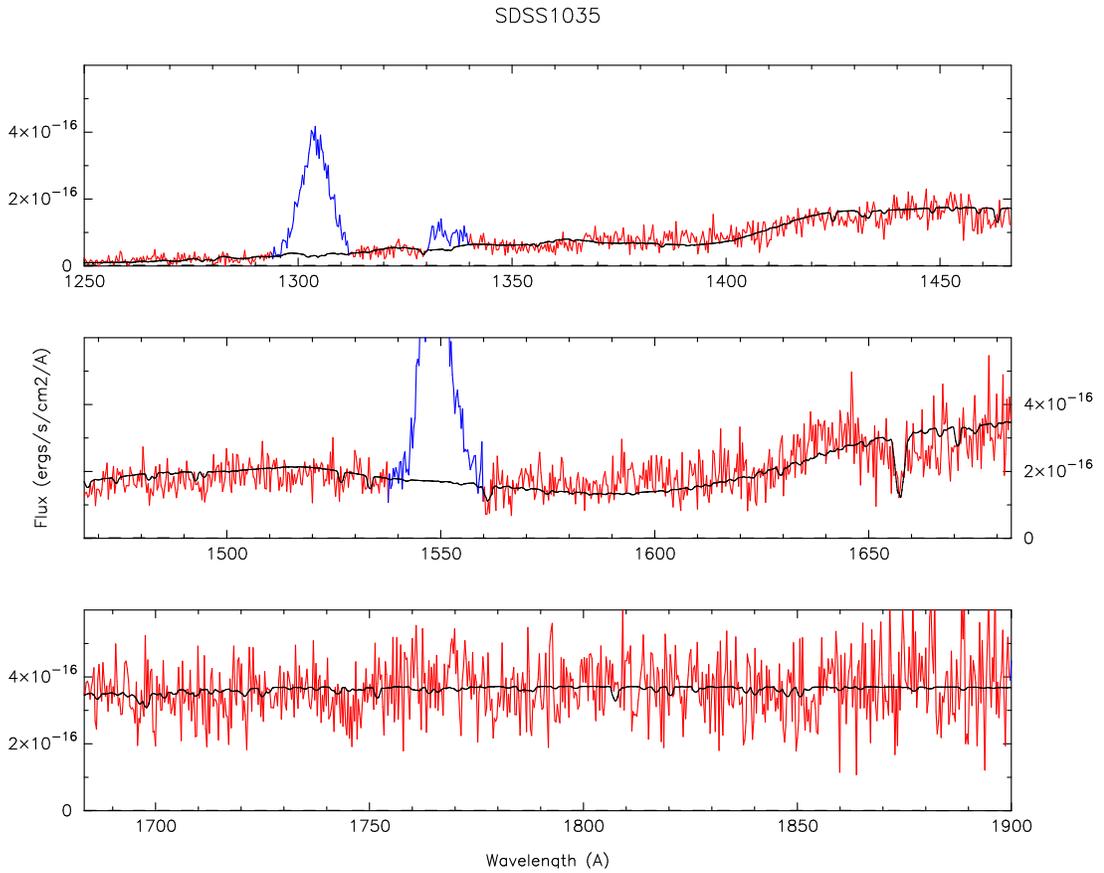}
\caption{ 
The spectral analysis of the COS spectrum of the WD in SDSS 1035
(in red) yields a solution (in black) with a temperature of $T=11,475 \pm 188$~K, 
with $log(g)=8.385 \pm 0.166$, metal abundance 
[Z]=0.03-0.05, and a broadening velocity $V=150 \pm 50$~km/s.  
Strong emission lines have been masked before the
fitting and are colored in blue. 
Below 1250~\AA\ , the COS spectrum exhibits a continuum flux 
much smaller than the noise (see Fig.\ref{sdss1035cos})  
and only shows emission lines from 
C\,{\sc iii} ($\sim$1175) and N\,{\sc v} ($\sim$1240), as well as geocoronal
emission O\,{\sc i} ($\sim 1300$).   
{\bf The Gaia distance to SDSS 1035 is $195^{+12}_{-10}$~pc.}  
\label{sdss1035wdfit}  
}
\end{figure}

Fitting the only clear line, C\,{\sc i} ($\sim$1657), we obtain 
[Z]=0.03-0.05, with a broadening
velocity $V=150 \pm 50$~km/s.  
This best fit model is presented in Fig.\ref{sdss1035wdfit}.
We note that this model also fits the C\,{\sc i} ($\sim$1561) 
absorption line.   

SDSS 1035 is an eclipsing high inclination system, just like IR Com, but
while the co-added COS spectrum of IR Com presents many WD absorption lines, 
the co-added COS spectrum of SDSS 1035 presents only
one clear absorption line. Therefore, the absence of lines in SDSS 1035 
is certainly not due to the broadening of the absorption lines 
from the WD orbital motion during the time of the observation 
(this is further discussed in Sec.6.2).  

\clearpage

\clearpage

\subsection{{\bf IY Ursae Majoris: Example of a Masked Eclipsing System}} 

Since IY UMa is an eclipsing system showing strong absorption due to 
veiling material, we give here details of the iron curtain modeling 
and pay particular attention to the orbital phases at which each 
of the 4 subexposures of the COS spectrum were obtained.

\begin{figure}[b!] 
\epsscale{1.05} 
\plottwo{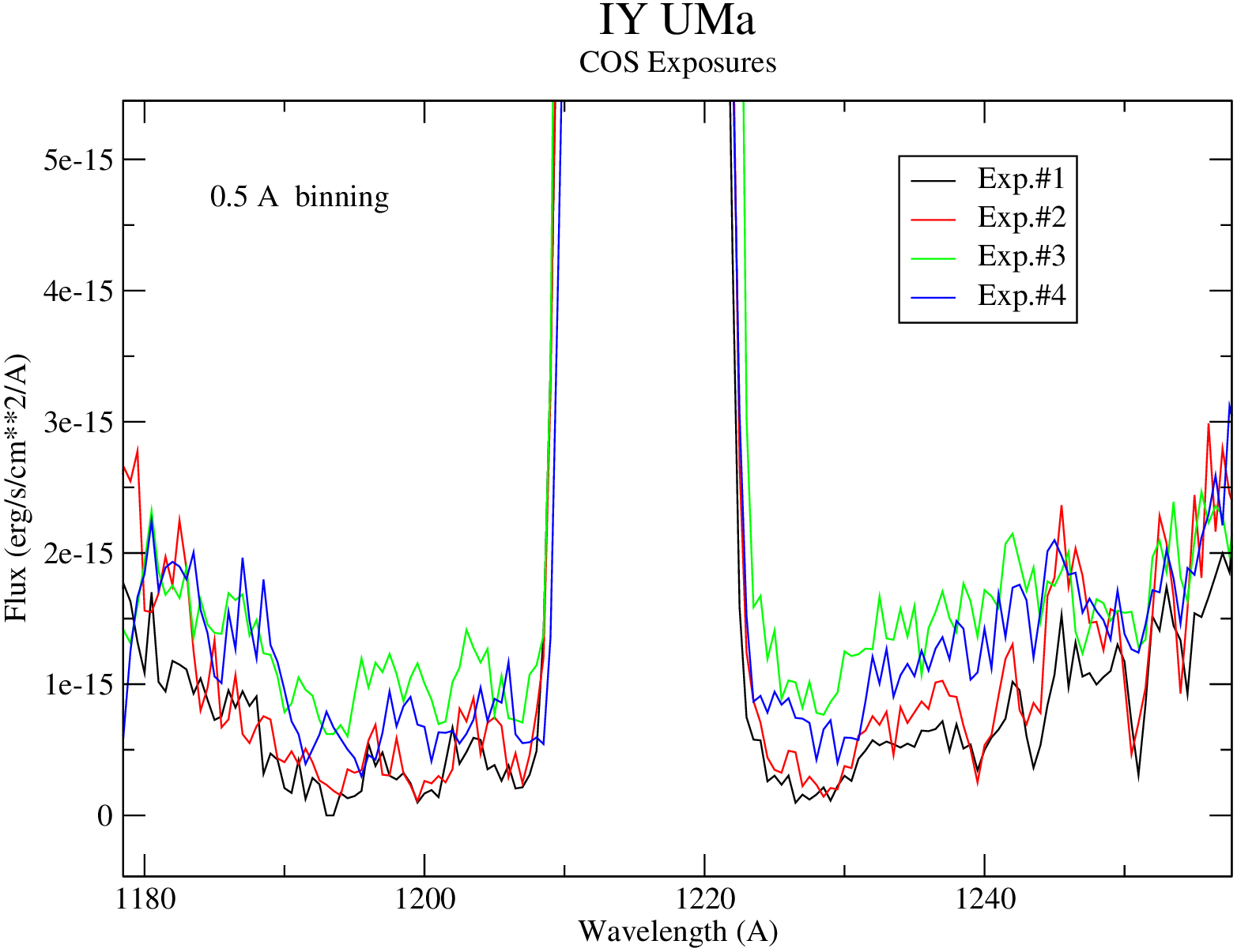}{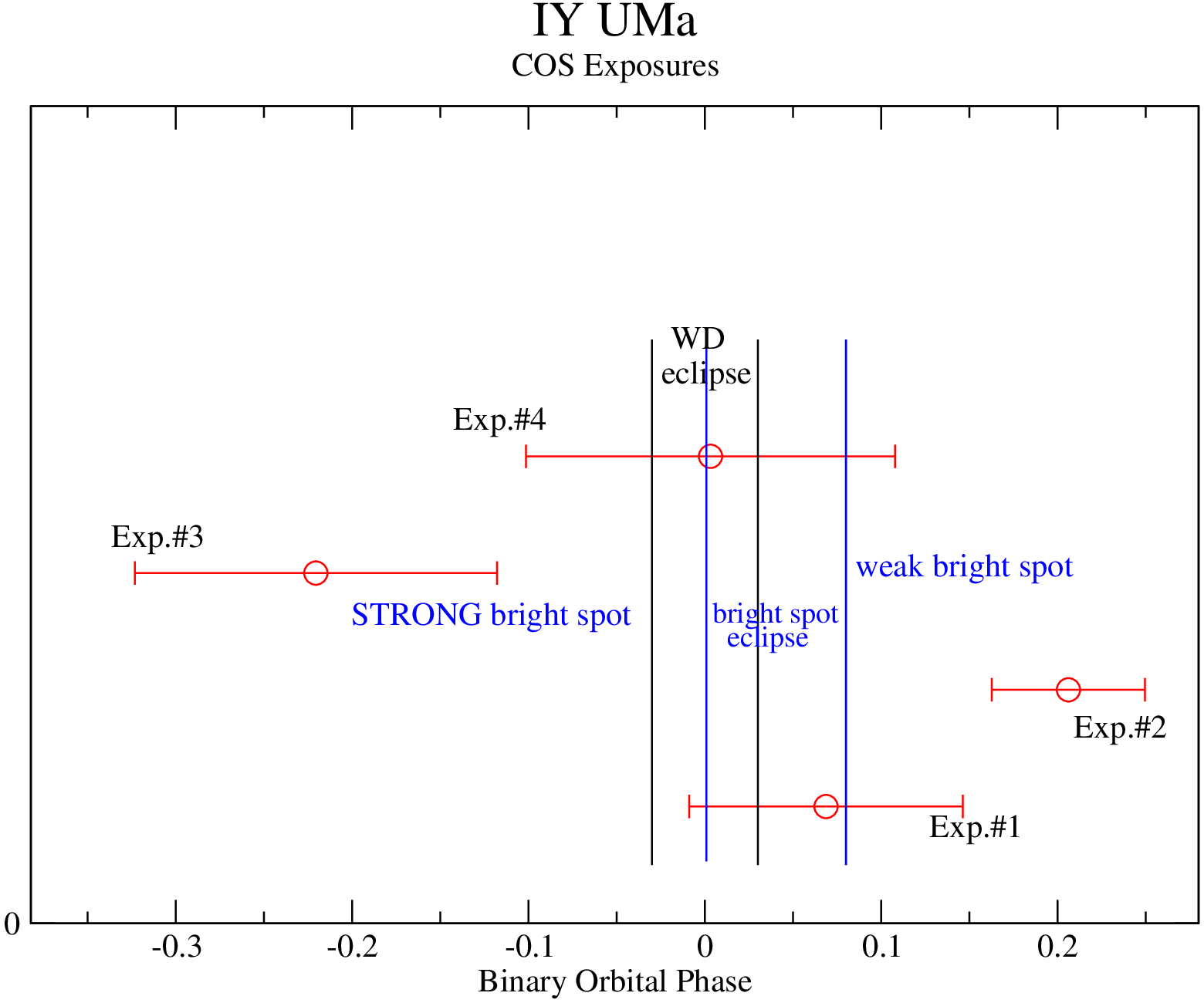} 
\caption{
The 4 individual subexposures of the COS spectrum of IY UMa are presented.
{\bf (a) Left.} 
The individual subexposure spectra are shown in the Ly$\alpha$ 
region, where the most noticeable change is taking place. 
The flux level at the bottom of the Ly$\alpha$ is of the order of 
$1 \times 10^{-15}$erg/s/cm$^2$/\AA\ 
in exposures \# 3 \& 4, and almost goes down to zero in 
exposures \# 1 \& 2. For convenience the exposures have 
been binned at 0.5~\AA , and the vertical ($F_{\lambda}$) 
axis has been extended to $<0$.   
\label{iyumasubex}
{\bf (b) Right.} 
The subexposure duration and timing (in red) are shown 
as a function of the binary orbital phase (X-axis). 
The Y-axis is only used to present the subexposures 
in a staggered  display.  
Exposures \# 3 \& 4 were obtained when the bright
spot was facing the observer; during exposure \# 1 \& 4  the
WD was eclipsed; only exposure \# 2 was obtained when the 
WD was not eclipsed and with little contribution from the bright spot. 
The duration of each exposure is indicated by the length of 
the horizontal red mark. The WD eclipse is between the two
vertical black lines and the bright spot eclipse is between
the two vertical blue lines, as assessed from Fig.1 in \citet{mca19}.     
}
\end{figure}

The first subexposure exhibits a continuum flux level
significantly lower than the other subexposures,
an indication that it was obtained during
eclipse. Subexposure \#3 has the highest
continuum flux level.   
The bottom of the Lyman$\alpha$ region undergoes the largest {\it relative}
change: it has the lowest flux
in exposure \#1 and highest flux in exposure \#3 (see Fig.\ref{iyumasubex}a).  
In addition, as expected for eclipsing systems, 
all exposures exhibit strong veiling, recognizable
by the strong Fe\,{\sc ii} absorption ``bands'' near $\sim 1565$~\AA\ 
and $\sim 1635$~\AA .  
We use the orbital period and ephemeris from \citet[][Table 1]{mca19} 
to carefully and precisely time the subexposures as a function of the orbital phase. 
We find that exposure \#3 (with the highest flux) was obtained 
at orbital phase $\Phi \sim$[-0.3,-0.1] (see Fig.\ref{iyumasubex}b), 
as the bright spot was facing the observer. During both exposures \#1 and 
\#4 the system went into eclipse, but exp.4 started before the eclipse
and also had strong contribution from the bright spot. 
Exposure \#2 is the only exposure obtained out of eclipse 
(at $\Phi \sim 0.2$) with little contribution from the bright spot.
This exposure, like exp.1, has also very little flux at 
the bottom of the Ly$\alpha$ region, a possible indication that the         
bright spot is the reason the bottom of the Ly$\alpha$ does
not go to zero in exp.3 \& 4.  
We therefore decide to use subexposure \#2 to carry out the spectral
analysis of the WD in IY UMa. Unfortunately, this is also the shortest
exposure with the lowest S/N.  

\clearpage

The spectrum of IY UMa presents all the absorption features 
associated with the presence of veiling material. 
However, the spectral analysis of the second exposure of the COS spectrum of 
IY UMa was first carried out without an iron curtain, but with 
a second flat component with a flux of $2.5 \times 10^{-16}$erg/s/cm$^2$/\AA\  
(since even in subexposure \#2 the bottom of the Ly$\alpha$ does not 
completely go down to zero).  
As explained explicitly in Sec.4.4, the continuum flux level of the second
component was found by trial and error. For IY UMa, we increased the 
continuum flux level of the second component in steps of $5 \times 10^{-17}$ 
and carried out the analysis, generating a $\chi^2$ map to find the 
best-fit for the given Gaia distance. We then chose the value of the
second component that yielded the lowest $\chi^2$. 

\begin{figure}[h!]
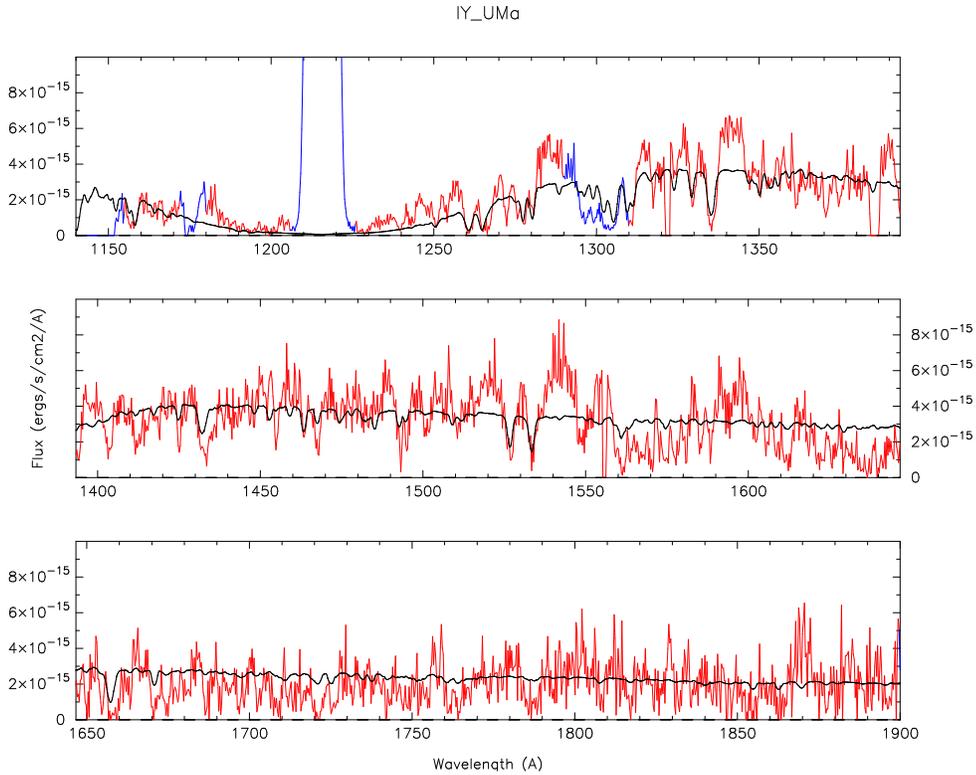

\vspace{-6.cm}   
\epsscale{0.8}  
\plotone{iyuma_wdfit.ps} 
\vspace{1.cm}   
\caption{
{\bf (a) Above.} 
The second exposure of the COS spectrum of IY UMa is modelled with a WD synthetic spectrum. 
Many of the absorption lines are not fitted at all. 
Such a modeling gives a WD temperature of about 17,000~K, with a
gravity $log(g) \approx 8.5$.  
The model has solar abundances and a broadening velocity of 200~km/s. 
{\bf (b) Below.} 
The addition of an iron curtain greatly improves the fit. 
The iron curtain model has solar abundances, except for 
[P]=0.01. Its temperature
is 10,000~K, with an electron density $n_e=10^{13}$cm$^{-3}$, turbulent
velocity of $75^{+25}_{15}$~km/s, and an atomic hydrogen column density of 
$N_{\rm H} =10\pm 2 \times 10^{21}$cm$^{-2}$.   
A second component with a constant flux of 
$2.5 \times 10^{-16}$erg/s/cm$^2$/\AA\ has been taken into account 
in both cases.   
{\bf The Gaia distance to IY UMa is $182\pm2$~pc.}   
\label{iyumawdfit} 
}
\plotone{iyuma_icfit.ps} 
\end{figure}

Following the same procedure as for SDSS 1538, we found the best fit
for the single WD model, which occurs for the grid model with 
$T=17,000$~K and $log(g)=8.5$. Because the model still needs the addition of an 
iron curtain we did not fine tune the model to a higher accuracy at this stage. 
This model is presented in Fig.\ref{iyumawdfit}a. 
That model has solar abundances and a projected WD stellar rotational velocity
of 200~km/s. As expected, many absorption features are not fitted, especially in the
longer wavelengths $\lambda > 1550$~\AA.   

In the next step, we added an iron curtain to the best fit WD model and varied
the parameters of the iron curtain to fit the absorption features
of the veiling material. For the iron curtain, we kept $T=10,000$~K 
and $n_e=10^{13}$cm$^{-3}$ constant, as suggested by \citet{hor94}. 
The best fit is  
obtained for a hydrogen column density $N_H=10 \pm 2 \times 10^{21}$cm$^{-2}$, 
and a turbulent velocity $V_{\rm turb}=75_{-15}^{+25}$~km/s. Both the
WD model and the iron curtain have solar abundances.  
We notice, however, that the addition of the iron curtain slightly 
changed to continuum flux level and the scaling 
to the Gaia distance of IY UMa. Therefore, we carried a new
spectral analysis iteration. Namely, we generated 
veiled (with the above best fit iron curtain model) WD models in the 
$log(g) -T_{\rm wd}$ parameter space, and carried out a spectral
analysis using these new veiled WD models (and with the above-mentioned
second flat component).
In other words,  we obtained a $\chi^2$-map (Fig.\ref{iyumachi}) in the 
$log(g)$ vs $T_{\rm wd}$ parameter space using a grid of veiled 
WD models: WD + absorbing slab, where the absorbing slab is the
one given above.
 
The final results of the veiled WD model fits (including all the uncertainties) 
yielded  $T=17,130 \pm 249$~K, $log(g)=8.480 \pm 0.125$, with solar abundances
and a broadening velocity of 150~km/s. 
Such a model fit is presented in Fig.\ref{iyumawdfit}b.  
The spectrum exhibits additional absorption from a much hotter
medium with lines from C\,{\sc iii} ($\sim$1175), Si\,{\sc iv} ($\sim$1400),
and C\,{\sc iv} ($\sim$1550). 
The spectrum, when compared to the model, presents some broad emission 
near 1550~\AA\ (C\,{\sc iv}) as well as some higher flux in the shorter
wavelength near 1250~\AA , 
1290~\AA , and 1340~\AA . 

Many of the absorption lines in the COS spectrum of IY UMa are 
reasonably well fitted. 
We note the presence of the P\,{\sc ii} (1452.89) 
absorption line in the model, which is not seen in the COS spectrum
of IY UMa. In order to remove this line from the model 
we had to lower [P] to 0.01
solar, while keeping all the other elements to solar abundances (=1.0).  
The iron  curtain fits pretty well
(in depth and shape) the absorption lines from carbon, silicon, 
sulfur and iron. 
As no other P lines can be unambiguously 
modeled, one can question whether the abundance of phosphorus (based
on a single line) is really subsolar or whether the iron curtain model needs further
improvement.  
It is clear that the approach of a single iron curtain 
is only an approximation, since the iron curtain is made of 
a gas that is not isothermal, does not have constant density,  
nor a constant turbulent (dispersion) velocity. 
Ideally, one needs to construct an iron curtain with multiple 
layers, each with a different density, temperature, and velocity. 
Such a modeling is well beyond the scope of the present work.

\clearpage

\clearpage

\subsection{{\bf DV Ursae Majoris}}  

\begin{figure}
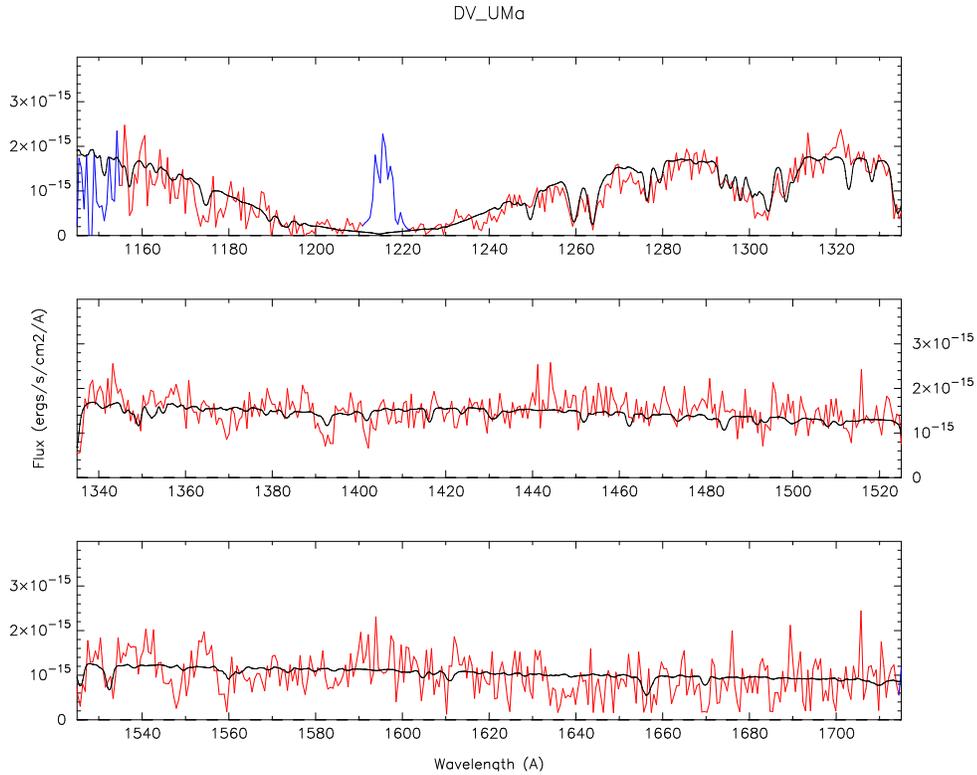
 
\vspace{-5.cm}   
\epsscale{0.8}  
\plotone{dvuma_T20000g8.70Solv150.ps} 
\caption{
{\bf (a) Above.} 
The STIS COS spectrum of DV UMa 
has been fitted with a WD model and taking into account 
a ``flat'' second component 
with a flux of $1.16 \times 10^{-16}$erg/s/cm$^2$/\AA . 
The WD model has a temperature
of 19,500~K, gravity $log(g)=8.6$, solar composition, 
and a broadening velocity of 150~km/s. 
{\bf (b) Below.} 
An iron curtain has been added to the model to account for the 
many Fe\,{\sc ii} absorption features. The iron curtain has a
temperature of 10,000~K, electron density of 
$n_e=10^{13}$cm$^{-3}$, a velocity dispersion
of 50~km/s, a hydrogen column density of $N_{\rm H}=3 \times 10^{21}$cm$^{-2}$
and solar abundance.  
{\bf The Gaia distance to DV UMa is $387^{+24}_{-22}$~pc.}   
\label{dvumawdfit} 
}
\plotone{dvuma_T20000g8.70Solv150ICT10000vtb50NH3e21.ps} 
\end{figure}

DV UMa is an eclipsing system with a STIS spectrum (a snapshot
lasting only 900~s) obtained around orbital phase $\Phi \approx 0.2$.
We carry out the analysis on this single exposure obtained out of eclipse.
Because of its high inclination ($\sim 83^{\circ}$), the spectrum is
heavily veiled. Consequently, we model a WD plus with the addition of
an iron curtain in an iterative manner as done for IY UMa. 
Here too, the bottom of the Ly$\alpha$ does not go down to zero and, 
since this cannot be accounted for with the WD model (it is not
hot enough with $T< 20,000$~K), we take into account a flat second component with 
an amplitude of $1.16 \times 10^{-16}$erg/s/cm$^2$/\AA\ (namely,
we remove this constant flux from the observed spectrum).   
The first iteration of the WD plus iron curtain model fits
yield a temperature of 19,500~K with $log(g)=8.6$ as shown 
in Fig. \ref{dvumawdfit}, where we present the WD model without
(a) and with (b) the iron curtain. 

The iron curtain, with a temperature of 10,000~K, electron density
of $10^{13}$cm$^{-3}$, a turbulent velocity dispersion of 50~km/s,
and a hydrogen column density of $3 \times 10^{21}$cm$^{-2}$, significantly
improves the fit in the longer wavelengths where the
{\it iron} bands form ($\sim 1560$~\AA\ and $\sim$1635~\AA ), but also in 
the shorter wavelengths near $\sim 1130$~\AA\ and $\sim$1150~\AA .    
However, some carbon lines appearing in the model spectrum do not show in 
the observed spectrum. 

\clearpage 

\clearpage

The iron curtain material is in all
likelihood accretion material (e.g. overflowing the disk rim at
the hot spot) and the WD surface composition is also made of
the same accretion material (which is being replenished by accretion
as diffusion takes place). 
Therefore, we assume that the WD photosphere and the iron curtain have the
same composition, and we vary this composition to try and match as many
absorption features as possible in the observed spectrum.   
We slightly vary the other parameters of the iron curtain in order
to fine tune the fit. 

\begin{figure}[b!]  
\vspace{-10.cm} 
\epsscale{0.9}  
\plotone{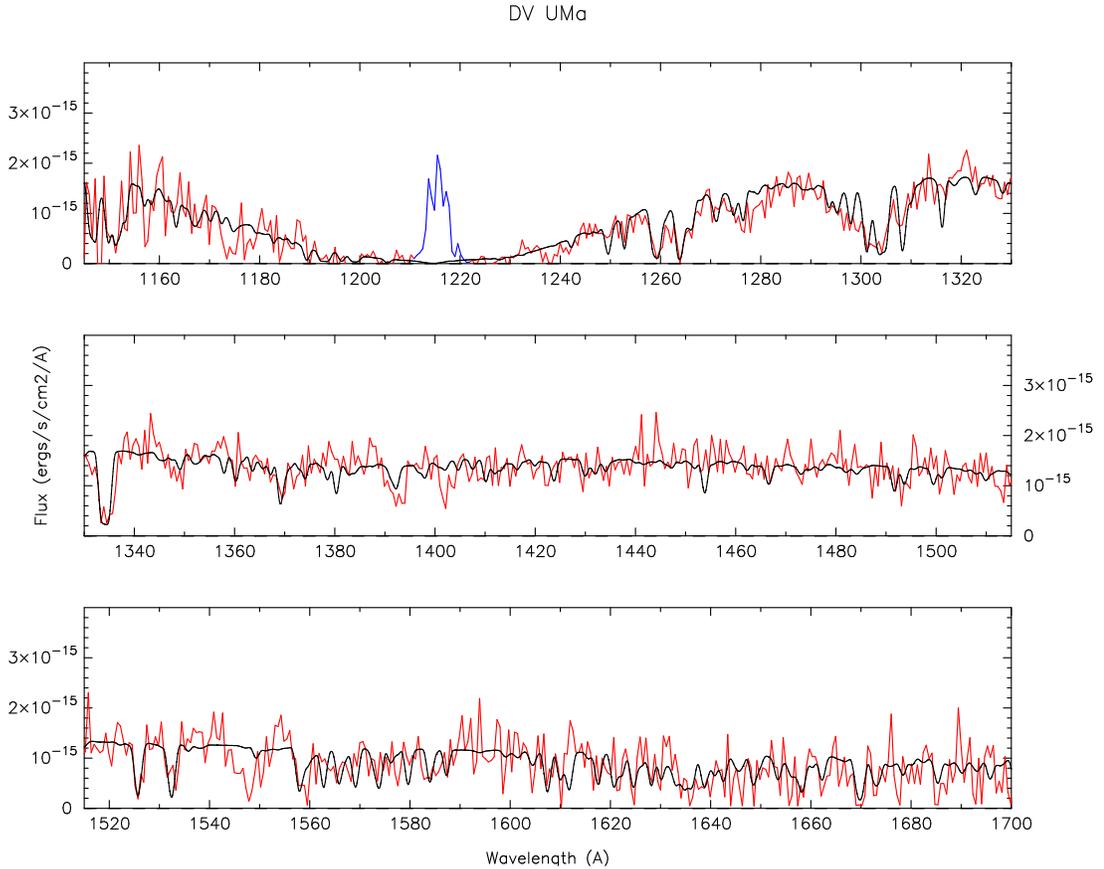} 
\caption{
The final fit to the STIS spectrum of DV UMa yields 
a WD with a temperature $T_{\rm wd} =19,325 \pm 481$~K, 
gravity $log(g)=8.565 \pm 0.177$, and a broadening velocity
 $V=150$~km/s. 
The iron curtain has a
temperature of 10,000~K, electron density of 
$n_e=10^{13}$cm$^{-3}$, a velocity dispersion
of 75~km/s, a hydrogen column density of $N_{\rm H}=2 \times 10^{21}$cm$^{-2}$. 
The iron curtain and the WD have the same abundances: 
[C]=0.1 solar, [Si]=0.3 solar, [P]$\lesssim$0.01 solar, with all the 
other elements having their solar value.  
A second component with a constant (flat) value of $1.16 \times 10^{-16}$erg/s/cm$^2$/\AA
was also taken into account.  
\label{dvumawdicfit} 
}
\end{figure}

We find that carbon, silicon and even phosphorus must be sub-solar,
while the other elements are kept at solar values as they do not 
affect the fit. We find a carbon abundance [C]=0.1 solar (within 
a factor of two), silicon abundance [Si]=$0.3\pm0.1$ solar, and 
phosphorus abundance [P]=0.01 solar or smaller.  
The low phosphorus abundance is needed since otherwise a strong
P\,{\sc ii} absorption line appears near 1452-3~\AA , 
as seen in Fig.\ref{dvumawdfit}b.   
This model is presented in Fig.\ref{dvumawdicfit}. 
The iron curtain in that model has a hydrogen column density
of $2 \times 10^{21}$cm$^{-2}$ and a turbulent velocity of 
75~km/s. The final result of the fine tuning of the fit yields 
(with all the error bars included) $T_{\rm wd}=19,325 \pm 481$~K, 
with $log(g)=8.565 \pm 0.177$.
Here too, it is difficult to fit all the lines, even of the same species.  
We note, a posteriori, that the left line of the sulfur  doublet, near 1250-1255~\AA , 
is not as strong in the STIS spectrum as in the model, an indication
that sulfur  is likely subsolar, i.e. [S]$< $1.0.

\clearpage

\subsection{{\bf IR Comae Berenices}} 

\begin{figure}[b!]
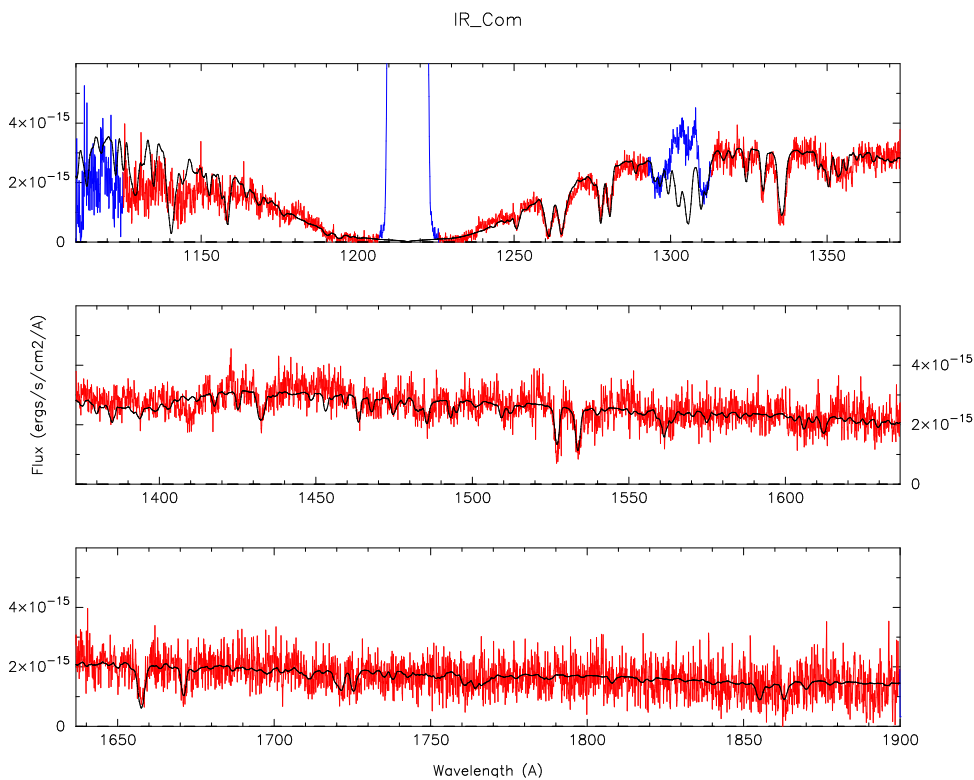
  
\epsscale{0.8}  
\vspace{-5.0cm} 
\plotone{ircomwdfit.ps}                   
\vspace{0.5cm} 
\caption{
{\bf (a) Above.}  
The analysis of the COS spectrum of IR Com yields a WD temperature of 
$17,860 \pm 180$~K, gravity $log(g)=8.613 \pm 0.090$, 
and elevated abundance of aluminum ([Al]$=5\pm1$ solar) 
and iron ([Fe]=$4\pm1$ solar) with 
a broadening velocity $V=150$~km/s. 
{\bf (b) Below.}  
The addition of a thin iron curtain 
increases the depth of aluminum and iron lines without the need
to have suprasolar aluminum and iron abundances in the WD model. 
{\bf The Gaia distance to IR Com is $217 \pm 3 $~pc.}   
\label{ircomwdfit} 
}
\plotone{ircomnewicfit.ps} 
\end{figure}

The COS data of IR Com is made of 5 subexposures. Subexposure \#4 was obtained
during eclipse (see Table \ref{obslog}) and has a significantly lower continuum flux level. 
The remaining 4 subexposures (\#1, 2, 3, \& 5) are very similar with the same
continuum flux level;  we combined them to generate a 
spectrum not affected by the eclipse, and we used that spectrum in our analysis.   
The bottom of the Ly$\alpha$ doesn't go to zero, and we   
take a flat flux second component of $3.5 \times 10^{-16}$erg/s/cm$^2$/\AA\ 
(at $E(B-V)=0.019$), 
which when subtracted from the spectrum brings the bottom of the Ly$\alpha$ to zero. 
The spectrum of IR Com does not exhibit any iron absorption bands characteristic
of the iron curtain, in spite of the fact that this is a high inclination
eclipsing system (which was also the case for SDSS 1035). 
Consequently, we carried out the analysis without iron curtain modeling. 
The results of the single WD analysis yield a WD temperature $T_{\rm wd}=17,860 \pm 180$~K,
and gravity $log(g)=8.613 \pm 0.090$. The fitting of the absorption lines
yields solar abundances, except for aluminum with [Al]=$5 \pm 1$ solar, 
based on 
Al\,{\sc ii} (1670.8) and Al\,{\sc iii} (1854.7 \& 1862.8) lines,
and iron [Fe]$\sim 4\pm1$ solar, which slightly improves the fit in the
very short wavelength and very long wavelength regions. 
We found a broadening velocity of $150 \pm 50$~km/s
(see Fig.\ref{ircomwdfit}a). 
Except for the very short wavelength region ($\lambda < 1160$~\AA ), the 
COS spectrum is relatively well fitted with the WD model spectrum. 

\clearpage

We note that in DV UMa and the other veiled systems (e.g. GY Cnc and BD Pav), 
the short wavelength region $\lambda < 1160$~\AA\  is also affected by Fe\,{\sc ii} 
absorption lines, creating two adjacent absorption bands at $\sim 1130$~\AA\ 
and $\sim 1150$~\AA\    
in addition to the Fe absorption bands seen near $\sim 1565$ and $1635$~\AA  , 
raising the possibility that the COS spectrum of IR Com might also be
affected by an iron curtain.  
Consequently, we added a cold absorbing slab/iron curtain model   
($T=10,000$~K, $n_e=10^{13}$cm$^{-3}$, and solar abundances)
to the stellar WD model,  
to check whether the short wavelength region fit could be improved.  

We obtained a better fit in the short wavelength region for an absorbing 
slab with a hydrogen column
density of $N_{\rm H}=5 \times 10^{20}$cm$^{-2}$ and a turbulent velocity of 
$\sim$100~km/s. However, such an iron curtain also creates strong absorption bands
at 1565 and 1635~\AA\ and strong C and Si lines which are not observed. Since the absorption bands 
at 1565 \& 1635~\AA\  and the absorption feature at $\lambda < 1160$~\AA\  
are all due to iron, one cannot generate one without generating the other
(even by changing the temperature and density of the iron curtain). 
Such an iron curtain had to be ruled out.  

However, we found that an {\it thin} absorbing slab with a hydrogen column density of 
{\it only} $N_{\rm H}=1 \times 10^{19}$cm$^{-2}$ and a turbulent velocity of 100~km/s
slightly improves the WD fit in the shorter wavelengths, and also 
helps fit the aluminum and iron lines without the need to increase [Al] and [Fe] abundances  
above solar in the WD model. 
This thin iron curtain model exhibits a slightly deeper
C\,{\sc ii} (1335) absorption line, and we therefore had to decrease the carbon abundance
in the curtain to [C]=0.4 while keeping solar abundance for the WD model
(in order to match the carbon 1330 line). 

We present this solar abundance WD with a thin iron curtain model 
in Fig.\ref{ircomwdfit}b.   
There is little
difference between the WD model (Fig.\ref{ircomwdfit}a) and the WD plus iron curtain model 
(Fig.\ref{ircomwdfit}b). Because of this, it is not clear whether the WD has increased
iron and aluminum abundances or just a thin iron curtain. 
The fact that the shorter wavelength region is not adequately modeled
may imply that our model might be inaccurate or incomplete, or might point out
to some calibration problems at the inner edge of the detector. 
Overall, the results indicate that the WD does have nearly solar abundances.   
We further discuss the thin iron curtain of IR Com in Sec.6.4.            

\clearpage

\subsection{{\bf GY Cancri}} 

The STIS snapshot spectrum of the eclipsing system GY Cnc was obtained around binary orbital
phase $\Phi \approx 0.25$, well out of eclipse ($i \sim 77^{\circ}$), 
in spite of that, it reveals a heavily veiled WD similar              
to the spectra of IY UMa and DV UMa (with $i> 80^{\circ}$).

\begin{figure}
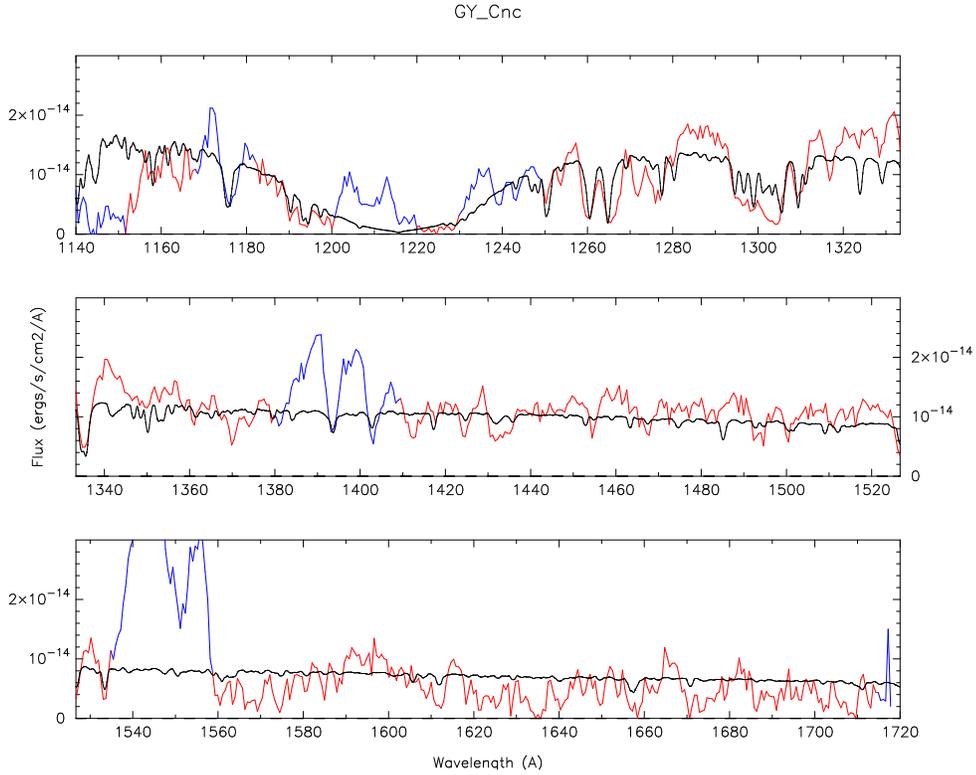
 
\vspace{-4.cm}   
\epsscale{0.8}  
\plotone{gycnc_wdfit.ps} 
\vspace{-0.5cm}   
\caption{
The final results of the modeling of the spectrum of GY Cnc is a WD 
with a temperature $T_{\rm wd}=22,515 \pm 292$~K, gravity $log(g)=7.876 \pm 0.108$,
and a broadening velocity of 150~km/s. 
{\bf (a) Above.} 
A single WD model alone is inadequate to fit most of the spectral features
especially in the longer wavelengths.  
{\bf (b) Below.} 
The addition of an iron curtain significantly improves the fit. 
Both the WD stellar model spectrum and the iron curtain have solar
abundances. 
{\bf The Gaia distance to GY Cnc is $275^{+5}_{-4}$~pc.}   
\label{gycncwdicfit} 
}
\plotone{gycnc_F2IC.ps} 
\end{figure}

We modeled the spectrum of GY Cnc with a WD plus
an iron curtain in the same iterative manner as we did for IY UMa and DV UMa,  
using first solar abundances. We also took into account a second flat component
of amplitude $2 \times 10^{-15}$erg/s/cm$^2$/\AA . 
The best fit yielded a gravity $log(g)=7.876 \pm 0.108$ with a temperature
$T_{\rm wd}=22,515 \pm 292$~K, and a WD projected stellar rotational (broadening) velocity 
of 150~km/s,  for a distance of $275 \pm 5$~pc, reddening 
$E(B-V)=0.023 \pm 0.013$, including the propagation of the uncertainties from
the statistical error, the modeling, and the instrumental/detector errors
(as was done for all other systems). 
The iron curtain model has a relatively large atomic hydrogen
column density, $5 \pm 1 \times 10^{21}$cm$^{-2}$, with a turbulent velocity of 
$75\pm25$~km/s. This model is presented in Fig.\ref{gycncwdicfit} with and
without an iron curtain.

The iron curtain does model relatively well the many iron absorption features
in the longer wavelengths $\sim 1500 - 1700$~\AA , since these are the absorption
features that we use in the fit to derive the iron curtain parameters. 
The strong emission lines of H\,{\sc i} (1216), Si\,{\sc iv} ($\sim$1400), C\,{\sc iv} ($\sim$1550)
are not fitted and are strongly blue-shifted, indicating they form in a wind. 
Strong absorption lines (not shifted) are present on top of the Si\,{\sc iv} and 
C\,{\sc iv} emissions, as well as N\,{\sc v} ($\sim$1240) absorption lines; 
they too form in a hotter gas and are not modeled here. 
While we do not attempt to fit these features forming in a hot gas, 
some low ionization sulfur  ($\sim 1250$), silicon ($\sim 1260$), carbon ($\sim 1330$),
and phosphorus ($\sim 1452$) absorption lines are too deep in the model. 

\clearpage

\clearpage 

\begin{figure}[b!]  
\epsscale{0.9}  
\vspace{-2.0cm} 
\plotone{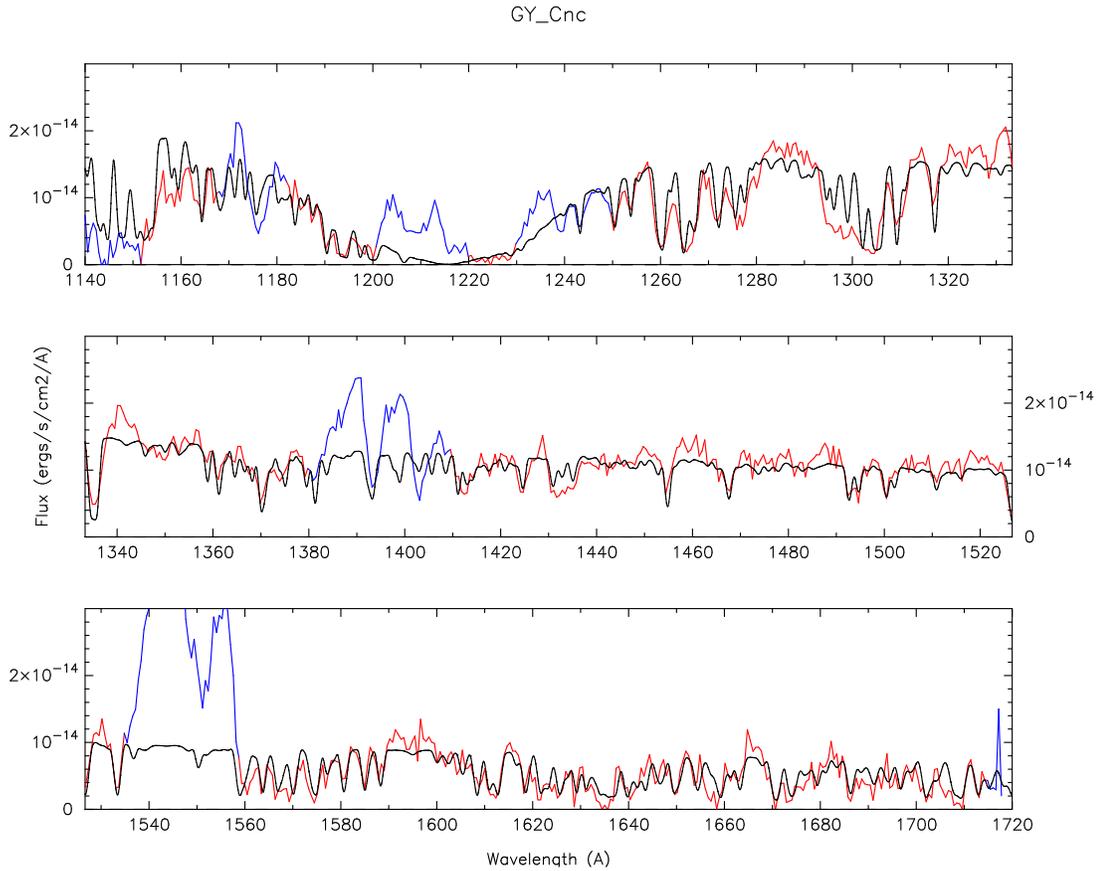} 
\vspace{-0.5cm} 
\caption{
The same WD and iron curtain model as in Fig.\ref{gycncwdicfit} is presented
here, but now both the WD and iron curtain have non-solar abundances: 
[C]=0.01, [Si]=0.1, [P]=0.01, and [S]=0.1. See text for details.  
\label{gycncfinalfit}
}
\end{figure}

In order to improve the fit, 
we lower the abundances of C, Si, P, and S in both the WD and the iron
curtain to better fit these lines. We find the best fit for 
[C]=0.01, [P]$\lesssim$0.01, [Si]=[S]=0.1, presented in the Fig.\ref{gycncfinalfit}. 
However, while the above mention line fit improved, the fit to some
of the other lines (in the $\sim 1160$~\AA\ and $\sim 1300$~\AA\ regions) 
is slightly degraded. From this new fit, it seems more apparent that 
C\,{\sc iii} ($\sim$1175) and N\,{\sc v} ($\sim$1240) have some broad emission. 
C\,{\sc ii} (1335) also presents some emission. Overall the non-solar
abundance model improves the fit and points to subsolar abundances.

\clearpage

\subsection{\bf{ BD Pavonis}} 

The COS data of BD Pav consists of 5 subexposures. 
Subxposure \#1  was collected from orbital phase $\Phi = 0.87$ 
to $\Phi=1.00$ with a continuum  flux level about 20\% 
lower than subexposures \#2 and \#3 (obtained at orbital phases 
$\Phi = 0.26$ and 0.36, respectively), but only 5 to 10\% lower than 
subexposures \#4 and \#5 (obtained at $\Phi = 0.61$ and 0.70).  
With an inclination not higher than $75^{\circ}$, BD Pav
shows signs that its WD is eclipsed \citep[][{\it barely} eclipsed]{kim18},  
explaining the lower flux in subexposure \#1. 
Exposures \#2 and \#3 have a higher flux as well as stronger
emission lines, pointing to the possibility of 
a (relatively) hot emitting component (which could be
the {\it second component} which we model with a flat continuum). 
It is not clear neither, whether subexposures \#4 and \#5 have a 
lower flux (than subexposures \#2 and \#3) due to stream-disk overflow material (as is often
the case around $\Phi=0.6-0.8$), or have an unaffected continuum
flux level. All the subexposures show signs of strong veiling, 
and all exhibit the same absorption lines. 
Since one cannot clearly identify which of the subexposures had a 
``clear shot'' at the WD,   
we decided to combined the 5 exposures and carry out the
analysis on the combined spectrum. We estimate that the continuum
flux level of the combined spectrum cannot be more than 10\% 
off from the unaffected spectrum of the WD itself. Since we 
perform the analysis taking into account a second component, it is more
likely that the combined spectrum continuum flux level is only a few \% off. 
For that reason, we included an additional error of 10\% in the continuum flux level to the final results, 
(for comparison such an error corresponds to an error of $\sim 3.16$\% 
in the distance).  

\begin{figure}
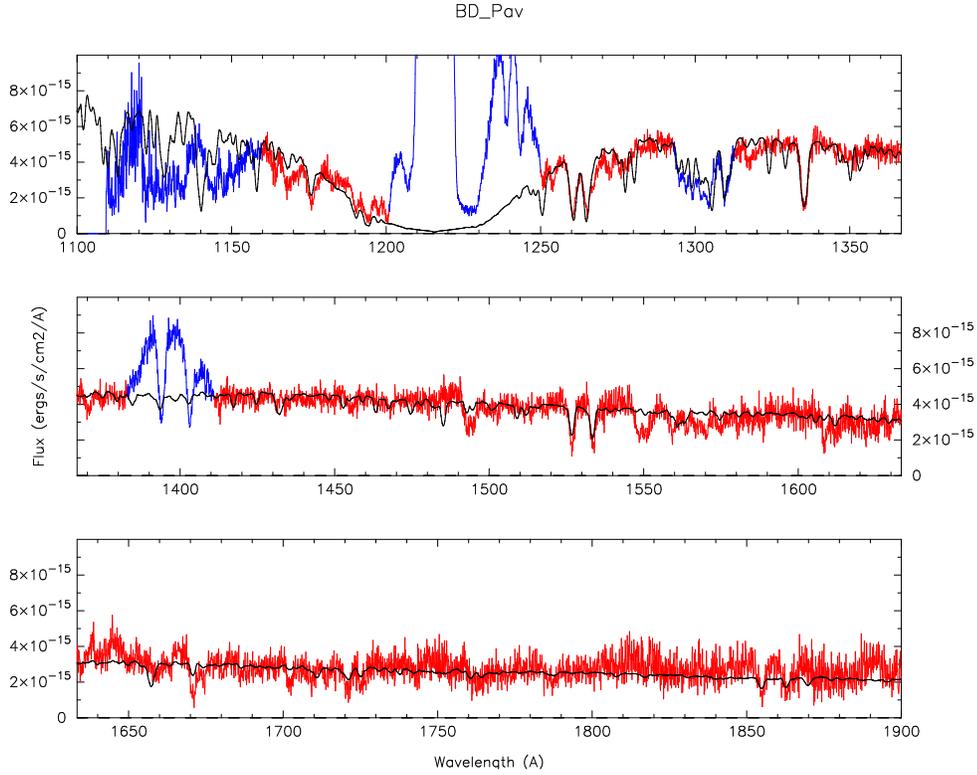
 
\vspace{-5.cm} 
\epsscale{0.8} 
\plotone{bdpav_T195g80Solv200ebv039.ps} 
\caption{
{\bf Above.(a)} 
The spectrum of BD Pav (in red)  
has been fitted with a WD model (in black), 
with a temperature of 19,500~K, gravity $log(g)=8.1$, with solar abundances
and broadening velocity of 200~km/s. The region below 1160~\AA\ 
could not be fitted and was masked together with the regions
contaminated by airglow and strong emission lines (in blue).  
{\bf Below. (b)} The addition of an ``iron curtain'' to the WD
model greatly improves the fit, especially in the shorter 
wavelengths region. Both the iron curtain and WD models
have subsolar carbon and silicon abundances.  
A second component was taken into account.   
{\bf The Gaia distance to BD Pav is $333 \pm 3 $~pc.}   
\label{bdpavwdfit} 
}
\plotone{bdpav_curtain.ps} 
\end{figure}

The final result of fitting a WD model to the COS spectrum of BD Pav 
yielded a temperature of $19,330 \pm 312$~K, gravity $log(g)=8.117 \pm 0.158$,
for a distance of $333 \pm 3$~pc, reddening $E(B-V)=0.057 \pm 0.018$. 
Two WD fits are presented in  Fig.\ref{bdpavwdfit}. 

In Fig.\ref{bdpavwdfit}a we present one of the solar abundance grid models
with $T=19,500$~K, $log(g)=8.1$, without an iron curtain, nor a second
component. Overall, the model fits the continuum flux level and some of
the absorption lines, but it fails to fit the very short wavelengths 
($\lambda < 1160$~\AA )
as well as many absorption features seen in the longer wavelengths 
($\sim 1500$~\AA ).  
This is a sign that an iron curtain is needed to better fit the spectrum. 

We also carried out a (iterative) WD plus iron curtain fit as we did for 
IY UMa and also included a second component of 
amplitude $7 \times 10^{-16}$erg/s/cm$^2$/\AA .  
We first assumed solar abundances for the WD and iron curtain,
but such a model also produced much more
pronounced Si\,{\sc ii} and C\,{\sc i}+C\,{\sc ii} absorption lines
that do not agree at all with the data. We noticed also that even the
single WD model produced carbon absorption lines (e.g. at 1140, 1160,
1325, and 1330~\AA ) that are not observed. We therefore decided to
vary the abundances of carbon and silicon  in both the iron curtain
and the WD model. 
As mentioned before, since the metals observed in the photosphere
of the WD are due to accretion, we set the WD and curtain abundances
to the same values in the model. 
We found that we have to lower the carbon and silicon abundances 
to fit most of the carbon and silicon lines. The WD with the iron curtain
model still produced
two strong lines that do not fit: one aluminum line (Al\,{\sc i} 1371.01~\AA ) 
and one phosphorus line (P\,{\sc ii} 1452.89~\AA ),  
both too strong in the model. 
Consequently, we also lower Al and P abundances to fit the data. 

The final result is presented in Fig.\ref{bdpavwdfit}b.  
The abundances we obtained  are 
[C]=$0.01_{-0.005}^{+0.01}$, [Al]=0.1$_{-0.05}^{+0.1}$, 
[Si]=0.1$_{-0.05}^{+0.1}$, and [P]$\lesssim0.01$ (assuming [Z]=1 for all other
species including Fe) in solar units for both the WD and its curtain.   
The broadening velocity is $200\pm50$~km/s. 
Here too, we notice that the sulfur  doublet is slightly deeper in the model
than in the observed spectrum, indicating that sulfur  is likely subsolar: 
$[S]<1.0$ solar. 
As with all other systems, the cold iron curtain has a temperature of 
10,000~K with an electron density of $n_{\rm e} = 10^{13}$cm$^{-3}$ \citep{hor94}.  
The turbulent velocity dispersion for the modeling of BD Pav iron curtain is  
$200\pm 50$~km/s, together with a hydrogen column density 
of $N_{\rm H} = 1.0_{-0.5}^+{1.0} \times 10^{21}$, which are needed to produce the strong iron 
absorption features. 

In the final model, the absorption that are not fitted are due to higher
ionization, i.e. C\,{\sc iii} ($\sim$1175), N\,{\sc v} ($\sim$1240), Si\,{\sc iv} 
($\sim$1400), and C\,{\sc iv} ($\sim$1550), all forming in a much hotter gas. 
We do not model this hotter gas, nor do we model
the emission lines.  
The spectrum also presents some broad emission lines of N\,{\sc v} ($\sim$1240),
Si\,{\sc iv} ($\sim$1400), and He\,{\sc ii} ($\sim 1640$) that we do not attempt to model. We also ignore
the Ly$\alpha$ and $\sim 1300$~\AA\ regions as they are contaminated
with daylight/airglow.

\clearpage

\clearpage

\subsection{{\bf HS 2214+2845}}

The COS data of HS 2214 consists of 4 subexposures obtained 
on 4 slightly different positions on the detector, 
slightly shifted relative to each other.  
The continuum flux level is the same in all the exposures, well
within the error bars, and the only difference is in the absorption lines.  
Therefore, we decided to start the analysis on the combined
(co-added) spectrum to derive the WD surface temperature and gravity, 
and to use the subexposures to derive the WD surface abundances and 
broadening velocity.  

\begin{figure}[b!] 
\epsscale{0.9}  
\vspace{-2.0cm} 
\plotone{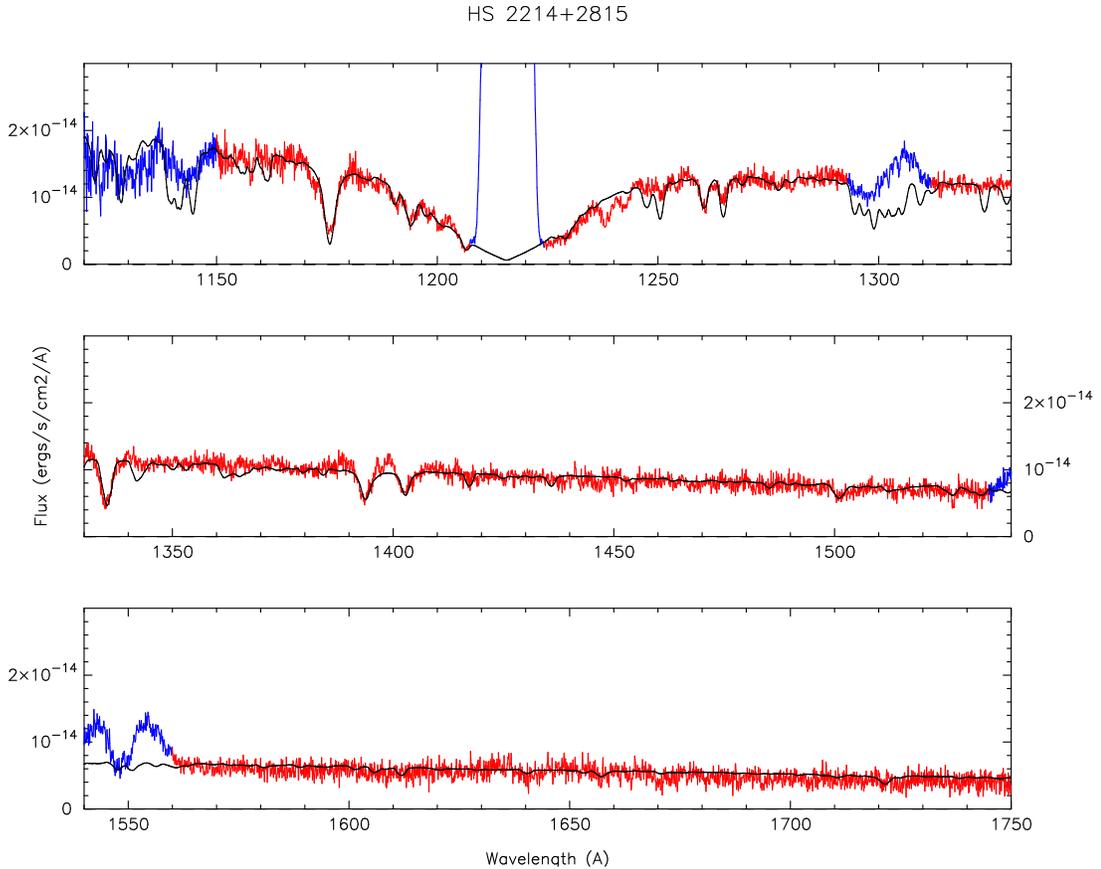}                 
\vspace{-0.5cm} 
\caption{
The COS spectrum of HS 2214+2845 (in red) has been fitted with a synthetic
WD model spectrum (in black) to derive the WD temperature and gravity,
for a distance of $400 \pm 8$~pc. The best fit yields a solution
$T_{\rm wd}=27,643 \pm 268$~K, $log(g)=8.30 \pm 0.16$. 
This models assumes a velocity broadening of 250~km/s and solar composition.
The regions in blue have been masked before the fitting. 
The fitting of the absorption lines is done separately in a 
different fit using individual subexposures.   
{\bf The Gaia distance to HS 2214+2815 is $400 \pm 8 $~pc.}   
\label{hs2214solarfit}
}
\end{figure}

The spectrum does not exhibit any sign of veiling and 
we carried out WD model fits without an iron curtain. 
We found that the addition of a small second component
(of amplitude $1.5 \times 10^{-15}$erg/s/cm$^2$/\AA ) slightly improves the fit.    
The WD temperature is $T_{\rm wd}=27,643 \pm 268$~K, with a gravity 
$log(g)=8.30 \pm 0.16$, for a distance of $400 \pm 8$~pc, and reddening
$E(B-V)=0.052 \pm 0.023$. Such a model fit is presented in Fig.\ref{hs2214solarfit},
with solar composition and broadening velocity
$V=250$~km/s. The model fits only a few absorption lines
and presents more lines than the observed spectrum. We then turned to the individual
subexposures to model the absorption lines.

\clearpage

From the timing of the subexposures (see Table \ref{obslog}), 
the first subexposure was obtained  at orbital phase $\Phi = 0.312$   
(the WD was moving toward the observer),  
subexposure \#2 at $\Phi = 0.522$ (the WD was facing the observer),  
subexposure \#3 at  $\Phi = 0.635$ 
(the WD started receding from the observer),  
and subexposure \#4 at $\Phi = 0.041$ (when the secondary was 
facing the observer). 
The shift observed in the absorption lines
is consistent with and confirms the above timing of the subexposures,
with one exception: in subexposure \#1 the C\,{\sc iii} ($\sim$1175) absorption
line is blue shifted by 3~\AA . Though, this could be due to hot material
ejected toward the observer around $\Phi \sim 0.3$, it is also the
very edge of the COS detector in the COS position in which this
exposure was set. We, therefore, ignore this 3~\AA\ blueshift.  

To derive the chemical abundances of the WD surface, 
we choose to fit subexposure \#2, since at the time the data was collected 
the WD was facing the observer. 
In addition, an important feature in the second exposure is the complete
absence of the N\,{\sc v} doublet ($\sim$1240) absorption lines, which are observed 
in the other three exposures. Also, the absorption
lines of C\,{\sc iii} ($\sim$1175), C\,{\sc ii} (1334), Si\,{\sc iv} ($\sim$1400),  
and C\,{\sc iv} ($\sim$1550) are not as pronounced in subexposure \#2 as in the other exposures. 
The N,{\sc v} (1240), and C\,{\sc iv} ($\sim$1550) absorption lines 
form in a much hotter gas and are not associated with the WD photosphere, 
and to some extent the C\,{\sc iii} ($\sim$1175) and Si\,{\sc iv} ($\sim$1400) lines, 
though they form in the WD photosphere at this temperature and gravity, 
might too be forming, in part, in the same hotter gas.    
Altogether, subexposure \#2 is less affected by these higher ionization
species absorption lines and is more representative of the WD spectrum. 
The continuum flux level in subexposure \#2 is otherwise the same
as in the other subexposures.  

\begin{figure}
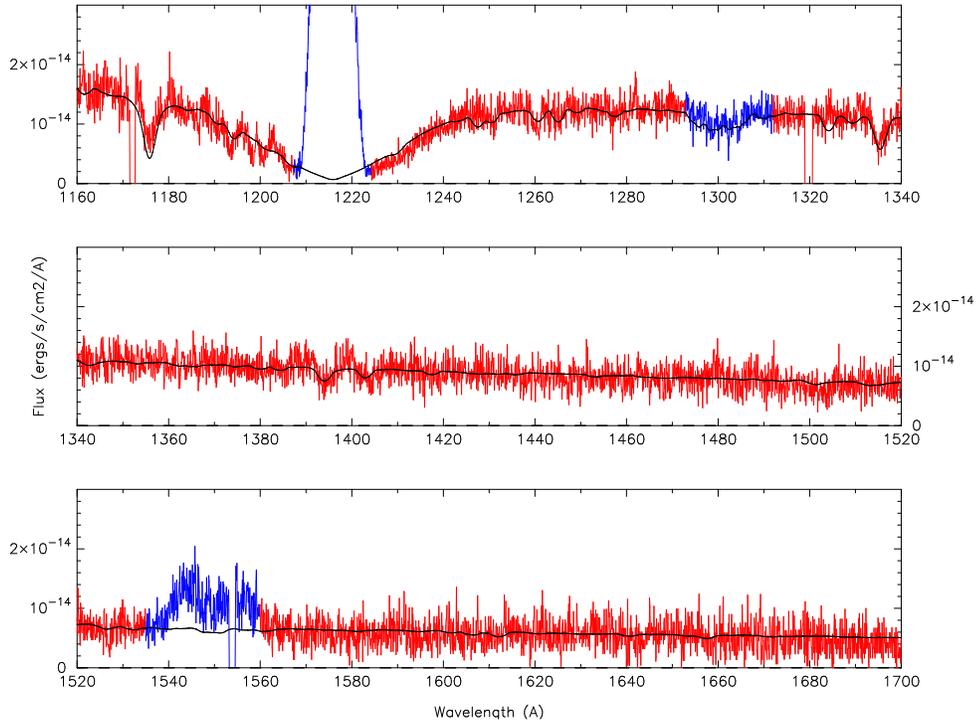
 
\vspace{-5.cm} 
\epsscale{0.8} 
\plotone{hs2214_s2.ps}                      
\caption{
{\bf (a) Above.} 
The subexposure \#2 (in red) of the COS spectrum of HS 2214+2845, obtained
at orbital phase $\sim 0.5$ (WD facing the observer), 
is used to derive the WD surface abundances and broadening velocity.  
The model (in black) is the same as presented in Fig.\ref{hs2214solarfit},     
except for a silicon abundance [Si]=0.2 solar (while
keeping the other species to solar) and a broadening velocity of 400~km/s. 
{\bf (b) Below.} 
For comparison, subexposure \#4, obtained near orbital phase 0.0 
(secondary facing the observer) shows stronger absorption lines, as well
as the apparition of additional lines, not seen in subexposure \#2,
but present in all other subexposures. 
The [Si]=0.2 model is also shown for comparison.  
\label{hs2214subfit} 
}
\plotone{hs2214_s4.ps}                      
\end{figure}

The fit to subexposure \#2 (Fig.\ref{hs2214subfit}a) 
gave a subsolar silicon abundance [Si]=$0.2\pm0.1 \times$solar, 
with a broadening velocity 
$V= 400 \pm 50$~km/s. All the other metals were kept
to solar abundances [Z]=1.  
The low silicon abundance together with the high broadening velocity
were needed to fit the shallow absorption line near 1194~\AA , 
1260~\AA\ and 1265~\AA . Even the 1300~\AA\ region, which 
can potentially be affected by airglow (as in exposure 4), 
is, too, well fitted. 

For comparison, in Fig.\ref{hs2214subfit}b, we show exactly the same model fit
together with subexposure \#4. One can clearly distinguish deeper absorption lines of 
C\,{\sc iii} ($\sim$1175), C\,{\sc ii} (1335), Si\,{\sc iv} ($\sim$1400) and 
C\,{\sc iv} ($\sim$1550), as well as the appearance of new absorption lines
of N\,{\sc v} ($\sim$1240) and Si\,{\sc ii} (1260 \& 1265).

The shift in the higher ionization species absorption lines follows
roughly that of the WD, indicating that they form near the WD, 
in (or more likely above) the hot inner disk. 
The reason these lines are attenuated or disappear near phase 0.5 (exposure \#2) 
is possibly be due to L1-stream  cold material
overflowing the edge of the disk and landing near phase 0.5-0.6 
\citep[see e.g.][]{lub89,god19} on the disk face near the WD. 
This material might reduce the scale height of the ionized material above the
inner disk face, pushing it down toward the disk mid-plane and out of the line 
of sight of the observer toward the WD, thereby removing from the spectrum
the absorption lines forming in the hot ionized material.  
Such a scenario requires a moderate inclination. 
The 4 exposures have the same continuum flux level, indicating that there
is not eclipse, occultation, or any other strong veiling of the WD.

\clearpage

\clearpage

\subsection{{\bf TT Crateris}}  

\begin{figure}[h!]
\epsscale{0.9}  
\vspace{-2.cm} 
\plotone{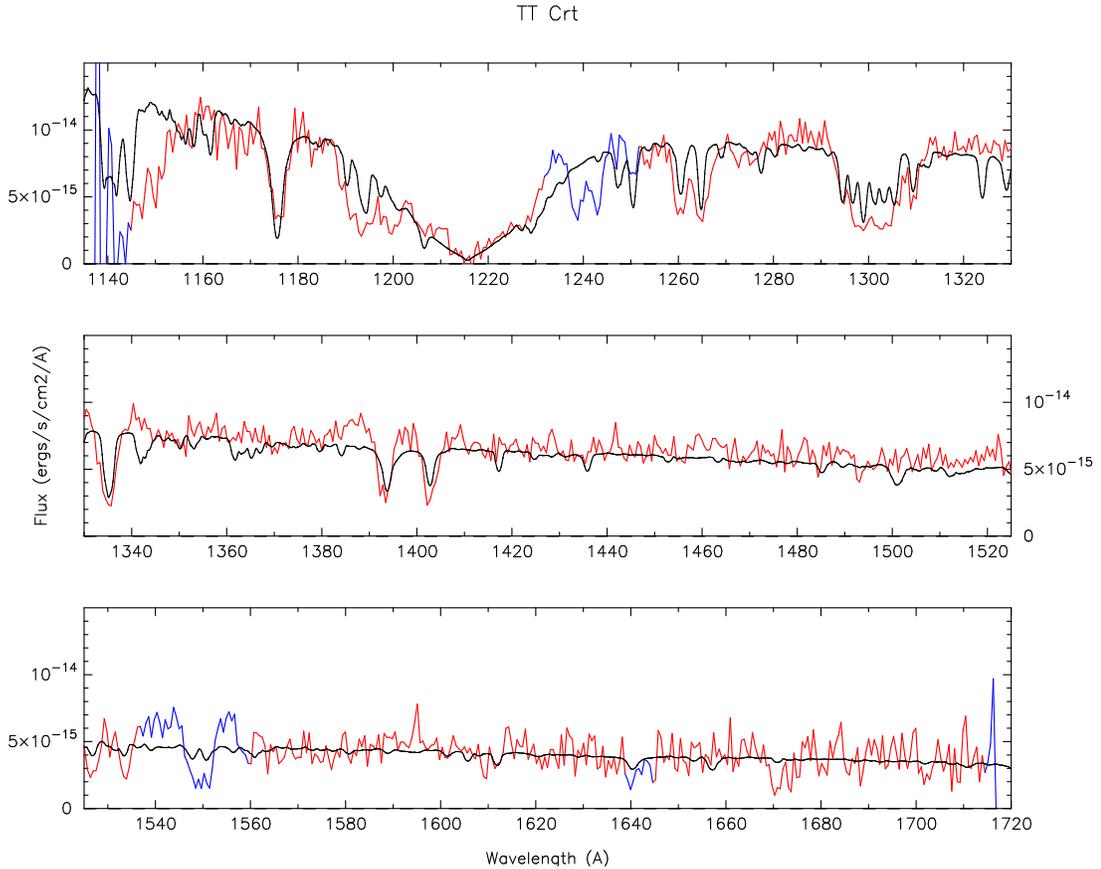}
\vspace{-0.5cm} 
\caption{
The results of the spectral fit of the STIS spectrum of TT Crt 
indicate a WD temperature $T_{\rm wd}= 26,990 \pm 557$~K with 
a gravity $log(g)=8.044 \pm 0.097$. This model has solar
abundances with a 200~km/s broadening velocity.  
{\bf The Gaia distance to TT Crt is $547^{+14}_{-12}$~pc.}   
\label{ttcrtwdsolfit}
}
\end{figure}

The STIS snapshot spectrum of TT  Crt was obtained near
binary orbital phase $\Phi = 0.29$, and with an inclination of 
$\approx 50-70^{\circ}$, it is not eclipsing nor is it   
expected to suffer from veiling. We therefore
carried out a single WD fit to the STIS spectrum, first assuming
solar composition. We found the need to add a second flat
component of amplitude $5\times 10^{-16}$~erg/s/cm$^2$/\AA .  
The analysis yielded a temperature of 
$26,990 \pm 557$~K with a gravity $log(g)=8.044 \pm 0.097$, 
for a distance $d=547_{-12}^{+14}$~pc and reddening 
$E(B-V)= 0.020 \pm 0.009$, including the propagation of all the  
errors. This model is presented in Fig.\ref{ttcrtwdsolfit}, from
which it is apparent that the absorption lines do not agree with  
solar abundances. The best matches, though far from being perfect,
are for the C\,{\sc iii} ($\sim$1175) line (which would be better fitted 
with a lower C abundance) and with the C\,{\sc ii} (1335) line 
(which would be better fitted with a higher C abundance
and higher velocity). All the silicon lines would be better
fitted with both a higher velocity and a higher Si abundance.

\clearpage 

\begin{figure}[h!]
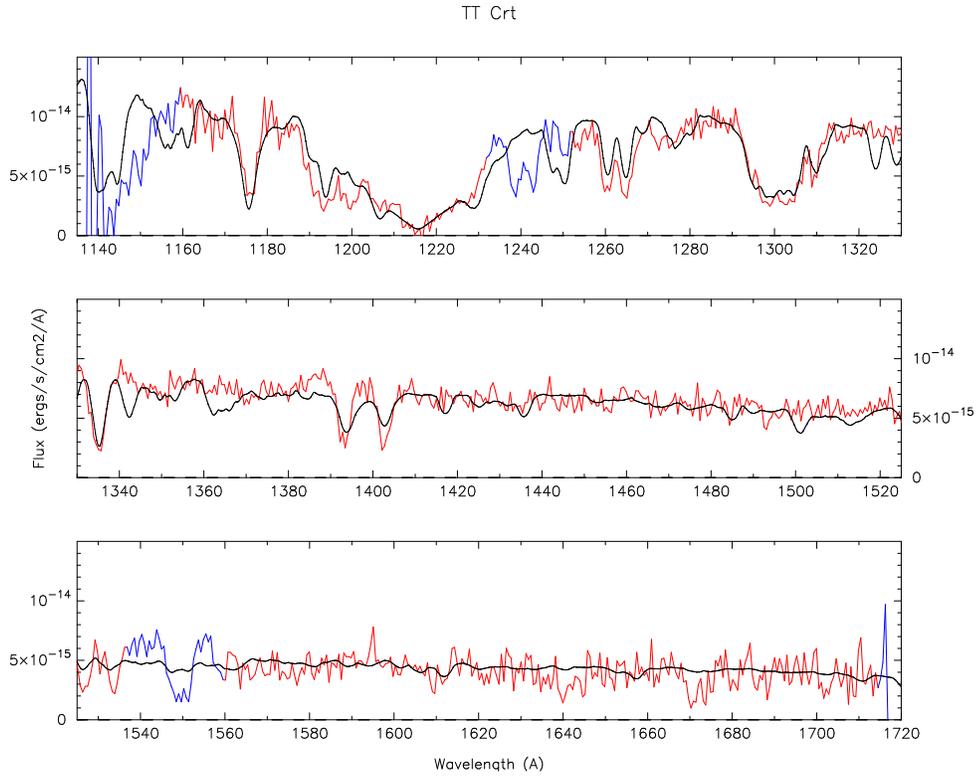

\vspace{-5.cm} 
\epsscale{0.8} 
\plotone{ttcrt_T28610g81Z5v400.ps}
\caption{
{\bf (a) Above.} 
Fitting all the absorption lines
simultaneously in the STIS spectrum of TT Crt with a WD model  
gives a metal abundance [Z]=5 solar with a 
broadening velocity of 400~km/s. See text for details.  
{\bf (b) Below.}  
Fitting a WD plus an iron curtain helps improve the fit in 
some regions but degrades the fit in the C\,{\sc iii} ($\sim$1175) line 
and Si (1300) feature. The (WD and iron curtain) abundances 
are subsolar: [C]=0.01, [Si=0.1], [P]=0.01, and [S]=0.1. 
The addition of the iron curtain significantly improves the fit
in the longer wavelengths (lower panel) and does not produce 
the C\,{\sc i} (1325 \& 1330) and Si\,{\sc iii} (1343 \& 1360) 
absorption lines which are not present in the STIS spectrum.  
See text for details. 
\label{ttcrtfits}
}
\plotone{ttcrt_ironcurtainfit.ps} 
\end{figure}

Consequently, next, we varied the chemical abundances
of the WD model as well as the 
broadening velocity in order to match the absorption lines. 
We found that a higher velocity together with higher    
abundances help fit the C\,{\sc ii} (1335) \& C\,{\sc iii} ($\sim$1175)
lines as well as the $\sim 1300$~\AA\ Si\,{\sc ii+iii} feature. 
A best fit was obtained for [Z]=5 (in solar units) with a broadening
velocity of 400~km/s. This model is presented in Fig.\ref{ttcrtfits}a.   
However, many of the other C and Si absorption lines are not well fitted. 
The Si\,{\sc ii} (1194, 1260, 1265), Si\,{\sc iv} ($\sim$1400) lines require  a 
higher Si abundance, while the absence of Si\,{\sc iii} (1343, 1360) lines  
require a lower Si abundance. Similarly, the absence of 
C\,{\sc i} (1325, 1330) lines also require a much lower C abundance. 
A possibility is that many of the lines that are observed
do not form in the stellar photosphere. 
Of course the C\,{\sc iv} ($\sim$1550) line with its emission wings form
in a much hotter gas, which could also partially contribute 
to the C\,{\sc iii} ($\sim$1175) and Si\,{\sc iv} ($\sim$1400) lines.  

The absence of  C\,{\sc i} (1325, 1330) lines in GY Cnc was due
to the fact that most lines formed in the iron curtain.  
In addition, we note that the spectrum of TT Crt shows 
absorption lines near 1610~\AA , 1670~\AA , and 1700~\AA , which are
prominent in the strongly veiled spectrum of DV UMa and BD Pav, 
and are due to the iron curtain. 
Hence, in the next step, we included an iron curtain in the modeling.
We assumed solar iron abundance for the curtain and varied its
parameters (i.e. turbulent velocity and hydrogen column density)
to match the spectrum of TT Crt in the longer wavelengths.  
A best fit gave a turbulent velocity of $100 \pm 25$~km/s
with a hydrogen column density if $5 \pm 2 \times 10^{20}$cm$^{-2}$.  
Keeping these two parameters constant, we then varied the silicon 
abundance of both the iron curtain and WD (assuming again that
they are equal) to match the Si\,{\sc ii} (1526.7 \& 1533.4) doublet
(which does not from in the WD stellar photosphere at this gravity
and temperature). We found that we have to lower it to [Si]=0.1. 
We then try and fit the C\,{\sc ii} (1335) feature, and find 
[C]=0.01. We lower [S] to 0.1  since the S\,{\sc i+ii} ($\sim 1250$)
lines are also very weak. And again we set [P]=0.01 as the 
P\,{\sc ii} (1452.89) is not observed. The need to lower all
these abundances is because many lines in the iron curtain
are very strong, even when assuming solar abundances.   
The low abundances of C, Si, P and S produce shallow absorption
lines in the WD stellar spectrum itself which are almost 
negligible when compared to the absorption lines due to the 
iron curtain.  This model is presented in Fig.\ref{ttcrtfits}(b).  
Here too, the fit is consistent with a hot gas (not modeled) contributing 
to the absorption lines of C\,{\sc iii} ($\sim$1175), Si\,{\sc iv} ($\sim$1400),
and C\,{\sc iv} ($\sim$1550). We also note that, while the iron curtain
provides a reasonable fit to the 1610~\AA ,1670~\AA , and 1700~\AA\
absorption features, it also fits the region 1560-1590~\AA\ pretty
accurately, which could easily be confused with noise. 
The small 1640~\AA\ absorption feature is due to He\,{\sc ii} and does
not form in the (cold) iron curtain nor in the WD stellar photosphere.     

\clearpage

\clearpage 

\subsection{{\bf V442 Centauri}} 

The STIS snapshot spectrum of V442 Cen, except for the characteristic hydrogen 
Ly$\alpha$ absorption, and N\,{\sc v} ($\sim$1240) and C\,{\sc iv} ($\sim$1550) broad emission lines, 
is rather featureless. It exhibits only three shallow
absorption lines: Si\,{\sc ii} ($\sim$1260),   C\,{\sc ii} ($\sim$1335), and    
Si\,{\sc iii} ($\sim$1502). 

\begin{figure}
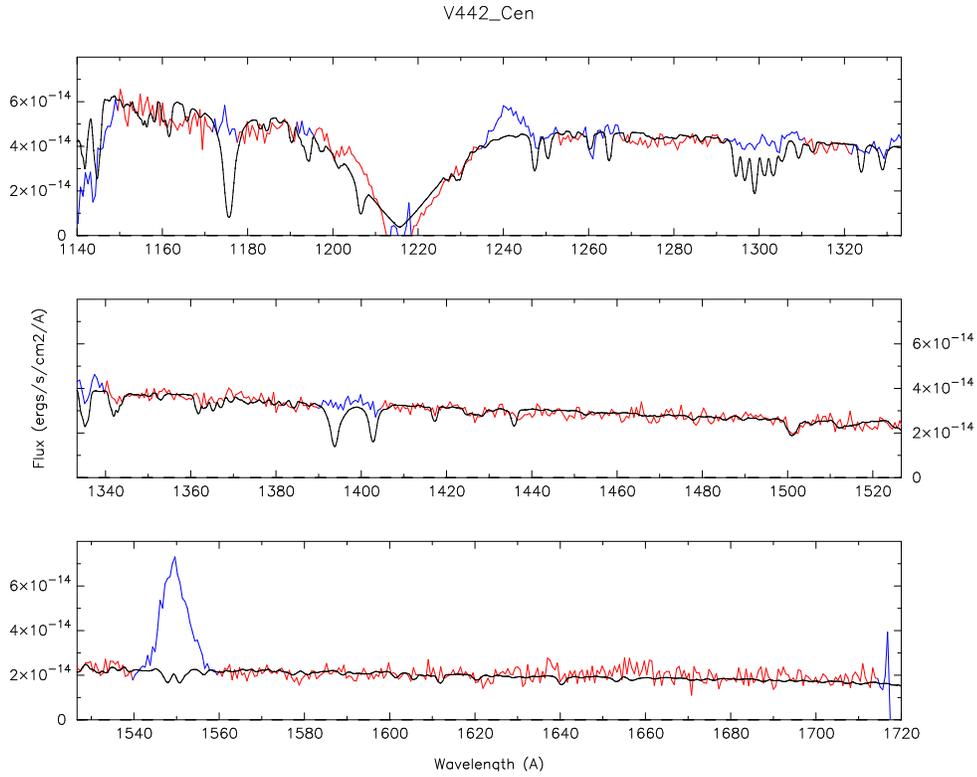
 
\vspace{-5.cm} 
\epsscale{0.8} 
\plotone{v442cen_T32g79Solv200ebv063.ps} 
\caption{
{\bf (a) Above.} 
The fit (in black) to the STIS spectrum (in red) of V442 Cen yields 
a WD temperature $T_{\rm wd}=31,032 \pm 308$~K with a gravity  
$log(g)=8.042 \pm 0.117$, for a distance 
$d=348 \pm 5$~pc, reddening $E(B-V)=0.048 \pm 0.015$.  
This model has solar abundances and a broadening                   
velocity of 400~km/s. The model has many absorption lines that are
not seen in the data, implying a low metallicity. 
{\bf (b) Below.}  
The the same as above in Fig.\ref{v442cenwdfit}a,  
but now the abundance of carbon has been lowered
to $10^{-3} \times$ solar, and that of silicon is 0.01, 
with a broadening  velocity of $V=250$~km/s. 
{\bf The dotted line represents the WD model only, the solid black
line includes ISM absorption modeling affecting the Ly$\alpha$ profile. 
The Gaia distance to V442 Cen is $348 \pm 5 $~pc.}   
\label{v442cenwdfit}
}
\plotone{v442cen_finalfit.ps} 
\end{figure}

We carry out a spectral fit without a second component
or an iron curtain, first with our grid of solar abundance models,
to find the temperature and gravity. 
The spectral fit results (Fig.\ref{v442cenwdfit}a) give 
a WD temperature of $31,032 \pm 308$~K, and gravity $log(g)=8.042 \pm 0.117$.  
The above-mentioned three shallow absorption lines can be fitted with nearly 
solar abundances as can be seen in Fig.\ref{v442cenwdfit}, but such a model
presents additional strong absorption lines of C\,{\sc iii} ($\sim$1175),
S\,{\sc ii} ($\sim$1250),  Si\,{\sc ii} (1265), C\,{\sc ii} (1330,
1335), Si\,{\sc iv} ($\sim$1400), which are not detected in the STIS spectrum.  

We note, however, that in Fig.\ref{v442cenwdfit}a the Ly$\alpha$ profile is rather 
poorly fitted. The bottom of the Ly$\alpha$ goes down to zero in
the observed spectrum with what seems to be some possible sharp emission
or noise in the middle. At such a high temperature, one does not expect
the bottom of the Ly$\alpha$ to go down to zero. 
The reddening toward V442 Cen is $E(B-V)=0.048 \pm 0.015$, while this is still
rather small, we decided to check how the ISM absorption affects the 
Ly$\alpha$ profile. We use the relation of \citet{boh78} and derive   
a hydrogen column density of $\approx 3 \times 10^{20}$cm$^{-2}$ 
from the value of the color excess $E(B-V)$.  
We then include modeling of the ISM absorption \citep[as described in][]{god07} 
taking $N(H_{\rm I} + N_2)=3 \times 10^{20}$cm$^{-2}$, a turbulent velocity
of 50km/s and a temperature of 170~K. This modeling does not include 
metals, but only hydrogen absorption. 
We find that we can better fit the bottom of the Ly$\alpha$ (bringing it to zero), 
but we can only fit the right wing of the Ly$\alpha$ profile.
The left wings of the profile itself seems to be affected by some weak
emission (near 1205~\AA ), as it slope is steeper than that of the right wing. 
We include the ISM modeling in the next step: modeling the absorption lines,
or rather modeling the absence of absorption lines. 

The fit (Fig.\ref{v442cenwdfit}b) to the absence of absorption lines 
(i.e. neglecting the shallow lines) gives a very small abundance
of carbon and silicon: [C]=0.001, [Si]=0.01, with [Z]=1.0 (in solar units),  
and a rotational broadening 
velocity of 250~km/s. The broadening velocity could be 
much smaller for a much smaller metallicity (of all metals), or could be 
larger too. Since we do not really fit a line, but rather the absence of
lines, we cannot derive the projected stellar rotational velocity.   
As discussed for SDSS 1538,  the Si\,{\sc ii} (1260) and C\,{\sc ii} (1335)
absorption lines are likely from the ISM,  
or from a very thin iron curtain (for C\,{\sc ii} (1335)). 
As to the the Si\,{\sc iii} (1502) absorption line, its origin is less 
clear. Again, as for SDSS 1538, the precise origin of these 3 lines
does not change our main result for the WD low 
abundances of carbon and silicon.

\clearpage 

\clearpage

\section{{\bf Summary, Further Considerations and Conclusions}} 
In Table \ref{spectralmodeling} we recapitulate the model fit assumptions  
made for each of the ten systems. 

Except for SDSS 1035, the temperature 
and gravity ($T,log(g)$) were modeled assuming solar abundances for 
different values of $E(B-V)$, with and/or without iron curtain, 
with and/or without a second component as indicated. The major
absorption lines were masked as indicated in the Figures. 
For SDSS 1035, $T$ and $log(g)$ were first assessed assuming solar abundance
(for $E(B-V)=0.0$), then they were re-assessed assuming $Z=0.03$ 
(for $E(B-V)=0.000$ and 0.034).  

For all the 10 systems, the abundances [Z] and broadening velocities $V$ were assessed for the best-fit
$T,log(g)$ found with a second component where needed, for one value of $E(B-V)$,
and without and/or without an iron curtain as detailed in Sec.5.   

In column (4) of Table \ref{spectralmodeling}: 
for clarity we did not list all the second component flux values associated with all the reddening values, 
but a larger reddening yielded a larger second component flux level. 
In column (5): 
if there is more than one reddening value, then the value of the second component flux (column 4)  
and the figures (column 7) are associated with the reddening values annotated with an asterisk $*$ 
in the same row.
Model fits were performed for different reddening values to assess the 
propagation of reddening uncertainties on the derived temperature and gravity. Since these effects 
were small, only one reddening value was assumed when deriving abundances and broadening velocities.

\begin{deluxetable}{lcccccc}[h!]  
\tablewidth{0pt}
\tablecaption{FUV Spectral Modeling Assumptions, Parameters, and Figures     
\label{spectralmodeling} 
} 
\tablehead{  
    (1)          &   (2)                  &   (3)           &    (4)                &     (5)          &  (6)     & (7)      \\ 
System           & Exposures              &  Parameters     & 2nd Component         &     $E(B-V)$     & Iron     & Fig.                \\ 
Name             &                        &  Modeled        & (erg/s/cm$^{-2}$/\AA) &                  & Curtain  &               \\ 
}
\startdata
{\bf SDSS1035}   &  Combined              & $T, log(g)$     &       ---             &  0.000           &  No     &                          \\
                 &  Combined              & $T,log(g),[Z],V$&       ---             &  0.000$^*$,0.034 &  No     & \ref{sdss1035wdfit},A\ref{sdss1035chi}       \\
{\bf SDSS1538}   &  Combined              & $T, log(g)$     &       ---             &  0.010$^*$,0.020 &  No     & \ref{sdss1538wdfit1},A\ref{sdss1538chifit2}  \\  
                 &  4th                   & [Z], $V$        &       ---             &  0.010           &  No     & \ref{sdss1538Si}  \\  
{\bf IY UMa}     &  2nd                   & $T, log(g)$     & $2.5\times10^{-16}$   &  0.000$^*$,0.030 &  No     & \ref{iyumawdfit}a   \\ 
                 &  2nd                   & $T,log(g),[Z],V$& $2.5\times10^{-16}$   &  0.000           &  Yes    & \ref{iyumawdfit}b,A\ref{iyumachi}       \\ 
{\bf DV UMa}     &  Snapshot              & $T,log(g)$      & $1.16\times10^{-16}$  &  0.000$^*$,0.020 &  No     & \ref{dvumawdfit}a,A\ref{dvumachi}    \\ 
                 &  Snapshot              & $T,log(g)$      & $1.16\times10^{-16}$  &  0.000           &  Yes    & \ref{dvumawdfit}b        \\ 
                 &  Snapshot              & $T,log(g),[Z],V$& $1.16\times10^{-16}$  &  0.000           &  Yes    & \ref{dvumawdicfit}        \\ 
{\bf IR Com}     &  1+2+3+5               & $T,log(g),[Z],V$& $3.5\times10^{-16}$   &  0.019$^*$,0.041 &  No     & \ref{ircomwdfit}a        \\
                 &  1+2+3+5               & $[Z],V$         & $3.5\times10^{-16}$   &  0.019$^*$,0.041 &  Yes    & \ref{ircomwdfit}b        \\
                 &  1+2+3+5               & $T,log(g)$      & $3.85\times10^{-16}$  &  0.030           &  No     & A\ref{ircomchiebv030}        \\
{\bf GY Cnc}     &  Snapshot              & $T,log(g)$      & $2.0\times10^{-15}$   &  0.010$^*$,0.036 &  No     & \ref{gycncwdicfit}a                      \\ 
                 &  Snapshot              & $T,log(g)$      & $2.0\times10^{-15}$   & 0.010$^*$,0.036  & Yes     & \ref{gycncwdicfit}b,A\ref{gycncchi} \\ 
                 &  Snapshot              & $[Z],V$         & $2.0\times10^{-15}$   &  0.010           &  Yes    & \ref{gycncfinalfit}       \\ 
{\bf BD Pav}     &  Combined              & $T,log(g)$      &       ---             &  0.039$^*$,0.075 &  No     & \ref{bdpavwdfit}a,A\ref{bdpavchifit}       \\
                 &  Combined              & $T,log(g),[Z],V$& $7.0\times10^{-16}$   &  0.039           &  Yes    & \ref{bdpavwdfit}b       \\
{\bf HS2214}     &  Combined              & $T,log(g)$      & $1.5\times10^{-15}$   &  0.029,0.052$^*$ &  No     & \ref{hs2214solarfit},A\ref{hs2214chi}       \\
                 &  2nd                   & $[Z],V$         & $1.5\times10^{-15}$   &  0.052           &  No     & \ref{hs2214subfit}a        \\
                 &  4th                   &                 & $1.5\times10^{-15}$   &  0.052           &  No     & \ref{hs2214subfit}b        \\
{\bf TT Crt}     &  Snapshot              & $T,log(g)$      &         ---           &  0.012$^*$,0.029 &  No     & A\ref{ttcrtchi}       \\
                 &  Snapshot              & $T,log(g),[Z],V$& $5.0\times10^{-16}$   &  0.012           &  No     & \ref{ttcrtwdsolfit},\ref{ttcrtfits}a       \\
                 &  Snapshot              & $[Z],V$         & $5.0\times10^{-16}$   &  0.012           &  Yes    & \ref{ttcrtfits}b       \\
{\bf V442 Cen}   &  Snapshot              & $T,log(g)$      &     ---               &  0.033,0.063$^*$ &  No     & \ref{v442cenwdfit}a,A\ref{v442cenchi}       \\
                 &  Snapshot              & $[Z],V$         &     ---               &        0.063     &  No     & \ref{v442cenwdfit}b       \\
\enddata
\end{deluxetable}

\clearpage

\subsection{\bf{ WD Masses and Temperatures.}} 

In Table \ref{mwdtwd} we list the WD temperature and gravity for the 
10 objects analyzed here, together with the corresponding WD mass and radius,    
which we derive using the mass-radius relation for non-zero temperature 
C-O WD from \citet{woo95}. 
For comparison, we also list the WD masses that were derived from the eclipse
light curves available for 4 systems \citep[SDSS 1035, IY UMa, DV UMa, and 
GY Cnc;][]{sav11,mca19}. 
Except for GY Cnc, we find that our FUV-analysis-derived WD masses
agree with the eclipse-light-curve-derived WD masses within the error bar
(1-$\sigma$). For GY Cnc the mass (or $log(g)$) agrees within 2.6$\sigma$.   
At the time of completion of this work, we were also able to compare
our WD masses and temperature with the work of \citet{pal21} for 6 systems. 
For comparison, we list In Table \ref{mwdtwd} the temperatures and WD masses 
obtained by \citet{pal21}.  
This enables us to confirm the results for 3 more systems:  
SDSS 1538, IR Com and V442 Cen for which we also find good WD mass agreement within the
error bars.   

While the WD mass we found for SDSS 1035 agrees well
with that derived from eclipse light curve, it does not agree with that derived
by \citet{pal21}.  A possible explanation for the discrepancy is that \citet{pal21}
includes a second component in the modeling of SDSS 1035, while we do not find this 
necessary. Another disagreement, though of less importance, is that for V442 Cen 
we obtained a temperature about 1200~K higher than \citet{pal21}. 
We recall, however, that we included ISM modeling due to the Ly$\alpha$ going down to zero, 
while the HST spectrum in \citet{pal21} does not appear to go down to zero in that region. 
We use the data calibrated by \textsc{calcos} from \textsc{mast} and 
\citet{pal21} likely re-calibrated the data differently. Furthermore, they do not take into consideration 
possible absorption from the ISM. 

For the rest, our results for $T_{\rm wd}$ and $M_{\rm wd}$ agree within the error bars with
the results of \citet{pal21}. Small differences are likely due to the different versions of 
\textsc{tlusty} (203 vs. 204n), 
and the running parameters we use in \textsc{synspec} such as for the NLTE approximation 
at high temperature and convection at low temperature.   
We also use different prescriptions for the hydrogen quasi-molecular satellite lines opacity.
Furthermore, the second component and the absorbing curtains are treated differently; 
the dereddening, masking, and calibration of the spectra are also performed differently.
However, overall, our results and \citet{pal21}'s results confirm each other \citep{pal21a}. 

The reason for the poor agreement found for GY Cnc with the elicpse light curve WD mass
may be due to the strong veiling.  
GY Cnc is the system the most affected by veiling after IY UMa. Its modeling required an iron curtain 
with a hydrogen column density of $5 \times 10^{21}$cm$^{-2}$,
10 times larger than for TT Crt. It is worth noting also that GY Cnc is the object for
which \citet{mca19} obtained the largest discrepancy (2$\sigma$) for the derived distance
based on their WD atmosphere fit: they obtained a distance of 320~pc with
a WD temperature of $\sim 25,900$~K. The exact reasons for the WD atmosphere fit 
discrepancies in the present work and in \citet{mca19} are not known, but, in addition
to strong veiling, GY Cnc has a prominent bright spot and its HST STIS spectrum was obtained
a few weeks only after outburst. 
We further emphasize that this is the first attempt to model the HST FUV spectrum of GY Cnc 
and that this object was excluded from all previous analyses (by us and by others).   

We can therefore confirm that FUV spectral fits supplemented with Gaia distances provide a 
robust method to derive CV WD masses and temperatures and open the path to constraining
the evolution of CVs \citep{pal21}.

\begin{deluxetable}{lcccccccc}[h!]  
\tablewidth{0pt}
\tablecaption{FUV Spectral Analysis Results: WD Masses and Temperatures   
\label{mwdtwd} 
} 
\tablehead{ 
  (1)          &    (2)           &     (3)       &   (4)           &    (5)               &   (6)            &     (7)          &   (8)          &  (9)    \\  
System         & $T_{\rm wd}$     & \citet{pal21} &  $log(g)$       &  $M_{\rm wd}$        & $M_{\rm wd}^E$   & \citet{pal21}    &  $R_{\rm wd}$  & Iron    \\  
Name           & (K)              &    (K)        &                 &  $(M_{\odot})$       & $(M_{\odot})$    & $(M_{\odot})$    &  (km)          & Curtain  
}
\startdata
{\bf SDSS1035} & $11,475\pm188$   & $11,876^{+108}_{115}$ & $8.385\pm0.166$ & $0.830^{+0.118}_{-0.102}$ & $0.835\pm0.009$  & $1.00^{+0.08}_{-0.10}$ & $6740^{-790}_{+890}$ & No \\[3pt]
{\bf SDSS1538} & $35,743\pm576$   & $35,284^{+600}_{-688}$ & $8.708\pm0.130$ & $1.073^{+0.066}_{-0.074}$ &                  & $0.97^{+0.09}_{-0.11}$ & $5284^{-600}_{+640}$ & No \\[3pt]
{\bf IY UMa}   & $17,130\pm249$   & $17,057^{+179}_{-79}$ &$8.480\pm0.125$  &$0.903^{+0.090}_{-0.088}$  &$0.955^{+0.013}_{-0.028}$& $0.99^{+0.04}_{-0.03}$ & $6300^{-575}_{+610}$ & Yes \\[3pt]
{\bf DV UMa}   & $19,325 \pm 481$ & $19,410^{+244}_{-400}$ & $8.565\pm0.177$ & $0.968^{+0.116}_{-0.127}$ & $1.09\pm0.03$    & $0.96^{+0.07}_{-0.10}$ & $5910^{-820}_{+850}$ & Yes \\[3pt]
{\bf IR Com}   & $17,860\pm180$   & $17,531^{+236}_{-271}$ & $8.613\pm0.090$ & $1.001^{+0.055}_{-0.066}$ &                  & $1.03^{+0.07}_{-0.09}$ & $5690^{-420}_{+410}$ & Yes \\[3pt]
{\bf GY Cnc}   & $22,515 \pm 292$ &                        & $7.876\pm0.108$ & $0.570^{+0.050}_{-0.050}$ & $0.881\pm0.016$  &                        & $10043^{-800}_{+830}$ & Yes\\[3pt]
{\bf BD Pav}   & $19,330 \pm 312$ &                        & $8.117\pm0.158$ & $0.683^{+0.087}_{-0.083}$ &                  &                        & $8320^{-960}_{+1035}$ & Yes\\[3pt]
{\bf HS2214}   & $27,643 \pm 268$ &                        & $8.301\pm0.160$ & $0.798^{+0.108}_{-0.086}$ &                  &                        & $7260^{-810}_{+1020}$ & No \\[3pt]
{\bf TT Crt}   & $26,990 \pm 557$ &                        & $8.044\pm0.097$ & $0.663^{+0.049}_{-0.049}$ &                  &                        & $8910^{-645}_{+690}$ & Yes \\[3pt]
{\bf V442 Cen} & $31,032 \pm 308$ & $29,802^{+211}_{-247}$ & $8.042\pm0.117$ & $0.671^{+0.057}_{-0.057}$ &                  & $0.64^{+0.06}_{-0.05}$ & $8980^{-785}_{+870}$ & No \\[3pt]
\enddata
\tablecomments{
In column (2) we list the WD temperatures from our spectral analysis, {\bf followed in column (3) 
by the WD temperatures from \citet{pal21}.} 
In column (5) we list the WD masses obtained from our spectral analysis together with the
WD masses derived from eclipse light curves (6) \citep{sav11,mca19} {\bf and from \citet{pal21} (7).}      
The masses (5) and radii (8) were computed 
from the values of $log(g)$ (4) using the mass-radius relation for non-zero temperature WD \citep{woo95}. 
Since for each spectrum the solution is a narrow diagonal band in the $T_{\rm wd}$ vs $log(g)$ parameter space, 
the larger temperature (+) is associated with the larger gravity (+), larger WD mass (+), 
and smaller WD radius (-), and {\it vice versa}. 
} 
\end{deluxetable}

\clearpage

\subsection{\bf{ WD Chemical Abundances and Stellar Rotational Velocities.}}

The abundances analysis, summarized in Table \ref{zwdvwd}, reveals that only two systems
(IY UMa and HS 2214) have a WD spectrum consistent with solar carbon abundance
[C]=1.  Seven systems have [C]$\sim$0.1-0.0001, 6 have subsolar silicon 
([Si]$\sim$0.1-0.01), and 4 have [P]$\lesssim$0.01 and [S]$<$1. 
Of all the 10 systems, only IY UMa has solar metal abundances ([C]=[Si]=[Z]=1; 
see also below). IR Com is considered separately in Sec.\ref{thinic}.  

All the dominant absorption lines in the spectra are due to carbon and
silicon, which makes the determination of the abundance of these species
more reliable. 
The low abundance of P was based solely on the absence of the single line
P\,{\sc ii} (1452.89) in the observed spectra which forms in the iron curtain
models. 
The S abundance was based solely on the sulfur  doublet near $\sim 1250$. 

Except for the N\,{\sc v} ($\sim$1240) doublet forming in a hot gas, no strong nitrogen absorption lines 
were detected in the spectra and nitrogen abundance could not be assessed.
Some nitrogen lines do form in the short wavelength region, but are either too shallow to give  
a reliable measurement (e.g. $\sim 1200$~\AA ), or are in that region near the edge of the detector
and where the iron curtain affects the spectra ($\sim 1140-1160$~\AA ).  
Nevertheless, nitrogen was kept solar, just as               
all the other metals marked in Tables \ref{zwdvwd} \& \ref{curtain} with [Z]=1.  

In Table \ref{zwdvwd}, we list the broadening velocities of the lines from the best fit
model for each system.  These velocities range from 150~km/s to 400~km/s,
these velocities are all sub-Keplerian, similar to the values obtained in earlier
studies with HST \citep{sio99}.   
The velocity $V$ is expected to correspond to the projected WD stellar rotational/spin velocity
$V_{\rm rot} sin(i)$ when the lines are not affected by the motion of the WD around the center
of mass during the duration of the exposure.
We list (column 10) the approximate velocity broadening $V_{\rm WD}$ due to the WD motion around the center of mass
during the duration of the observation. 
For the systems for which the absorption lines are not due to the absorbing slab, 
one can assume a projected WD rotational velocity $V_{\rm rot} sin(i) \approx V-V_{\rm WD}$. 
Doing so we obtain $V_{\rm rot} sin(i)$ for SDSS 1035 ($\approx 100$~km/s), SDSS 1538 ($\approx 300$~km/s), 
IR Com ($\approx 50$~km/s, or less; the size of the velocity uncertainty), HS 2214 ($\approx 375$~km/s),
and V442 Cen ($\le 250$~km/s). 
For the other systems (IY UMa, DV UMa, GY Cnc, BD Pav, TT Crt), where the absorption lines are due mainly to the
iron curtain, the resulting velocity $V-V_{\rm WD}$ is more representative of the actual  
broadening of the absorption lines in the veiling material, which depends on both the
turbulent velocity and column density (see next subsection).

\begin{deluxetable}{lcccccccccc}[h!]  
\tablewidth{0pt}
\tablecaption{FUV Spectral Analysis Results: Abundances and Velocities    
\label{zwdvwd} 
} 
\tablehead{  
    (1)            &   (2)                 &   (3)                 &    (4)                &   (5)           &    (6)                &   (7)     &   (8)         &     (9)             &  (10)     & (11)  \\ 
System             &   [C]                 &    [Al]               &    [Si]               &    [P]          & [S]                   &  [Fe]     &  [Z]          & $V$   & $V_{\rm WD}$ &  N\,{\sc h}  \\ 
Name               & ($\odot$)             & ($\odot$)             & ($\odot$)             & ($\odot$)       & ($\odot$)             & ($\odot$) & ($\odot$)     &   (km/s)            & (km/s)   & ($10^{20}$cm$^{-2}$)      
}
\startdata
{\bf SDSS1035}   &  $0.03\pm0.02$          &                       &                       &                 &                       &           & $0.03\pm0.02$ & $150\pm50$          &    57     & ---  \\[3pt] 
{\bf SDSS1538}   & $\sim10^{-4}$           &                       & $1_{-0.5}^{+1}$       &                 &                       &           & 1.0           & $400\pm100$         & $\sim$100 & ---  \\[3pt]
{\bf IY UMa}     &                         &                       &                       & $\lesssim 0.01$ &                       &           & 1.0           & $150\pm50$          &    44     & 100  \\ [3pt]  
{\bf DV UMa}     & $0.1_{-0.05}^{+0.1}$    &                       & $0.3\pm0.1$           & $\lesssim 0.01$ & $\lesssim1.0$         &           & 1.0           & $150\pm50$          &    40     & 20   \\ [3pt]  
{\bf IR Com}     & $1.0_{-0.5}^{+1.0}$     & $5\pm1$               & $1\pm0.5$             &                 &                       & $4\pm1$   & 1.0           & $150\pm50$          &    154    & ---  \\[3pt]
{\bf IR Com}     & $0.4\pm0.1$             &                       &                       &                 &                       &           & 1.0           & $150\pm50$          &    154    & $\lesssim 0.1 $ \\ [3pt]  
{\bf GY Cnc}     & $0.01_{-0.005}^{+0.01}$ &                       & $0.1_{-0.05}^{+0.10}$ & $\lesssim 0.01$ & $0.1_{-0.05}^{+0.10}$ &           & 1.0           & $150\pm50$          &    28     & 50   \\ [3pt]  
{\bf BD Pav}     & $0.01_{-0.005}^{+0.01}$ & $0.1_{-0.05}^{+0.10}$ & $0.1_{-0.05}^{+0.10}$ & $\lesssim 0.01$ & $\lesssim1.0$         &           & 1.0           & $200\pm50$          &    190    & 10   \\ [3pt]  
{\bf HS2214}     &                         &                       & $0.2\pm0.1$           &                 &                       &           & 1.0           & $400\pm50$          & $\sim$26  & ---  \\[3pt]
{\bf TT Crt}     & $0.01^{+0.01}_{-0.005}$ &                       & $0.1^{+0.10}_{-0.05}$ & $\lesssim 0.01$ & $0.1^{+0.10}_{-0.00}$ &           & 1.0           & $400\pm100$         &    25     & 5    \\[3pt]     
{\bf V442 Cen}   & $\lesssim0.001$         &                       & $\lesssim0.01 $       &                 &                       &           & 1.0           & $\gtrsim250$        & $\sim$10  & ---  \\[3pt]
\hline 
\enddata
\tablecomments{
When a specific absorption line was used to model the abundance of a given element, 
the resulting abundance is written directly the column of that element. 
The remaining of the chemical elements were set to the value listed in column (8) 
marked with [Z]. 
In column (9) we list the broadening velocity
of the absorption lines from the analysis. 
In column (10) we list the approximate velocity broadening due to the motion of the WD 
around the center of mass during the time the exposure was obtained,
computed using the $K_1$ values (see Table \ref{syspar};  assumed to be 100~km/s if unknown)
and the ratio of the exposure time (Table \ref{obslog}) to the orbital period
(Table \ref{syspar}).  
In the last column (11) we indicate the hydrogen column density of the iron curtain.   
Assuming that IR Com is affected by an iron curtain implies that the WD photospheric 
abundances are all solar (1x), except carbon (0.4x), for that reason results are shown 
with and without an iron curtain - see text for a detailed discussion.  
}
\end{deluxetable}

\clearpage

\subsection{\bf{The Iron Curtain.}} 

The effect of veiling material is obvious in IY UMa, DV UMa, GY Cnc, 
and BD Pav, and, while less noticeable, it is still detectable in TT Crt.  
For these 5 systems, the abundances were derived using an iron curtain 
masking the WD, and assuming that the WD and iron curtain have the same
abundances. The iron curtain models are listed in Table \ref{curtain}
for each veiled system.  
  
In the iron curtain modeling, we first fitted the forest of iron lines
in the longer wavelengths assuming solar iron abundance for the 
curtain (and WD). We then found that  
even low abundances of C, Si, S,..  ($<< 1$ solar) in the curtain   
generate deep and wide absorption lines, which forced us to reduce these
abundances to subsolar. 
Since we assumed that the WD and iron curtain had the same abundances, 
we reduced the WD photosphere C, Si, S, .. abundances as well, which removed from
the model lines that were not observed (sulfur  lines at $\sim$1250~\AA ,
silicon lines at 1260,1265~\AA , $\sim$1340~\AA , $\sim$1360~\AA, 
carbon lines at $\sim$1330~\AA , etc..; these lines do not form in the iron curtain). 
Therefore, our assumption that the WD and iron curtain have the same abundances is  
self-consistent. We did not change, however, the iron abundance in the curtain
and kept it to solar, since it is used to match the iron forest in the longer wavelengths 
and derive the iron curtain basic parameters (turbulent dispersion velocity and 
hydrogen column density).  

Assuming solar iron abundance for the iron curtain is reasonable, but may
introduce an uncertainty.  
While the iron abundance in a CV WD or secondary 
star cannot be large enough to produce the kind of Fe\,{\sc ii} absorption lines 
seen in the FUV spectra of these 5 systems, the iron abundance of stars 
(at all age [0-10 Gyr])  in the solar neighborhood shows a scatter of 
[Fe/H] abundances of up to about $\pm 0.5$~dex \citep{reb21}.  
Metal-rich ([Fe/H] $\sim 0.2$~dex or 1.6$\times$solar) and 
super-metal-rich ([Fe/H] $\sim 0.5$~dex, or 3.2$\times$solar) 
stars in the solar neighborhood are not distinct in terms of their kinematics 
or ages, and seem to form a continuous distribution; this does not exclude that some of the 
super-metal-rich stars in the solar neighborhood might originate
from the inner disk and even the Galactic bulge \citep{fel13}. 
Therefore, we do not expect the iron abundance in a CV WD to be larger than $\sim3\times$solar,
and the probability of it reaching $\sim 3$ is certainly very small.  
It is more likely that the iron abundance in a CV WD is nearly 1.    
Consequently, the abundances of C, Al, Si, P, and S derived from the solar [Fe/H] assumption 
in the iron curtain (Table \ref{curtain}) is a good assumption and would be (in the extreme case)
accurate to within a factor not exceeding $\sim 3$ ($\sim 0.5$~dex).

\begin{deluxetable}{ccccccccc}[h!] 
\tablewidth{0pt}
\tablecaption{Iron Curtain Modeling Parameters  
\label{curtain} 
} 
\tablehead{
System    &    N\,{\sc h}            & $V_{\rm turb}$ & [C]     & [Al]     &  [Si]    &  [P]           & [S]         &  [Z]    \\ 
Name      &    (cm$^{-2}$)           &   (km/s)       & $\odot$ & $\odot$  & $\odot$  & $\odot$        & $\odot$     & $\odot$     
}
\startdata
IY UMa    & $1\pm0.2\times10^{22}$   &  75$_{-15}^{+25}$ &      &          &          & $\lesssim0.01$ &             &  1.0   \\   
DV UMa    & $2\pm1\times10^{21}$     &  75$\pm25$     & 0.1     &          & 0.3      & $\lesssim0.01$ & $\lesssim1$ &  1.0   \\   
IR Com    & $\lesssim1\times10^{19}$ & 100$\pm25$     & 0.4     &          &          &                &             &  1.0   \\ 
GY Cnc    & $5\pm2\times10^{21}$     &  75$\pm25$     & 0.01    &          & 0.1      & $\lesssim0.01$ &  0.1        &  1.0   \\   
BD Pav    & $1\pm0.3\times10^{21}$   & 200$\pm50$     & 0.01    & 0.1      & 0.1      & $\lesssim0.01$ & $\lesssim1$ &  1.0   \\   
TT Crt    & $5\pm2\times10^{20}$     & 100$\pm25$     & 0.01    &          & 0.1      & $\lesssim0.01$ &  0.1        &  1.0   \\   
\enddata
\tablecomments{
All the iron curtain models here have the same temperature ($T=10,000$~K) and 
electron density ($n_e=10^{13}$cm$^{-3}$) as in \citet{hor94}, and iron is set to solar ([Fe]=1). 
The error bars on the abundances are the same as for the WD in Table \ref{zwdvwd}.  
Here too, when a specific absorption line was used to model the abundance of a given element, 
the resulting abundance is written directly the column of that element. 
The remaining of the chemical elements were set to the abundance in the column
marked with [Z]. 
} 
\end{deluxetable} 

It is true that 
a single isothermal iron curtain model with a given electron density, hydrogen column density, 
and turbulent velocity is only an approximation, since the veiling material is made of 
a gas that is not especially isothermal, nor has a constant density,  
nor a constant turbulent velocity. As such, we do not expect the fit to be perfect 
and the iron curtain only comes to improve the WD fit. 
Furthermore, one has to be aware that 
a lower iron curtain density requires a lower iron curtain temperature 
for the iron features to form
and the hydrogen column density inversely correlates with the turbulent
velocity to match the strength of the curtain absorption features in the
final spectrum \citep{hor94}. 

The low carbon abundance we found in 7 systems cannot be due to the iron curtain
simplistic modeling, since  
for the systems that did not include an iron curtain modeling, we find 
3 systems out of 5 also have a low carbon abundance (for the systems with an
iron curtain we find 4 out of 5 with a low carbon abundance). The same
is true for the low silicon abundance. However, the same cannot be said
of the low phosphorus abundance. The P\,{\sc ii} ($\sim$1453) absorption line  only
appeared in the iron curtain modeling and no P lines appear in the WD 
stellar model. Since none of the 10 HST WDs spectra analyzed here exhibit  the
P\,{\sc ii} absorption line, we cannot rule out that the P\,{\sc ii} line  
is the product of an inaccurate iron curtain modeling, 
albeit able to reproduce all the other spectral features due to C, Al, Si, S, and Fe. 
More iron curtain modeling is needed, as well as a rigorous inspection of all the
CV WDs FUV spectra available to determine whether this line is the sole product of an inaccurate
iron curtain modeling or whether it is also an observed feature.  

We note that the absorbing slab turbulent velocity listed in Table \ref{curtain} 
does not especially correspond to the broadening velocity of the absorption lines
in the veiling material, since the broadening of the iron curtain lines (for a given
temperature and electron density) depends also on the hydrogen column density. 
For that reason, and except for IR Com where most of the lines are from the WD photosphere,
we cannot derive the projected rotational stellar velocity for the systems listed
in Table \ref{curtain}.   

\subsection{\bf{The Thin Iron Curtain: IR Com with or without a Mask?} \label{thinic}} 

In a system like HS 2214+2845, the spectra reveal 
stronger absorption lines as well as the appearance of new lines
at orbital phases $\sim 0.0$, 0.31, and 0.64, but not when 
the WD is facing the observer near orbital phase 0.5. 
While it is known that veiling material, likely due to the L1-stream
overflowing the edge of the disk, can affect the spectrum
at orbital phases near 0.25 and 0.6-0.8 \citep[see][and references therein]{god19}, 
it is not clear that some partial veiling does not take place 
at other orbital phase or even in systems with a low inclination.  

From the IR Com thin iron curtain ($N_{\rm H}=10^{19}$cm$^{-2}$)  
modeling,
we find that {\it seemingly} one cannot differentiate between 
a thin curtain veiling a solar abundances (with [C]=0.4 solar) 
WD photosphere and a non-veiled WD photosphere with increased (supra-solar)  
aluminum ([Al]$\sim$5) and iron ([Fe]$\sim$4) abundances.   
Both produce the observed Al\,{\sc ii} lines and
Fe\,{\sc ii} saw tooth pattern (in the longer wavelengths). 
The question then arises as to whether IR Com  should be 
modeled with or without a mask.  
However, first, for reason we explained above,  
it is unlikely that iron abundance in the WD or donor
star is larger than 3 (in solar units), and, second, these  
Al and Fe absorption features were  
observed in U Gem at orbital phases $\Phi=0.25,0.67-0.81$ 
and were attributed to veiling material 
\citep[][note that the line identified as a WD argon line 
in that work is actually the Al\,{\sc ii}  1670.79~\AA\  curtain line]{god17}.  
Consequently, we conclude that the WD in IR Com has Al and Fe solar abundances and
is veiled by a thin iron curtain.

\subsection{{\bf Conclusion.}} 
We confirm that with the availability of accurate Gaia distances, 
FUV spectral fits  provide a new tool to derive the WD mass and temperature 
in CVs, even for systems where the WD is significantly veiled.   
We made the assumption that the veiling material and the WD have the same abundances, 
and found that this assumption is self-consistent.  
Overall, our results further strengthen previous findings that DNe in quiescence
have WDs with subsolar abundances of carbon and silicon, as well as possibly subsolar
abundances of phosphorus and sulfur.  All the derived projected stellar rotational 
velocities we found are well below the Keplerian break-up velocities. 
We also raise the possibility that 
WD suprasolar abundance of aluminum, nitrogen  and iron, derived from fitting
FUV absorption lines in some systems, may rather be due to  the presence of a  
thin iron curtain.  Alas, nitrogen abundance could not be derived   
to assess the N/C ratio, as dominant nitrogen lines only form at shorter wavelengths, 
requiring FUSE or e.g. COS FUV G140L (1280~\AA ) spectral analyses. 
At these wavelengths ($\lambda < 1170$~\AA ) veiling material can also produce strong absorption
bands of Fe\,{\sc ii} as well as deeper absorption lines of nitrogen, making such 
a task more challenging. 

The analysis of CV white dwarf metal abundances and their 
implications represents a relatively new frontier in CV white 
dwarf research. The principal objective is to determine the 
abundances of accreted metals for a statistically significant 
large sample of CVs. Overabundances of metals especially, 
odd-numbered species like aluminum and phosphorus are built 
up through proton captures outside the explosive CNO Bi-Cycle 
burning of past nova-explosions, contaminating the donor star 
during the nova common envelope stage which envelops the donor 
star \citep{sio14,spa21}. 
Whether this process is responsible for the N/C composition anomaly,  
or whether metal overabundances were carried over by the WD after 
the AGB thermal pulsing stage, these nuclei 
are likely the ancient relics of hot CNO Burning.

\begin{acknowledgements}
PG is pleased to thank William (Bill) P. Blair at the 
Henry Augustus Rowland Department of Physics \& Astronomy at The 
Johns Hopkins University, Baltimore, Maryland, USA, for his 
indefatigable, kind hospitality. 
This research is based on observations made with the NASA/ESA Hubble Space Telescope,
obtained from the data archive at the Space Telescope Science Institute.
Support for this research is provided by NASA through grant number 
HST-AR-16152 to Villanova University from the Space Telescope Science
Institute. 
STScI is operated by the Association of University for Research in 
Astronomy, Inc. under NASA contract NAS 5-26555.
In our research we use the AAVSO International Database and we are  
thankful to the AAVSO observers worldwide, professionals as well 
as amateurs.  
To assess distances, this work used data from the European Space Agency (ESA) mission
{\it Gaia} (\url{https://www.cosmos.esa.int/gaia}), processed by the {\it Gaia}
Data Processing and Analysis Consortium (DPAC,
\url{https://www.cosmos.esa.int/web/gaia/dpac/consortium}). Funding for the DPAC
has been provided by national institutions, in particular the institutions
participating in the {\it Gaia} Multilateral Agreement.
To assess dust reddening, 
this research has also made use of the NASA/IPAC Infrared Science Archive
which is funded by the National Aeronautics and 
Space Administration and operated by the California Institute of 
Technology. 
\end{acknowledgements} 

\facilities{
HST (COS, STIS), AAVSO, GAIA, IRSA, IRAS, COBE).
} 
\\ 
   
\software{
IRAF \citep[NOAO PC-IRAF Revision 2.12.2-EXPORT SUN;][]{tod93}, 
\textsc{tlusty} (v203) \textsc{synspec} (v48) Rotin(v4) \citep{hub17a,hub17b,hub17c}, 
\textsc{circus} \citep{hub96}, PGPLOT (v5.2), Cygwin-X (Cygwin v1.7.16),
xmgrace (Grace v2), XV (v3.10) } 
\\

\clearpage

\appendix

\section{{\bf The $\chi^2$ Maps}} 

\begin{figure}[h!] 
\epsscale{0.8} 
\plotone{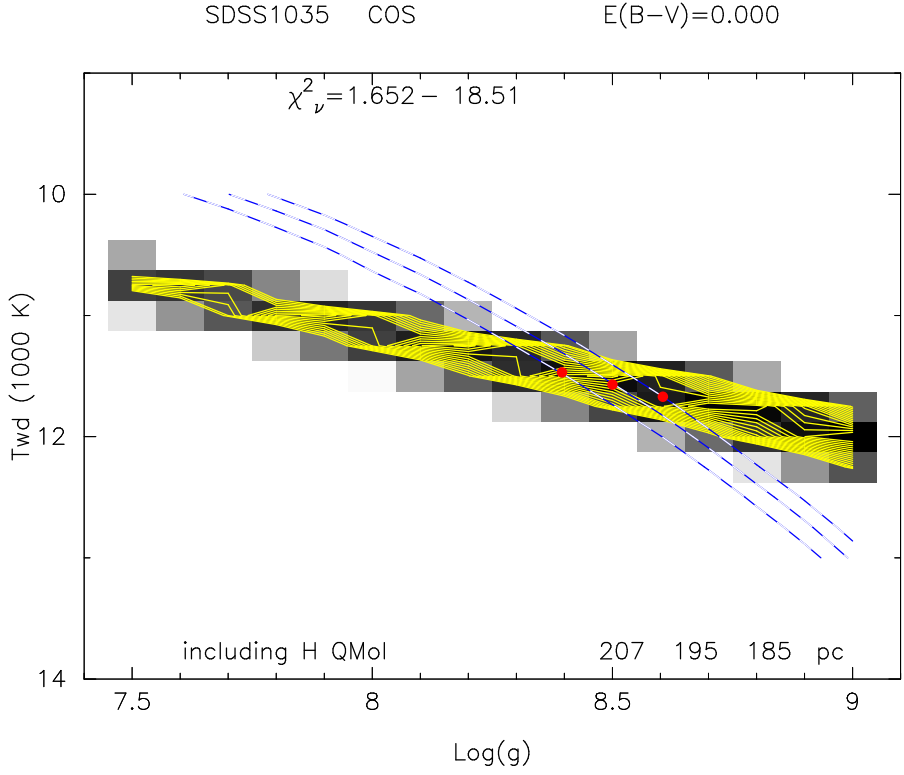}
\caption{
The results of the fitting of the COS spectrum of SDSS 1035 with a 
WD model spectrum are displayed as a map of the $\chi^2_{\nu}$
value in the $log(g)$ vs. decreasing $T_{\rm wd}$ parameter space. 
For clarity, only the lowest $\chi^2_{\nu}$ models are shown
with the gray scale and yellow contour lines (darker grey  
indicates smaller $\chi^2_{\nu}$) forming a diagonal.  
The $\chi^2_{\nu}$ value increases from $\sim 1.652$ 
(within the diagonal) to 18.51 (upper right and lower
right corner - in white) as indicated in the panel. 
The three white-blue dashed lines correspond to the distances: 
207~pc (left dashed line), 
195~pc (middle dashed line), and 185~pc (right dashed line). 
The least chi square along the line d=195 pc gives the solution
$log(g)=8.50$ and  $T_{\rm wd}=11,570$~K (middle red dot). 
The solution for 207~pc is marked with the left red dot
and that for 185~pc is marked with the right red dot.   
The solution is summarized as $log(g)=8.500 \pm 0.105$ with 
$T_{\rm wd}= 11,570 \pm 100$~K for a distance  
$d=195_{+12}^{-10}$~pc. 
The WD model has a metal abundance of 0.02 solar.   
The COS spectrum was not dereddened.  
Similar $\chi^2$ maps were also computed by running model fits
assuming $E(B-V)=0.034$, and assuming solar abundances
(separately).  
\label{sdss1035chi} 
}
\end{figure}

\begin{figure}[h!]
\epsscale{0.8}  
\plotone{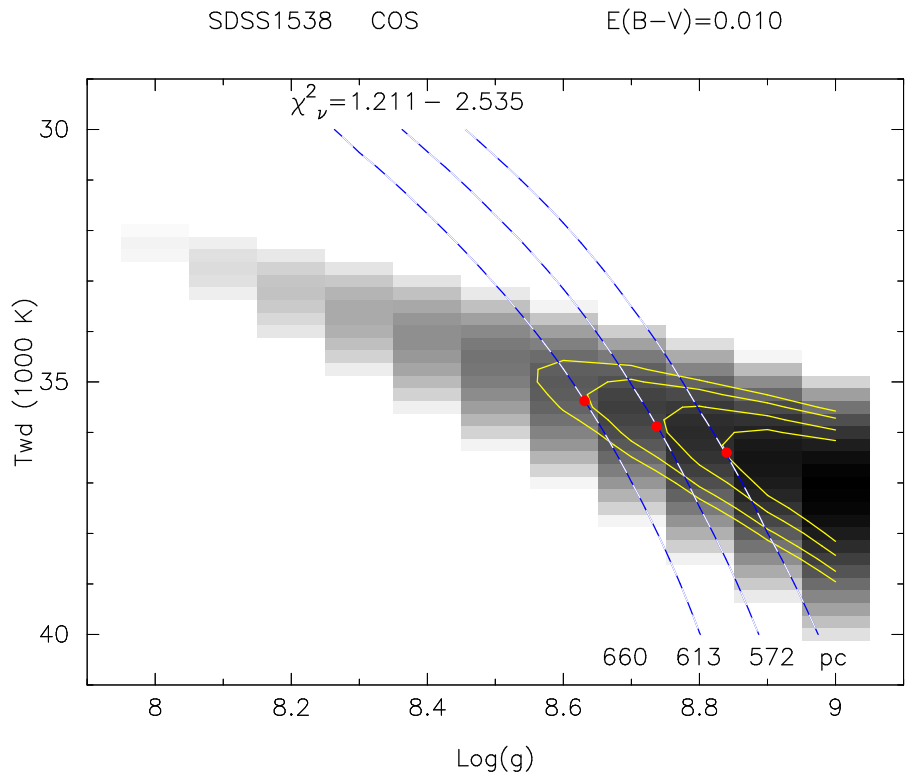}
\caption{
The $chi^2$-map of the modeling of the COS spectrum of 
SDSS 1538 (CRTS J153817.3+512338) is shown
assuming $E(B-V)=0.01$.  
The least $\chi^2_{\nu}$ model scaling to the Gaia distance of 613~pc yields  
$log(g)=8.74$ and  $T_{\rm wd}=35,875$~K (middle red spot).  
The solutions are also shown for upper (left red spot) and lower (right
red spot) limits of the Gaia distance.  
\label{sdss1538chifit2} 
}
\end{figure}

\begin{figure}[h!] 
\vspace{-2.cm} 
\epsscale{0.8} 
\plotone{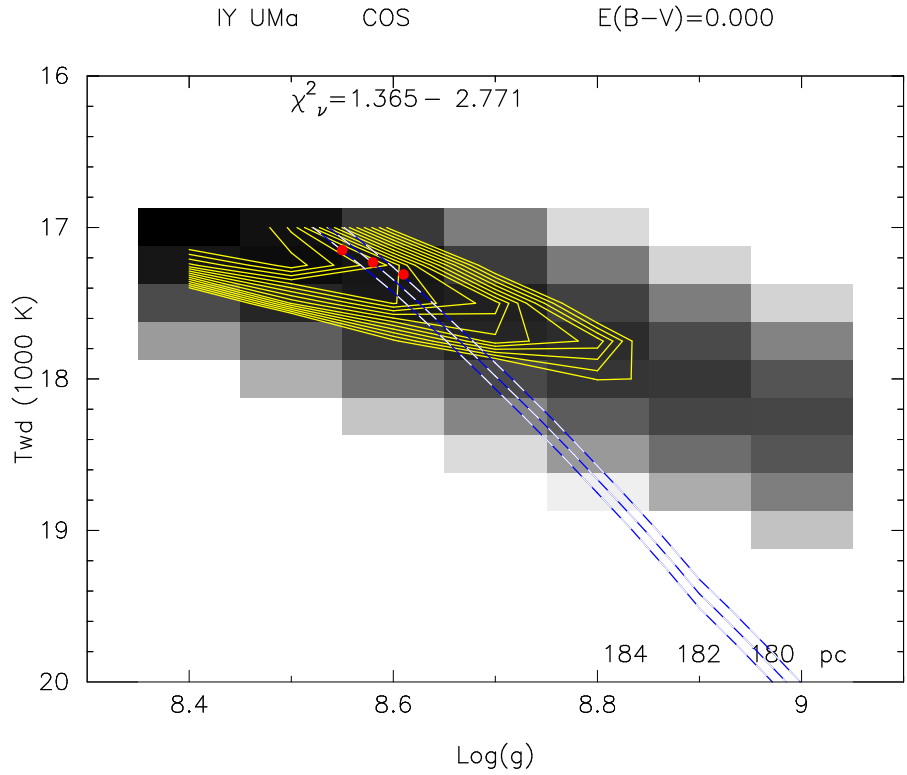}
\caption{
The $\chi^2$-map of the modeling of the second exposure of the COS spectrum 
of IY UMa is displayed. The modeling assumed  a reddening of $E(B-V)=0.0$ 
and the WD model include an iron curtain and a second flat component.
The solutions for the Gaia distance ($\pm$error) are shown with the
3 red dots. 
\label{iyumachi} 
}
\end{figure}

\begin{figure}[h!] 
\vspace{-2.cm} 
\epsscale{0.8} 
\plotone{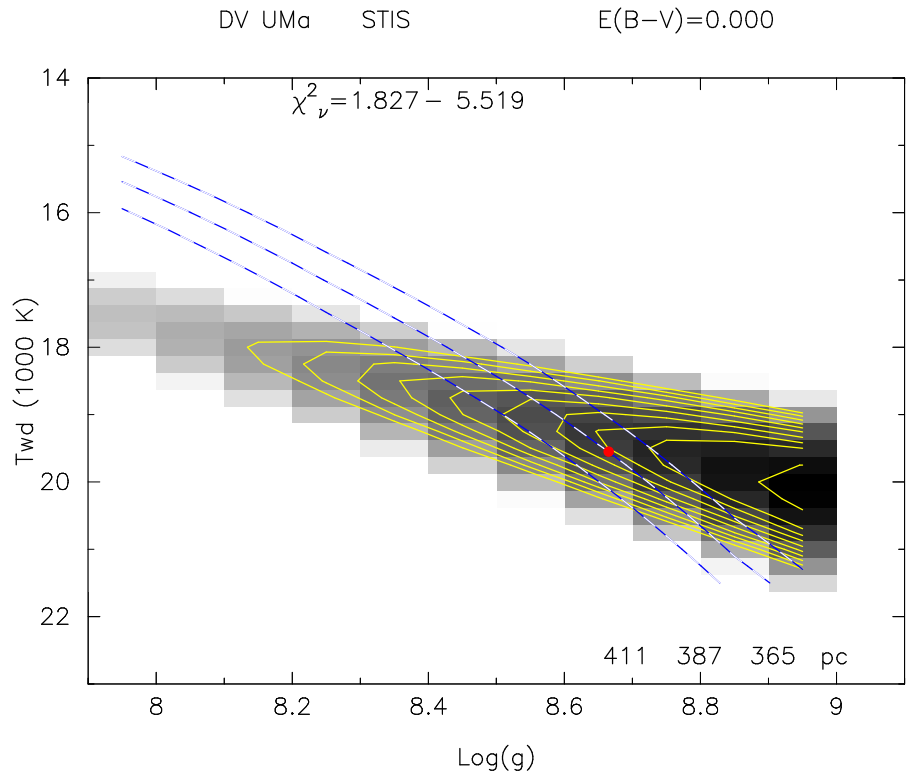}
\caption{
The $\chi^2$-map of the fitting of the STIS spectrum of 
DV UMa is shown assuming $E(B-V)=0$ and including a second
flat component.  
The solution for the Gaia distance of 387~pc is marked with the red dot:
$T_{\rm wd}=20,050$~K with $log(g)=8.715$.
\label{dvumachi} 
}
\end{figure}

\begin{figure}[h!]  
\epsscale{0.8} 
\plotone{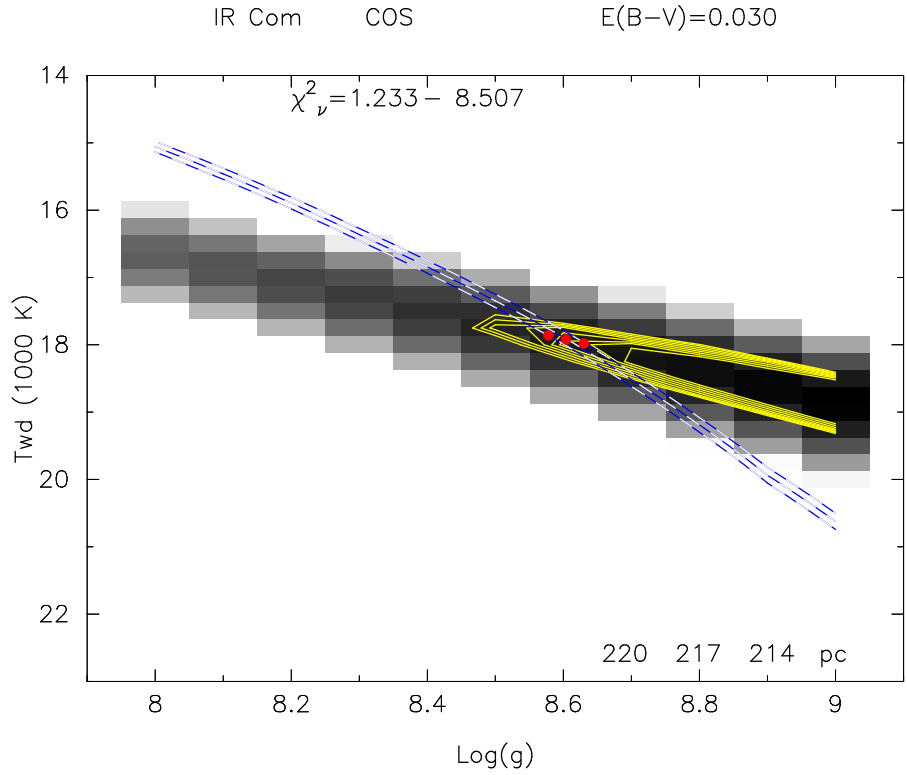}  
\caption{ 
The $\chi^2$-amp of the results of the spectral fit to the COS spectrum of IR Com
is presented assuming $E(B-V)=0.03$ and a second flat component.  
The models fitting the Gaia distance ($\pm$error) are found along the triple
blue dashed line and achieve a minimum $\chi^2_{\nu}$ 
denoted by the three red dots. 
\label{ircomchiebv030}
}
\end{figure}

\begin{figure}[h!] 
\epsscale{0.8} 
\plotone{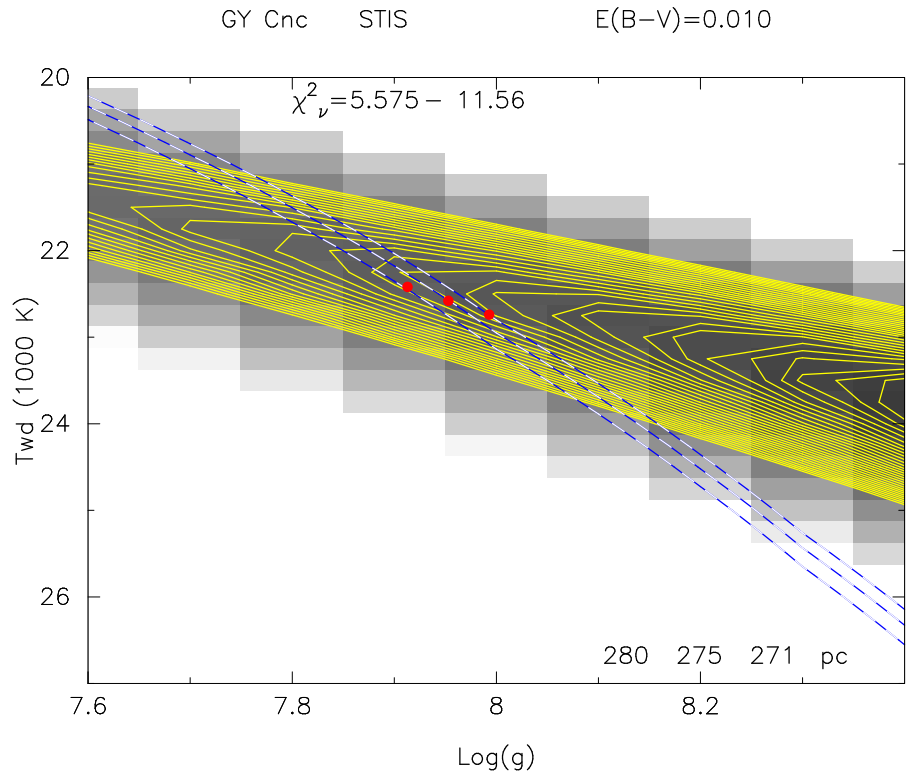} 
\caption{
The results of the fitting of the STIS spectrum of GY Cnc with a 
veiled WD model spectrum is displayed as a map of the $\chi^2_{\nu}$
value in the $log(g)$ vs. decreasing $T_{\rm wd}$ parameter space. 
We assume a reddening of $E(B-V)=0.010$ and include 
a second flat component as well as an absorbing slab.
\label{gycncchi} 
}
\end{figure}

\begin{figure}[h!]
\epsscale{0.8}  
\plotone{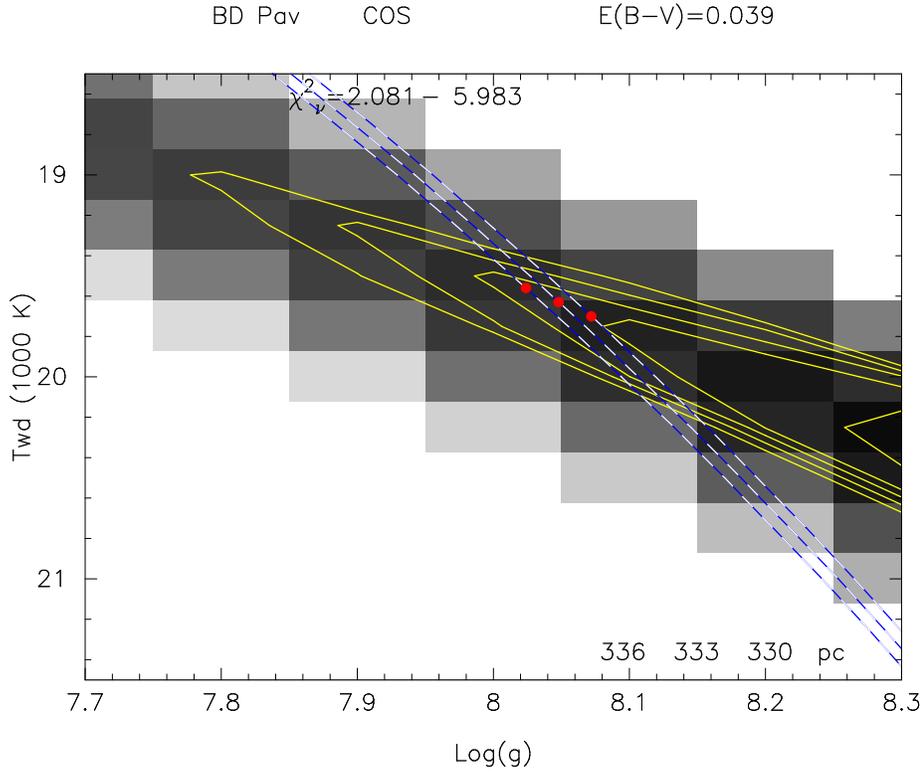} 
\caption{ 
The results of the first step in the  modeling of the COS spectrum of 
BD Pav are shown as a map of the $\chi^2_{\nu}$ in the parameter
space $T_{\rm wd}$ vs. $log(g)$. No second component and no 
iron curtain are included in this first step, the spectrum
was dereddened assuming $E(B-V)=0.039$.  
The best fit scaling to the Gaia distance yields  
$log(g)=8.048$ and  $T_{\rm wd}=19,630$~K (middle red spot). 
After the inclusion of a second component and an absorbing slab,
and taking into account the upper limit of the reddening
($E(B-V)=0.075$), 
the best fit temperature and gravity change noticeably 
(see the results section).  
\label{bdpavchifit} 
}
\end{figure}

\begin{figure}[h!] 
\epsscale{0.8}  
\plotone{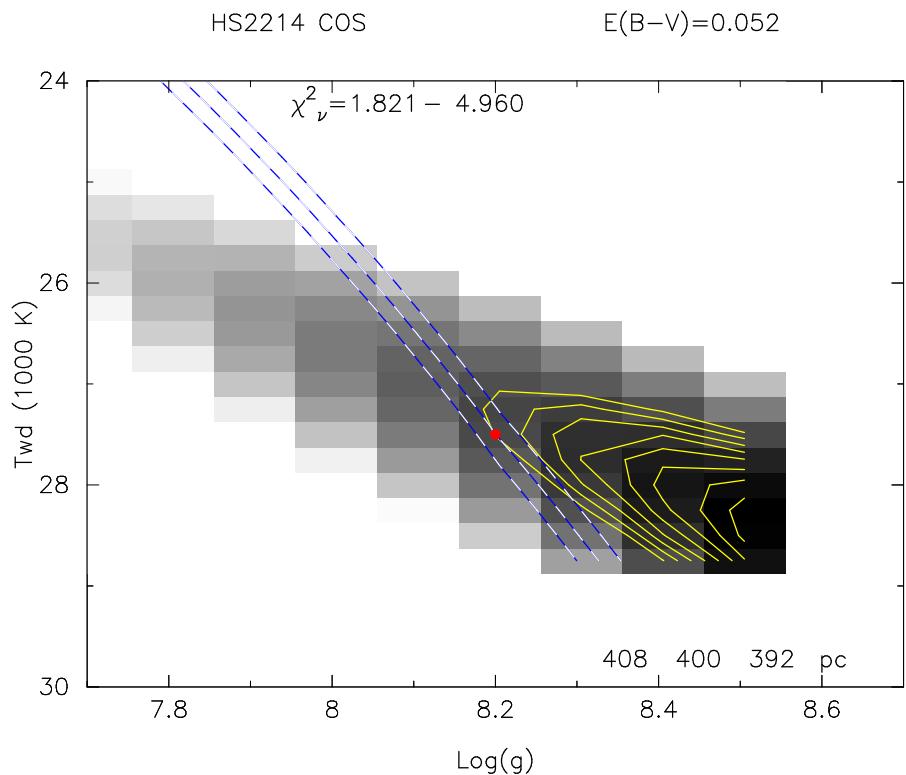}
\caption{
The results of the spectral fit of the COS spectrum of HS 2214+2845,
assuming a reddening of $E(B-V)=0.052$,are
summarized in this map of the values taken by the reduced 
$\chi^2_{\nu}$. 
The least chi square along the line d=400 pc gives the solution
$T_{\rm wd}=27,500$~K with $log(g)=8.20$ (red dot).  
Note that the distance lines (blue dashed lines) 
achieve a minimum $\chi^2_{\nu}$ slightly off the center
of the gray diagonal. 
\label{hs2214chi}
}
\end{figure}

\begin{figure}[h!]
\epsscale{0.8}  
\plotone{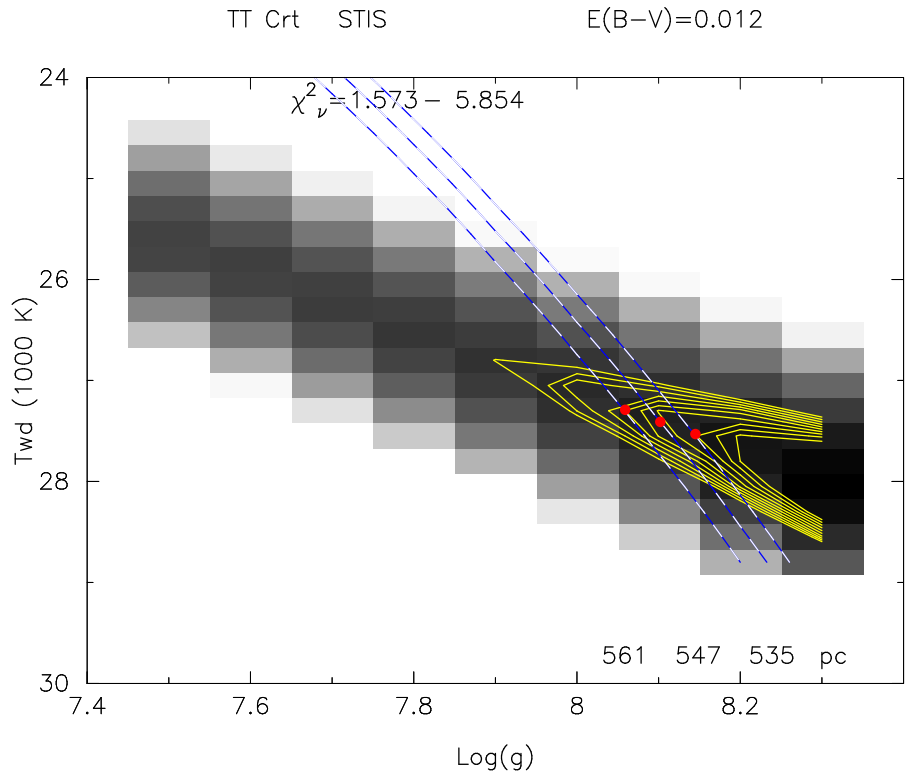}
\caption{
The $\chi^2$-map of results of the spectral fit of the STIS spectrum of TT Crt 
is displayed assuming $E(B-V)=0.012$.  
The least chi square along the line d=547 pc gives the solution
$T_{\rm wd}=27,290$~K with $log(g)=8.102$ (middle red dot).  
\label{ttcrtchi}
}
\end{figure}

\begin{figure}[h!]
\epsscale{0.8} 
\plotone{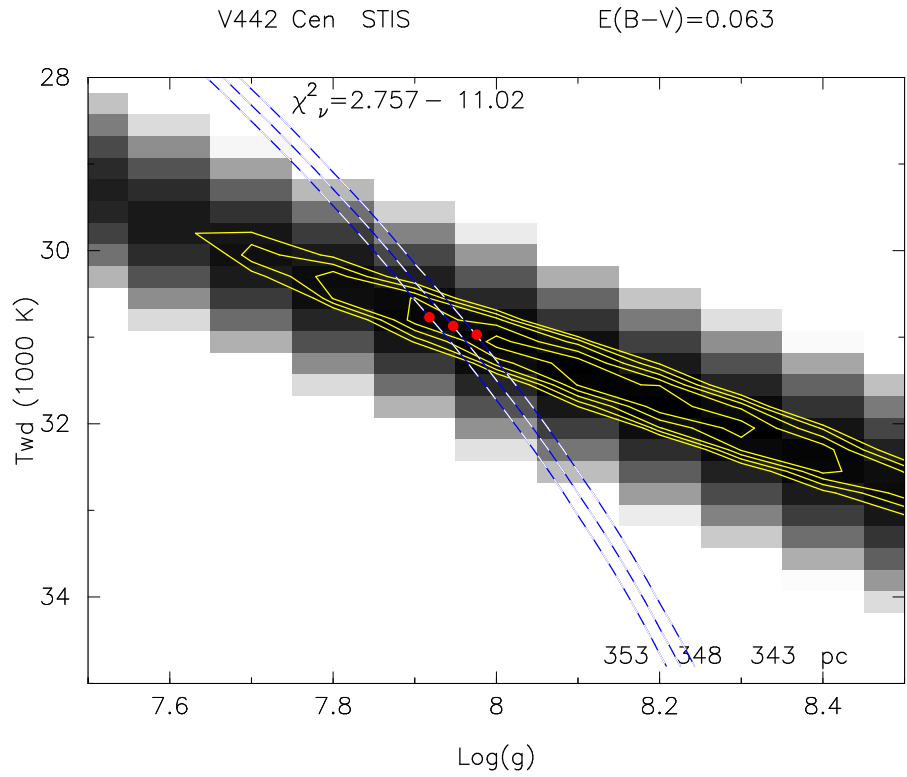}
\caption{
The $\chi^2$-map of the spectral fit to the STIS spectrum of 
V442 Cen assuming E(B-V)=0.063 is displayed. 
Solar abundances were assumed in the modeling, which did
not include a second component nor an absorbing slab.  
\label{v442cenchi} 
}
\end{figure}

We generate a grid of WD model spectra in a confined 
region of the gravity-temperature parameter space,  
ranging from $log(g)=7.5$ to $log(g)=9.0$ and $T_{\rm wd}=10,000$
to $T_{\rm wd}=40,000$, 
in steps of 250~K in temperature and $0.1$ in gravity ($log(g)$). 
In order to derive the WD temperature and gravity,   
each HST spectrum is fitted against a portion of this
grid, a subgrid, typically including approximately 100 to 400 WD models. 
For each single model fit (with a given WD temperature and gravity) 
within a subgrid, a $\chi^2_{\nu}$ value as well as a scaling distance is obtained.  
The results for all the model fits within that subgrid, i.e. for a given 
DN system, are then summarized  as a map of $\chi^2_{\nu}$ in the parameter
space $log(g)$ vs. $T_{\rm wd}$.
Such a $\chi^2$-map is used to visually display and derive the 
best-fit WD model for the Gaia distance.  
In this section of the Appendix we present such $\chi^2$-maps 
for the 10 DNe systems in Figs.\ref{sdss1035chi}-\ref{v442cenchi}. 

In each $\chi^2$-map, 
we draw yellow contour lines of the $\chi^2_{\nu}$ values superposed to
a gray scale of the least $\chi^2_{\nu}$ 
in the (decreasing) WD effective surface temperature ($T_{\rm wd}$) versus 
(increasing) WD surface gravity ($log(g)$) parameter space.
Each theoretical stellar spectrum model occupies a 
250 K $\times 0.1 (log(g))$ rectangle area as shown in the figures. 
For clarity, only the lowest $\chi^2_{\nu}$ models are shown
with the gray scale (darker indicates smaller $\chi^2_{\nu}$) 
and yellow contour lines,  forming a diagonal in the 
($log(g),T_{\rm wd}$) parameter space. 
The range of the $\chi^2_{\nu}$ values is indicated in each figure. 

Since a distance is also obtained for each model (rectangle), 
we draw lines of constant distance corresponding to the 
Gaia distance, including its upper and lower limits: 
the three blue dashed lines.

In each $\chi^2$-map, the temperature of the WD is then found 
where the $\chi^2_{\nu}$ takes a minimum along the Gaia distance
line. This minimum is achieved in the region where the blue 
dashed distance line intersects the least-$\chi^2_{\nu}$ gray diagonal
and is marked with red dot.   

The $\chi^2$ maps presented here, however, are only a small sample representative 
of all the model fits that were run. Only one $\chi^2$ map is presented
for each DN system, but many more were computed for different values of
the reddening and of the second component flux level, as well as for different
iron curtain models (where applicable). As such the temperature and 
gravity solutions displayed here in these $\chi^2$-map are not the final results presented in
Table \ref{mwdtwd}, but they contributed to them (i.e. the temperature
and gravity solutions in the following $\chi^2$ are within the error
bars of the final results presented in Table \ref{mwdtwd}).

In all the $\chi^2$-maps, low $\chi^2$
values are obtained on a diagonal starting at a colder temperature 
at the lower limit of $log(g)$ (near the upper left corner of the maps),  
and crossing to a hotter temperature at the upper limit of $log(g)$
(near the lower right corner in the maps). 
This diagonal corresponds to the fitting of the hydrogen Lyman$\alpha$ 
absorption feature, and is a result of the well-known degeneracy of the solution
\citep[e.g.][]{gan05}, due to the fact that increasing the temperature 
and increasing the gravity have opposite effects on the Ly$\alpha$ 
profile (narrowing vs. broadening, respectively).  
 
We note, however, that 7 of the 10 $\chi^2$-maps exhibit the lowest $\chi^2$ 
(absolute minimum, darkest rectangle) at the very end (higher temperature/gravity) of that
diagonal. 
A reasonable cause could be that  a ``hot'' second component (rather than a flat one) 
is responsible for this effect, but it appears that this effect also occurs in
cases where the slope of the observed spectrum is shallower (colder) than the
best-fit model, and while such a second component slightly reduces the effect
in a system like HS 2214+2815, it does not provide a satisfactory explanation. 
We also notice that some systems (e.g. DV UMa) exhibit some extra flux in the very short wavelengths
($< 1200$~\AA ), which could partially explain this effect. 
For IY UMa, the opposite is certainly true: the COS spectrum presents some extra flux 
on the right wing of Ly$\alpha$ when compared to the models, and the least $\chi^2_{\nu}$
is obtained near the low temperature/gravity of the diagonal in its $\chi^2$-map.  
Last, we find the higher temperature/gravity models have in general shallower absorption 
lines, and since most systems have rather low abundances, a slightly better fit
is obtained at higher temperature (in spite of the fact that we mask the dominant
absorption feature in the first place). 
Except for SDSS 1035, all the $chi^2$-maps 
were generated using solar abundance WD models [generating non-solar abundance $\chi^2$-map requires
generating grid of models for all the different abundances obtained in this work,
which would be prohibitively CPU-expensive], and the $\chi^2$-map of SDSS 1035  
almost does not show that effect.  
Consequently, we conclude that this effect is most likely a combination of the above causes, 
and it is further affected by the veiling material strongly masking half of the systems.  
In any case, the relative difference in the $\chi^2_{\nu}$ 
obtained for the Gaia distance best-fit is only of the order of 1\% larger than for the absolute
minimum $\chi^2_{\nu}$.

\subsection{{\bf SDSS1035 }}

\clearpage

\subsection{{\bf SDSS 1538+5123 }}

\clearpage

\subsection{{\bf IY UMa }}

\clearpage

\subsection{{\bf DV UMa }}

\clearpage

\subsection{{\bf IR Com }}

\clearpage

\subsection{{\bf GY Cnc }}

\clearpage

\subsection{{\bf BD Pav }}

\clearpage

\subsection{{\bf HS 2214+2845 }}

\clearpage

\subsection{{\bf TT Crt }}

\clearpage

\subsection{{\bf V442 Cen}}

\clearpage

\begin{center} 
{\bf{ ORCID iDs}} 
\end{center} 
Patrick Godon \url{https://orcid.org/0000-0002-4806-5319}  \\ 
Edward M. Sion \url{https://orcid.org/0000-0003-4440-0551}


\begin{thebibliography}{}

\bibitem[Avni(1976)]{avn76}
Avni, Y. 1976, \apj, 210, 642  

\bibitem[Bailer-Jones(2015)]{bai15}
Bailer-Jones, C.A.L. 2019, \pasp, 127, 994 


\bibitem[Bohlin et al.(2014)]{boh14}
Bohlin, R.C., Gordon, K.D., Tremblay, P.-E. 2014, \pasp, 126, 711 

\bibitem[Bohlin et al.(1978)]{boh78}
Bohlin, R.C., Savage, B.D., \& Drake, J.F. 1978, \apj, 224, 132 


\bibitem[Capitanio et al.(2017)]{cap17}
Capitanio, L, Lallement, R., Vergely, J.-L. et al.  2017, \aap, 606, 65 

\bibitem[Debes et al.(2016)]{deb16} 
Debes, J.H., Becker, G., Roman-Duval, J. et al. 2016, 
Instrument Science Report COS 2016-15 (v1) 


\bibitem[La Dous(1991)]{lad91}
La Dous, C. 1991, \aap, 252, 100 

\bibitem[Feline et al.(2005)]{fel05}
Feling, W.J., Dhillon, V.S., March, T.R., Watson, C.A. Littlefair, S.P.
2005, \mnras, 364, 1158 

\bibitem[Feltzing \& Chiba(2013)]{fel13} 
Feltzing, S., \& Chiba, M. 2013, New Astronomy Reviews, 57, 80. 

\bibitem[Fitzpatrick \& Massa(2007)]{fit07}
Fitzpatrick, E.L., \& Massa, D. 2007, \apj, 663, 320 

\bibitem[Frank et al.(2002)]{fra02}
Frank, J., King, A., \& Raine, D.J. 2002, Accretion Power in Astrophysics
(3rd ed.; Cambridge Univ. Press) 

\bibitem[Friend et al.(1990)]{fri90}
Friend, M.T., Martin, J.S., Smith, R.C., \& Jones, D.H.P. 1990,
\mnras, 246, 637 

\bibitem[Brown et al.(2021)]{bro21}
Gaia Collaboration; Brown A.G.A., Vallenari, A., Prusti, T. et al. 
(and 422 more) 2021,\aap, 649, 1 

\bibitem[G\"ansicke et al.(2003)]{gan03}
G\"ansicke, B.T., Szkody, P., de Martino, D. et al. 2003, \apj, 594, 443 

\bibitem[G\"ansicke et al.(2005)]{gan05}
G\"ansicke, B.T., Szkody, P., Howell, S.B. \& Sion, E.M. 2005, \apj, 629, 451   

\bibitem[Godon(2019)]{god19}
Godon, P. 2019, \apj, 870, 112 

\bibitem[Godon \& Sion(2021)]{god21}
Godon, P., \& Sion, E.M. 2021, \apj, 908, 173 

\bibitem[Godon et al.(2007)]{god07}
Godon, P., Sion, E.M., Barrett, P.E., \& Szkody, P. 2007, \apj, 656, 1092 

\bibitem[Godon et al.(2017)]{god17}
Godon, P., Shara, M.M., Sion, E.M., \& Zurek, D. 2017, \apj, 850, 146 

\bibitem[Godon et al.(2012)]{god12}
Godon, P., Sion, E.M., Levay, K. et al. 2012, \apjs, 203, 29 


\bibitem[Hack \& La Dous(1993)]{hac93}
Hack, M., \& La Dous C. 1993, {\it Cataclysmic Variables \& Related Objects},
(NASA SP-507)/US Gov.Printing Office 

\bibitem[Harrison et al.(2004)]{har04} 
Harrison, T.E., Osborne, H.L., Howell, S.B. 2004, \aj, 127, 3493  

\bibitem[Harrison et al.(2005)]{har05} 
Harrison, T.E., Osborne, H.L., Howell, S.B. 2005, \aj, 129, 2400  

\bibitem[Horne et al.(1994)]{hor94}
Horne, K., Marsh, T.R., Cheng, F.H., Hubeny, I., Lanz, T. 1994, \apj, 426, 294 

\bibitem[Howell et al.(2010)]{how10}
Howell, S.B., Harrison, T.E., Szkody, P., \& Silverstri, N.M. 2010, \aj, 139, 1771 

\bibitem[Hubeny (1988)]{hub88}
Hubeny, I. 1988, CoPhC, 52, 103 

\bibitem[Hubeny \& Heap(1996)]{hub96} 
Hubeny, I. \& Heap S.R. 1996, \apj, 470, 1144 

\bibitem[Hubeny \& Lanz(1995)]{hub95}
Hubeny, I., \& Lanz, T. 1995, \apj, 439, 875 

\bibitem[Hubeny \& Lanz(2011)]{hub11}
Hubeny, I., \& Lanz, T. 2011, SYNSPEC, Astrophys. Source Code 
Library, 1189, 822                   

\bibitem[Hubeny \& Lanz(2017a)]{hub17a}
Hubeny, I., \& Lanz, T. 2017a, A Brief Introductory Guide to TLUSTY
and SYNSPEC, arXiv:1706.01859 

\bibitem[Hubeny \& Lanz(2017b)]{hub17b}
Hubeny, I., \& Lanz, T. 2017b, TLUSTY User's Guide II: Reference Manual, 
arXiv:1706.01935 

\bibitem[Hubeny \& Lanz(2017c)]{hub17c}
Hubeny, I., \& Lanz, T. 2017c, TLUSTY User's Guide III: Operational Manual, 
arXiv:1706.01937 

\bibitem[Hubeny et al.(1994)]{hub94}
Hubeny, I., Lanz, T., \& Jeffery, C.S. 1994, in Newsletter on Analysis of 
Astronomical Spectra No. 20, ed. C.S. Jeffery, St. Andrews Univ., p.30  

\bibitem[Iliadis(2018)]{ili18}    
Iliadis, C. 2018, private communication

\bibitem[Kimura et al.(2018)]{kim18} 
Kimura, M., Kato, T., Maehara, H., 
et al. 2018, \pasj, 70, 78 

\bibitem[Lampton et al.(1976)]{lam76}
Lampton, M., Margon, B., Bowyer, S. 1976, \apj, 208, 177 
 
\bibitem[Lauffer et al.(2018)]{lau18}    
Lauffer, G., Romero, A.D., Kepler, S.O. 2018, \mnras, 480, 1547

\bibitem[Lindergren et al.(2018)]{lin18}
Lindergren, L, Hern\'andez, J., Bombrun, A. et al. 2018, \aap, 616, 2 

\bibitem[Lubow(1989)]{lub89}
Lubow, S.H. 1989, \apj, 340, 1064 

\bibitem[Littlefair et al.(2008)]{lit08}
Littlefair, S.P., Dhillon, V.S., Marsh, T.R., G\"ansicke, B.T., 
Southworth, J. et al. 2008, \mnras, 388, 1582  

\bibitem[Livio \& Pringle(1998)]{liv98}    
Livio, M. \& Pringle, J. 1998, \apj, 503, 339   

\bibitem[Long \& Gilliland(1999)]{lon99}
Long, K.S., \& Gilliland, R.L. 1999, \apj, 511, 916 

\bibitem[Long et al.(2006)]{lon06}
Long, K.S., Brammer, G., \& Fronning, C.S. 2006, \apj, 648, 558 

\bibitem[Long et al.(2009)]{lon09}
Long, K.S. G\"ansicke, B.T., Knigge, C. et al. 2009, \apj, 697, 1512 

\bibitem[Luri et al.(2018)]{lur18}
Luri, X., Brown, A.G.A., Sarro, L.M. et al. 2018, \aap, 616, 9 

\bibitem[Manser \& G\"ansicke(2014)]{man14}
Manser, C.J., \& G\"ansicke, B.T. 2014, \mnras, 442, L23 

\bibitem[McAllister et al.(2019)]{mca19}
McAllister, M., Littlefair, S.P., Parsons, S.G. et al. 2019, \mnras, 486, 5535 

\bibitem[Narayan \& Popham(1989)]{nar89} 
Narayan, R.\& Popham, R. 1989, \apj, 346, L25 
 
\bibitem[Nelemans et al.(2016)]{nel16}
Nelemans, G., et al. 2016, \apj, 817, 69 

\bibitem[Pala(2021)]{pal21a} 
Pala, A.F. 2021, private communication 

\bibitem[Pala et al.(2017)]{pal17}
Pala, A.F., G\"ansicke, B.T., Townsley, D., Boyd, D., Cook, M.J. 
et al. 2017, \mnras, 466, 2855 

\bibitem[Pala et al.(2021)]{pal21}
Pala, A.F., G\"ansicke, B.T., Belloni, D., Parsons, S.G., Marsh, T.R. 
et al. 2022, \mnras, 510, 611  

\bibitem[Ramsay et al.(2017)]{ram17}
Ramsay, G., Schreiber, M.R., G\"ansicke, B.T. et al. 2017, \aap, 604, 107 

\bibitem[Rebassa-Mansergas et al.(2021)]{reb21} 
Rebassa-Mansergas, A., Maldonado, J., Raddi. R., Knowles, A.T., Torres, S. 
et al., 2021, \mnras, 505, 3165 

\bibitem[Redfield \& Linsky(2004)]{red04}
Redfield, S., \& Linsky, J.L. 2004, \apj, 602, 776 

\bibitem[Ritter \& Kolb(2003)]{rit03} 
Ritter, H., Kolb, U. 2003, \aap, 404, 301 (update RKcat7.24, 2016)  

\bibitem[Sasseen et al.(2002)]{sas02}
Sasseen, T.P., Hurwitz, M., Dixon, W.V., Airieau, S. 2002, \apj, 566, 267 

\bibitem[Savage \& Mathis(1979)]{sav79} 
Savage, B.D., \& Mathis, J.S. 1979, \araa, 17, 73 

\bibitem[Savoury et al.(2011)]{sav11}
Savoury, C.D.J., Littlefair, S.P., Dhillon, V.S. et al. 2011, \mnras, 415, 2025 

\bibitem[Selvelli \& Gilmozzi(2013)]{sel13}
Sevelli, P., \& Gilmozzi, R. 2013, \aap, 560, 49 

\bibitem[Schaefer(2018)]{sch18}
Schaefer, B.E. 2018, \mnras, 481, 3033 

\bibitem[Schlegel(1998)]{sch98} 
Schlegel, D.J., Finkbeiner, D.P., \& Davis, M. 1998, \apj, 500, 525  

\bibitem[Schlafly \& Finkbeiner(2011)]{sch11}
Schlafly, E.F., \& Finkbeiner, D.P. 2011, \apj, 737, 103 

\bibitem[Schreiber et al.(2016)]{sch16}
Schreiber, M. et al. 2016, \mnras, 455, L16   

\bibitem[Sion(1998)]{sio98} 
Sion, E.M. 1998, in Chan K.L., Cheng K.S., Singh H.P., eds, 
ASP Conf. Ser. Vol.138, 1997 Pacific Rim Conferesnce on Stellar
Astrophysics. Astron. Soc. Pac., San Francisco, p. 317 

\bibitem[Sion(1999)]{sio99}
Sion, E.M. 1999, \pasp, 111, 532 

\bibitem[Sion et al.(1997)]{sio97}
Sion, E.M., Cheng, F.H., Sparks, W.M. et al. 1997, \apj, 480, L17  

\bibitem[Sion et al.(1998)]{sioetal98} 
Sion, E.M., Cheng, F.H., Szkody, P. et al. 1998, \apj, 496, 449  

\bibitem[Sion et al.(2008)]{sio08}
Sion, E.M., G\"ansicke, B.T., Long, K.S. et al. 2008, \apj, 681, 543 

\bibitem[Sion et al.(1995b)]{sio95b}
Sion, E.M., Huang, M., Szkody, P., \& Cheng, F.H. 1995, \apj, 445, L31  

\bibitem[Sion \& Sparks (2014)]{sio14}
Sion, E.M. \& Sparks, W.M. 2014, \apj, 796, L10

\bibitem[Sion et al.(1995a)]{sio95a} 
Sion, E.M., Szkody, P., Cheng, F.H., \& Huang, M. 1995, \apj, 444, L97 

\bibitem[Sparks \& Sion(2021)]{spa21}
Sparks, W.M. \& Sion, E.M. 2021, \apj, 914, 5 

\bibitem[Szkody et al.(2021)]{szk21}
Szkody, P., Godon, P., G\"ansicke, B.T., et al. 2021, \apj, 914, 40 

\bibitem[Thorstensen et al.(2004)]{tho04}
Thorstensen, J.R., Fenton, W.H., Taylor, C.J. 2004, \pasp, 116, 300 
 
\bibitem[Tody(1993)]{tod93} 
Tody, D. 1993, in ASP Conf. Ser. 52, Astronomical Data Analysis
Software and Systems II, ed. R.J. Hanisch, R.J.B. Brissenden, 
\& J. Barnes (San Fransisco, CA;ASP), 173 


\bibitem[Warner(1995)]{war95}
Warner, B. 1995, {\it Cataclysmic Variable Stars} 
Cambridge Astrophysics Series, 28 (Cambridge University Press: Cambridge) 

\bibitem[Wood(1995)]{woo95}
Wood, M.A. 1995, in Proc. 9th Europ. Workshop on WDs, 
443, White Dwarfs, ed. D. Koester \& K. Werner (Berlin: Springer), 41 

\bibitem[Yoo \& Langer(2004)]{yoo04} 
Yoon,S.-C., Langer, N. 2004, \aap, 419, 645  

\bibitem[Zorotovic et al.(2020)]{zor20} 
Zorotovic, M. \& Schreiber, M. 2020, Advances in Space Research, 
66, 1080 

\end{thebibliography}
\end{document}